\RequirePackage{fix-cm}
\RequirePackage{fixltx2e}
\documentclass[12pt,twoside,english,aip, jcp,floatfix]{revtex4-1}

\usepackage[T1]{fontenc}
\usepackage[latin9]{inputenc}
\usepackage{geometry}
\usepackage{xcolor}
\usepackage{babel}
\usepackage{float}
\usepackage{amsmath}
\usepackage{amssymb}
\usepackage{graphicx}
\usepackage{wasysym}
\usepackage[unicode=true,pdfusetitle,
 bookmarks=false,
 breaklinks=true,pdfborder={0 0 0},pdfborderstyle={},backref=false,colorlinks=true]
 {hyperref}
\hypersetup{
 citecolor = blue, linkcolor = blue,  urlcolor  = blue}

\makeatletter

\let\SF@@footnote\footnote
\def\footnote{\ifx\protect\@typeset@protect
    \expandafter\SF@@footnote
  \else
    \expandafter\SF@gobble@opt
  \fi
}
\expandafter\def\csname SF@gobble@opt \endcsname{\@ifnextchar[
  \SF@gobble@twobracket
  \@gobble
}
\edef\SF@gobble@opt{\noexpand\protect
  \expandafter\noexpand\csname SF@gobble@opt \endcsname}
\def\SF@gobble@twobracket[#1]#2{}
\providecommand{\tabularnewline}{\\}
\floatstyle{ruled}
\newfloat{algorithm}{H}{loa}
\providecommand{\algorithmname}{Algorithm}
\floatname{algorithm}{\protect\algorithmname}

\renewcommand{\textendash}{--}

\usepackage{orcidlink}

\makeatother

\usepackage{listings}

\begin{document}

\title{Prediction uncertainty validation for computational chemists}

\author{Pascal PERNOT \orcidlink{0000-0001-8586-6222}}

\affiliation{Institut de Chimie Physique, UMR8000 CNRS,~\\
Universit\'e Paris-Saclay, 91405 Orsay, France}
\email{pascal.pernot@cnrs.fr}

\begin{abstract}
\noindent Validation of prediction uncertainty (PU) is becoming an
essential task for modern computational chemistry. Designed to quantify
the reliability of predictions in meteorology, the \emph{calibration-sharpness}
(CS) framework is now widely used to optimize and validate uncertainty-aware
machine learning (ML) methods. However, its application is not limited
to ML and it can serve as a principled framework for any PU validation.
The present article is intended as a step-by-step introduction to
the concepts and techniques of PU validation in the CS framework,
adapted to the specifics of computational chemistry. The presented
methods range from elementary graphical checks to more sophisticated
ones based on local calibration statistics. The concept of \emph{tightness},
is introduced. The methods are illustrated on synthetic datasets and
applied to uncertainty quantification data issued from the computational
chemistry literature.
\vskip 1cm
\noindent\em{This is the author's version of the manuscript accepted for publication in The Journal of
Chemical Physics (AIP). The published version can be accessed at} \url{https://doi.org/10.1063/5.0109572} 
\end{abstract}
\maketitle

\section{Introduction}

Uncertainty quantification (UQ) is becoming a major issue for chemical
machine learning (ML),\citep{Weymuth2022} notably for the prediction
of molecular and material properties.\citep{Janet2019,Musil2019,Scalia2020,Tran2020,Wang2021,Zhan2021,Imbalzano2021,Busk2022,Hu2022}
This is also the case for quantum chemistry, when a level of confidence
on predictions is sought out.\citep{Pernot2015,DeWaele2016,Proppe2016,Pernot2017,Proppe2017,Pernot2017b,Proppe2019a,Lejaeghere2020,Proppe2022,Pernot2022a,Reiher2022,Weymuth2022}
In these contexts, the validation of UQ outputs is essential to enable
their use in applications such as active learning or actionable predictions
for the industry. 

The practice of validation methods in the computational chemistry
(CC) UQ literature is quite diverse: from absent to elaborate through
inappropriate. Even appropriate methods are found in several variants.
There is clearly a lack of uniformity and of well-defined reference
methods. The \emph{calibration-sharpness} (CS) framework\citep{Gneiting2014}
provides a principled approach to ML-UQ validation.\citep{Tran2020,Scalia2020,Busk2022}
Scalia \emph{et al. }\citep{Scalia2020} distinguish two validation
settings: (i) \emph{confidence}- or \emph{intervals}-based calibration,\citep{Kuleshov2018}
comparing the empirical coverage of prediction intervals to their
intended confidence level; and (ii) \emph{error}-based calibration,\citep{Levi2020}
comparing errors to their predicted dispersion (\emph{variance}-based
calibration would be more appropriate,\citep{Kuppers2022} as both
validation settings are based on error statistics, and I will use
this denomination below).

In a recent article, noted thereafter PER2022,\citep{Pernot2022a}
I explored the application of the CS framework to the validation of
CC-UQ. My goal was to derive a practical set of validation tools adapted
to the specifics of CC-UQ, notably (i) the frequent use of statistical
summaries (standard or expanded uncertainties),\citep{Ruscic2014}
instead of the prediction distributions expected by the CS framework,
(ii) the possible presence of uncertainty on the reference data used
for validation, (iii) the small size of most validation datasets when
compared to ML applications, which limits the power of statistical
tests, and (iv) the non-normality of the error distributions due to
the frequent predominance of model errors.\citep{Pernot2015,Pernot2018,Pernot2020b,Pernot2021}

Considering these constraints, I was driven into considering two validation
options for calibration, based on the available information. When
\emph{expanded uncertainties} are available, such as $U_{95}$ (the
half-range of a 95\% confidence interval, as recommended in thermochemistry\citep{Ruscic2014}),
calibration should be tested by comparing the effective coverage of
the corresponding prediction intervals to the target probability.
But when \emph{standard uncertainties} are available, the best option
to avoid undue distribution hypotheses is to test the variance of
\emph{z}-scores (errors normalized by the corresponding uncertainty),
which should be 1. This dichotomy maps perfectly the settings of Scalia\emph{
et al.} \citep{Scalia2020}, although implementation details may differ.

However, average calibration of a prediction uncertainty scheme over
a validation set does not guarantee its small-scale reliability.\citep{Kuleshov2018}
When designing a prediction method, this is typically addressed by
the consideration of \emph{sharpness}, a statistic quantifying the
concentration of predictive distributions. Within a set of calibrated
method, one should prefer the sharpest one. However, even the sharpest
one might\emph{ }still fail at small-scale reliability. This led me
in PER2022 to propose \emph{local calibration} analysis schemes (LCP
and LZV methods), where calibration is assessed within subsets of
the validation set. This is a form of \emph{group calibration}\citep{Chung2021}
or \emph{multicalibration}\citep{HebertJohnson2017}. I will show
below how the LZV analysis relates also to the \emph{reliability diagrams}
introduced recently by Levi \emph{et al.}\citep{Levi2020}

Being hampered by the lack in the CS framework of a concept qualifying
small-scale or local calibration, I introduce below the \emph{tightness}
concept. As I found few to no use for \emph{sharpness} in a pure validation
context (it is mostly useful in the benchmarking or design of probabilistic
prediction methods), I mostly refer in the following to a calibration-tightness
(CT) framework.

A point that was not treated in PER2022 is the case where prediction
statistics, typically mean and standard deviation, are based on small
prediction ensembles. This is a frequent scenario in ML-UQ.\citep{Smith2018,Zheng2022}
The robustness of the calibration and tightness validation methods
in presence of this source of statistical noise needed to be studied.
Moreover, as the ML-UQ literature makes an abundant use of \emph{ranking}-based
statistics I also evaluated the interest of the \emph{correlation
coefficients} between uncertainty and absolute errors\citep{Tynes2021}
and the so-called \emph{confidence curves}\citep{Scalia2020} for
CC-UQ validation.

The next section (Sect.\,\ref{sec:Summary-of-the}) presents a short
overview of the concepts and validation methods. Its aim is to provide
a step-by-step approach to the calibration-tightness UQ validation
framework and enable, as far as possible, its use by non-statisticians.
After this, readers new to the field might like to skip the technical
sections (\ref{sec:Notations-and-definitions}-\ref{sec:Quantitative-methods})
and go directly to Section\,\ref{sec:Examples} for examples of application
to a variety of CC-UQ datasets. 

Sect.\,\ref{sec:Notations-and-definitions} introduces general definitions
and notations used throughout the study and also the synthetic datasets
used to illustrate the methods. Sect.\,\ref{sec:Graphical-methods}
presents simple graphical validation checks that do not require statistical
testing procedures. These might be used for screening out problematic
UQ outputs. Unfortunately, quantitative validation is often necessary
to conclude in situations where rejection of calibration or tightness
is not clear cut. Quantitative methods for ranking-, intervals- and
variance-based methods are presented in Sect.\,\ref{sec:Quantitative-methods},
with the necessary statistical tools and derived graphics. Sect.\,\ref{sec:Examples}
presents applications of these tools to datasets from the CC-UQ literature.
Available software implementing the CS and CT frameworks, or parts
of them, is presented in Sect.\,\ref{sec:Available-software}. A
conclusive discussion is presented in Sect.\,\ref{sec:Discussion-and-conclusion}.

\section{A short guide to CC-UQ validation\label{sec:Summary-of-the}}

This section provides a brief introduction to the concepts and methods
for UQ validation in computational chemistry, by guiding potential
users to the choice of methods best adapted to their data. Without
further delving into the technical details, readers new to this topic
should then be able to understand the case studies presented in Section\,\ref{sec:Examples},
and to appreciate the interest and usefulness of these tools. Links
to the main text are provided for each topic. For bibliographic references,
please consult the relevant sections.

To begin with, one needs a validation set, which can be as minimal
as a set of errors ($E$) and the corresponding dispersion statements.
Errors are the differences between predicted values of a quantity
of interest (QoI) $V$ and reference values, and dispersion statements
on errors can take the form of predictive distributions, prediction
ensembles or statistical summaries, typically uncertainties. Note
that these should account for the dispersion of reference values,
if any. {[}Sect.\,\ref{subsec:Notations}{]}

The goal of calibration validation is to check the \emph{statistical
consistency} between the errors and their dispersion statements. One
considers two complementary validation levels: \emph{average} \emph{calibration}
(simply referred-to below as calibration), where the statistical consistency
is checked over the whole validation set, and \emph{tightness}, where
the statistical consistency is checked at a finer scale, typically
in relevant subsets of the validation set. Calibration alone does
not guarantee the reliability of individual prediction uncertainties,
and tightness should be sought for. Note that a set of predictions
cannot be considered to be tight if it is not calibrated. {[}Sect.\,\ref{subsec:Concepts-and-definitions}{]}

Many validation methods are proposed in the literature. The most important
decision criterion to choose a pertinent method is based on the available
dispersion information. The main types occurring in CC-UQ (uncertainty,
expanded uncertainty, prediction ensembles and predictive distributions)
are considered now to present the available options.

\paragraph{Uncertainty. }

Let us consider first a very common scenario, where one has a set
of $M$ errors and uncertainties, noted $E=\left\{ E_{i}\right\} _{i=1}^{M}$
and $u_{E}=\left\{ u_{E,i}\right\} _{i=1}^{M}$. Without further characterization,
an uncertainty has to be understood as a \emph{standard }uncertainty,
i.e. the standard deviation of the possible values of the corresponding
property. Hence, the basic probabilistic model linking an error $E_{i}$
to an uncertainty $u_{E,i}$ is $E_{i}\sim D(0,u_{E,i})$, meaning
that $E_{i}$ is a random realization from an unspecified distribution
$D$, centered on 0 (errors are assumed to be corrected of systematic
bias) with scale/dispersion parameter $u_{E,i}$. One should thus
consider distribution-free validation methods, and simply answer the
question: ``Does uncertainty correctly quantify the dispersion of
the errors ?''. {[}Sect.\,\ref{subsec:Calibration}{]}
\begin{itemize}
\item If one deals with a constant value of $u_{E}$ (homoscedastic case),
one cannot do much better than to check that $u_{E}^{2}$ correctly
describes the variance of the errors set, i.e. $\mathrm{Var}(E)\simeq u_{E}^{2}$
or $\mathrm{Var}(Z)\simeq1$ for $Z_{i}=E_{i}/u_{E,i}$, within the
limits allowed by the size of the validation set $M$. Note that this
is a test for (average) calibration and this simple scenario does
not enable to assess tightness. For this, one would need to have additional
data, such as the set of predicted values $V$ from which $E$ is
derived, and check that $\mathrm{Var}(Z)\simeq1$ in relevant subsets
of $V$. This is called a \emph{local} \emph{Z-variance (LZV) analysis}.
{[}Sect.\,\ref{subsec:Local-Z-variance}{]}
\item If $u_{E}$ is not constant (heteroscedastic case), a simple graphical
method, where one plots $E$ vs $u_{E}$ can help to answer the main
question (Fig.\,\ref{fig:01}): if on such a plot the dispersion
of $E$ does not increase linearly with $u_{E}$, the statistical
consistency can be rejected without further trial. In the opposite
case, additional tests are necessary to assess calibration and tightness.
{[}Sect.\,\ref{subsec:Heteroscedastic-validation-sets}{]} For calibration,
one should check that $\mathrm{Var}(Z)\simeq1$. For tightness, validation
methods use the estimation of error statistics within subsets of sorted
$u_{E}$ values: 
\begin{itemize}
\item In the \emph{LZV analysis} {[}Fig.\,\ref{fig:05}(a){]}, one makes
bins of $u_{E}$ and one plots the value of $\mathrm{Var}(Z)$ for
each bin against the central value of the bin. For tight predictions,
all $\mathrm{Var}(Z)$ values should be close to 1. {[}Sect.\,\ref{subsec:Local-Z-variance}{]}
\item \emph{Reliability diagrams} {[}Fig.\,\ref{fig:05}(b){]} are based
on a similar setup, but for each bin, one plots the standard deviation
of errors ($\mathrm{SD}(E)$) vs the root mean squared uncertainties
($\mathrm{RMS}(u_{E})$). For tight predictions, the points should
lie near the identity line. {[}Sect.\,\ref{subsec:Tightness}{]}
\item In\emph{ confidence curves} {[}Fig.\,\ref{fig:05}(c){]}, one makes
sets of errors iteratively pruned from the values associated with
an increasing percentage of the largest uncertainties. The mean absolute
error (MAE) for those sets is plotted against the percentage of pruning.
A uniformly decreasing curve reveals that the larger absolute errors
are associated with larger uncertainties, but it does not inform us
on a proper scaling. To assess tightness, one needs to compare the
confidence curve to a \emph{probabilistic reference curve} obtained
by the same procedure using a synthetic dataset of errors generated
using the $E_{i}\sim D(0,u_{E,i})$ probabilistic model. {[}Sect.\,\ref{subsec:ranking-based}{]}
\end{itemize}
In complement to a calibration test (for instance $\mathrm{Var}(Z)=1$),
a reliability diagram and a LZV analysis provide basically the same
information and enable to validate tightness. In the case of a confidence
curve, a probabilistic reference is required to reach the same goal,
which implies a choice of distribution for $D$. As in the homoscedastic
case, if the predicted values ($V$) are also available, tightness
can be tested by a LZV analysis using subsets of $V$.
\end{itemize}

\paragraph{Expanded uncertainty.}

A less common scenario is based on \emph{expanded} uncertainties ($U_{E,P}$),
which are the half range of a probability interval (typically at the
$P=95$\,\% level). Without information on the distribution $D$
of prediction errors, one cannot reliably estimate a standard uncertainty
from an expanded uncertainty, and the variance-based validation methods
proposed above cannot be used. One should then have recourse to intervals-based
validation methods. {[}Sect.\,\ref{subsec:Calibration}{]}

The \emph{prediction interval coverage probability} (PICP) $\nu_{P}$
is estimated as the percentage of errors $E_{i}$ lying within the
corresponding interval $\left[-U_{E,P,i},U_{E,P,i}\right]$. For a
good calibration, one should have $\nu_{P}=P$, within the statistical
uncertainty due to the size of the dataset. Applied to the whole validation
set, this approach enables to validate calibration. For tightness,
the same test is performed within subsets of the validation set, either
along $U_{E,P}$ if it is not constant, and/or $V$, if available,
resulting in a \emph{local coverage probability (LCP) analysis} {[}Fig.\,\ref{fig:04}(a,b){]}.
{[}Sect.\,\ref{subsec:Local-coverage-probability}{]}

\paragraph{Prediction ensembles.}

Let us now consider ensemble-based dispersion assessments, which are
common in ML-UQ. One has then an ensemble of errors for each prediction,
from which to extract statistics. 

For small ensembles (less than several hundred points), it is illusory
to get reliable prediction intervals, and it is recommended to estimate
$u_{E}$ as the \emph{standard error} of an ensemble and use variance-based
validation methods as described above. Note that for very small ensembles
(smaller than 10 points) further complications arise, as the estimation
of $u_{E}$ is itself very uncertain, and getting calibration/tightness
diagnostics might be unrealistic. {[}Sect.\,\ref{subsec:Small-ensembles}{]}

For large ensembles, one has the choice to use either intervals- and/or
variance-based validation methods. In the case of intervals-based
validation, a set of target probability levels $P$ can be tested
in order to validate the shape of the prediction distribution. This
multiple intervals-based calibration test is much more stringent than
a variance-based validation. 

Often, ML prediction ensembles are used for active learning more than
for estimating prediction uncertainty. In such cases, the confidence
curve is an interesting tool: a continuously decreasing confidence
curve is sufficient to validate that a ML algorithm enables reliably
to identify predictions with potentially large errors.

\paragraph{Predictive distributions.}

Some ML methods provide for each prediction a distribution (typically
normal) with is mean and dispersion parameters. As for large prediction
ensembles, the full panel of variance- and intervals-based validation
methods is accessible. Additionally, \emph{calibration curves} are
commonly used in this scenario to assess average calibration, but
they do not enable to test tightness {[}Sect.\,\ref{subsec:Graphical-representations}{]}.
Note that a failure of intervals-based validation might be due either
to the choice of distribution and/or to its estimated parameters,
which complicates the diagnostic. 

\section{Concepts, definitions and notations\label{sec:Notations-and-definitions}}

\subsection{Concepts and definitions\label{subsec:Concepts-and-definitions}}

In order to validate the calibration of a prediction model or algorithm,
one needs a \emph{validation set}, composed of predicted values, their
dispersion assessments, and reference values to compare with. Dispersion
assessments can take the form of predictive distributions, prediction
ensembles or statistical summaries, typically uncertainties. 

\paragraph{Uncertainty.}

Let us first recall the definition of uncertainty in metrology\citep{GUM}:\emph{
``a non-negative number that quantifies the dispersion of the values
being attributed to a quantity of interest'' }(QoI, noted $V$).
Depending on the statistic used to quantify the dispersion, one distinguishes
between \emph{standard~uncertainty} (noted $u_{V}$ thereafter),
for which the dispersion is estimated by a \emph{standard deviation}\citep{GUM},
and \emph{expanded~uncertainty} (noted $U_{V,P}$ thereafter), for
which the dispersion is estimated by the \emph{half range of a probability
interval}, typically at the 95\,\% level ($U_{V,95}$).\citep{Ruscic2014}
It is important to note that designing a probability interval from
a \emph{standard }uncertainty requires information on the QoI distribution,
while no additional information is required for an \emph{expanded}
uncertainty.\footnote{Recently introduced,\citep{Cox2022} \emph{characteristic uncertainty}
is estimated by $U_{V,95}/2$ . This proposition covers the gap between
$u$ and $U_{V,p}$ in terms of information needed for prediction
interval design.}

\paragraph{Error.}

In the UQ validation framework, the quantity of interest is the \emph{prediction
error}, i.e. the difference between a predicted value and a reference
value. Different error sources might be characterized by specific
uncertainties (numerical, parametric, model, aleatoric, epistemic...).\citep{Lejaeghere2020}
The prediction error should aggregate all the underlying error sources
and, in absence of ambiguity, will be referred to simply as the error.
The \emph{prediction uncertainty}, which is the uncertainty on the
prediction error, should thus provide a scale for the dispersion of
prediction errors. This offers us a rationale for its validation,
as presented in Sect.\,\ref{subsec:Heteroscedastic-validation-sets}. 

\paragraph{Calibration.}

The calibration-sharpness (CS) framework\citep{Gneiting2014} provides
definitions for major concepts. \emph{Calibration} estimates the\emph{
``statistical compatibility of probabilistic forecasts and observations;
essentially, realizations should be indistinguishable from random
draws from predictive distributions''}\citep{Gneiting2014} where
a\emph{ probabilistic forecast} or \emph{probabilistic prediction},
provides a distribution over the values that can be taken by a QoI.
In this framework, calibration is generally understood as \emph{average}
calibration, i.e. the calibration estimated over the full validation
set. It is well understood that average calibration is insufficient
to guarantee useful predictions.\citep{Kuleshov2018,Pernot2022a} 

\paragraph{Sharpness.}

In the \emph{optimization} framework of UQ methods, \emph{sharpness}
metrics are used to identify more concentrated predictive distributions.
\emph{Sharpness} is defined as\emph{ }``\emph{the concentration of
a predictive distribution in absolute terms; a property exclusive
to the forecasts}''\emph{.}\citep{Gneiting2014} Sharpness metrics
are typically average dispersion statistics (mean prediction uncertainty
or variance,\citep{Kuleshov2018,Tran2020} or mean prediction interval
width\citep{Chung2021,Lai2022}), that do not involve the reference
values. As such, sharpness is barely relevant to UQ validation.

\paragraph{Tightness.}

Stronger calibration concepts have been introduced to palliate the
limitations of average calibration to describe the small-scale reliability
of predictions to observations: \emph{individual calibration} (calibration
over each element of the validation set);\emph{\citep{Chung2020}}
\emph{group calibration} (calibration over pertinent groups of the
validation set);\emph{\citep{Chung2021}} \emph{adversarial group
calibration} (calibration over randomly generated groups of the validation
set);\emph{\citep{Chung2021} }and \emph{local calibration} (calibration
over groups mapping a pertinent feature).\citep{Pernot2022a} Local
calibration has to be understood as a form of group calibration \citep{Chung2021}
or \emph{multicalibration} \citep{HebertJohnson2017}, based on the
sub-setting of a continuous feature into adjacent or overlapping intervals.
Its purpose is to identify local or \emph{small-scale} departures
from calibration which might have a diagnostic interest. When the
mapping feature is prediction uncertainty, local calibration is tightly
related to \emph{reliability diagrams} (agreement of uncertainty with
the dispersion of errors), which is also referred to as \emph{perfect}
calibration.\citep{Levi2020}

As sharpness cannot be used to characterize this small-scale reliability,
I propose to use instead \emph{tightness }as\emph{ }a dedicated concept
to characterize the small-scale adaptation of UQ predictions to reference
values. More widely, a set of predictions can be considered to be
\emph{tight} if it satisfies the requirements of any of the stronger
calibration concepts (individual, group, local or perfect calibration).
This offers a convenient shortcut for propositions such as \emph{group
calibrated}, \emph{locally calibrated or perfectly calibrated}. 

Note that it is tempting to assume that tightness implies average
calibration. However, statistical uncertainty on calibration statistics
for small groups might lead to scenarios where one accepts tightness
while rejecting average calibration. As for sharpness, it is therefore
important for tightness to be conditional to average calibration:
\emph{a probabilistic prediction method cannot be tight if it is not
(average) calibrated}. 

\subsection{Notations\label{subsec:Notations}}

\paragraph{Prediction.}

Let us consider a QoI, $V$, for which one wants to make predictions
with some form of confidence assessment. For a probabilistic prediction,
the predictive distribution on $V$ can be characterized by its quantile
function $q_{V}(p)$, where $p$ is a probability (the quantile function
is the inverse of the cumulative distribution function).

However, few UQ methods provide predictive distribution functions,
and empirical approximations $\tilde{q}_{V}$ are more often accessible
from\emph{ ensembles} or samples, representative of the predictive
distribution. In such cases, the standard uncertainty $u_{V}$ is
estimated by the standard deviation of the sample,\citep{GUM} and
the expanded uncertainty $U_{V}$ from the empirical quantiles\citep{Wilcox2012,Wilcox2018}
\begin{equation}
U_{V,P}=\frac{1}{2}\left(\tilde{q}_{V}\left((1+p)/2\right)-\tilde{q}_{V}\left((1-p)/2\right)\right)\label{eq:Uv,p}
\end{equation}
where, to conform with usual notations, $P$ is the percentage corresponding
to $p$ ($P=100p$).

The most frequent scenario in the computational chemistry UQ literature
is to have a single statistical summary \textendash{} very commonly
the standard uncertainty $u_{V}$ and more rarely the expanded uncertainty
$U_{V,95}$.\citep{Ruscic2014} The consequences of the absence of
predictive distribution on the CS validation framework are explored
below.

\paragraph{Validation.}

For the sake of validation, predictions of $V$, $\{V_{i}\}_{i=1}^{M}$,
are made for a series of $M$ test systems for which one has reference
values $\{R_{i}\}_{i=1}^{M}$. For each validation system $i$, one
needs at least one UQ object among $q_{V}$, $\tilde{q}_{V}$, $u_{V}$
or $U_{V,P}$ as defined above. Validation is made by assessing the
statistical compatibility between the errors $E_{i}=R_{i}-V_{i}$
and the corresponding dispersion statements. 

The minimal validation set is thus composed of errors and the corresponding
UQ estimators, for instance $\{E_{i},u_{E_{i}}\}_{i=1}^{M}$, where
$u_{E_{i}}=u_{V_{i}}$ if the reference data are not uncertain. When
the reference values are themselves uncertain, this has to be propagated
to the errors. For instance, if $V_{i}$ has a standard uncertainty
$u_{V_{i}}$ and $R_{i}$ a standard uncertainty $u_{R_{i}}$, the
uncertainty on $E_{i}$ is obtained by combination of variances $u_{E_{i}}=\sqrt{u_{V_{i}}^{2}+u_{R_{i}}^{2}}$
(considering that $V_{i}$ and $R_{i}$ are statistically independent).\citep{GUM}
Alternative combination schemes have to be considered for other types
of uncertainty.\citep{Pernot2022a} Note that if the uncertainty on
the reference values contributes significantly to the uncertainty
on $E$, failure of validation tests might be difficult to interpret,
as they might occur as well from the predictor as from the reference
data. 

\emph{Prediction intervals} are at the center of the CS validation
framework. A $P$\% error prediction interval can be estimated from
the $q_{E}(p)$ quantile function or its empirical variant ($\tilde{q}_{E}(p)$)
as
\begin{equation}
I_{E_{i},P}=\left[q_{E_{i}}\left((1-p)/2\right),q_{E_{i}}\left((1+p)/2\right)\right]\label{eq:pred-int-1}
\end{equation}
If one assumes the symmetry of intervals around $E_{i}$, expanded
uncertainties can also be used directly, i.e.
\begin{equation}
I_{E_{i},P}=\left[-U_{E_{i},P},U_{E_{i},P}\right]
\end{equation}
A contrario, it is not possible to design a prediction interval form
a standard uncertainty $u_{E_{i}}$ without making hypotheses on the
error distribution. Being mostly dominated by model errors, computational
chemistry error distributions are often non-normal,\citep{Pernot2020b,Pernot2021}
which prevents the use of simple recipes (such as the $2\sigma$ rule).
Making unsupported distribution hypotheses would add a fragility layer
to the validation process, complicating the interpretation of negative
validation tests. This prevents the application intervals-based validation
to sets of standard uncertainties, and the use variance-based methods,
such as reliability diagrams\citep{Levi2020} of $z$-scores ($z_{i}=E_{i}/u_{E_{i}}$)
statistics\citep{Pernot2022a}, has been proposed as an alternative.

Intervals- and variance-based CT validation methods are presented
in Sect.\,\ref{sec:Quantitative-methods}.

\subsection{Synthetic datasets}

The methods presented below are illustrated on synthetic validation
sets $\left\{ V_{i},E_{i},u_{E_{i}}\right\} _{i=1}^{M}$, with $M=1000$.
$V$ is sampled uniformly in the interval $[-2,2]$ . The SYNT01 errors
are obtained from a probabilistic model 
\begin{equation}
E_{i}\sim D(0,u_{E_{i}})\label{eq:probmod-1}
\end{equation}
where $D(\mu,\sigma)$ is a probability density function with mean
$\mu$ and standard deviation $\sigma$. This dataset is tagged as
\emph{consistent}, as errors and uncertainties are statistically consistent
and should provide positive calibration and tightness tests. The other
sets (SYNT02 and SYNT03) do not derive directly from this probabilistic
model and are labeled as \emph{non-consistent}. 
\begin{description}
\item [{SYNT01}] heteroscedastic \emph{consistent} set, where the errors
are generated from a zero-centered normal distribution $E_{i}\sim N(0,u_{E_{i}})$
with a standard deviation depending on $V$, through $u_{E_{i}}=0.01(1+V_{i}^{2})$. 
\item [{SYNT02}] heteroscedastic \emph{non-consistent} set, with errors
sampled from a normal distribution $E_{i}\sim N(0,<u_{E}>)$ and the
same uncertainties as in SYNT01. 
\item [{SYNT03}] homoscedastic \emph{non-consistent} set, with errors taken
from SYNT01 and constant $u_{E_{i}}=<u_{E}>$.
\end{description}

\section{Basic graphical methods\label{sec:Graphical-methods}}

Considering a minimal validation set $\left\{ E_{i},u_{E_{i}}\right\} _{i=1}^{M}$,
it is possible to draw simple graphs to check that uncertainty quantifies
correctly the dispersion of errors. One has to consider two cases:
(1) $u_{E}$ varies notably over the validation set (heteroscedastic
set); and (2) $u_{E}$ is (nearly) constant (homoscedastic set).

\subsection{Heteroscedastic validation sets\label{subsec:Heteroscedastic-validation-sets}}

The consistency between errors and uncertainties (Eq.\,\ref{eq:probmod-1})
is based on an asymmetrical relation, which can be summarized as follows
\begin{align}
\textrm{large}\,|E| & \Longrightarrow\textrm{large}\,u_{E}\\
\textrm{small}\,u_{E} & \Longrightarrow\textrm{small}\,|E|
\end{align}
i.e. large errors should occur only from predictions with large uncertainties
and predictions with small uncertainties should be associated with
small errors. The asymmetry results from the fact that small errors
might arise as well from predictions with small uncertainty as from
predictions with large uncertainty. In consequence, one should not
expect a strong correlation between absolute errors and uncertainties
(see Sect.\,\ref{subsec:ranking-based}) and there is not much to
learn from plots of $|E|$ vs $u_{E}$.

\paragraph{Basic plot.}

\begin{figure}[t]
\noindent \begin{centering}
\includegraphics[height=6cm]{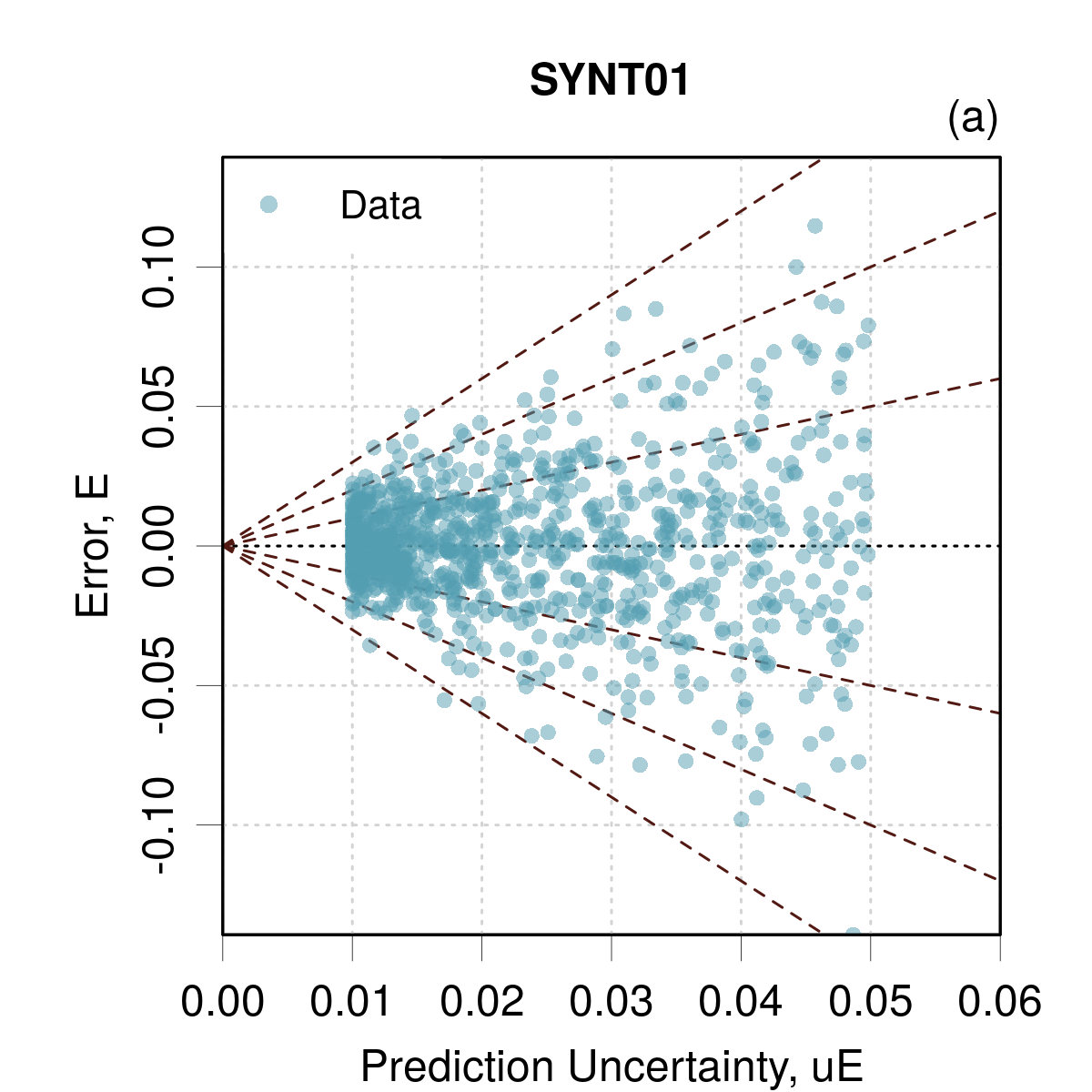}\includegraphics[height=6cm]{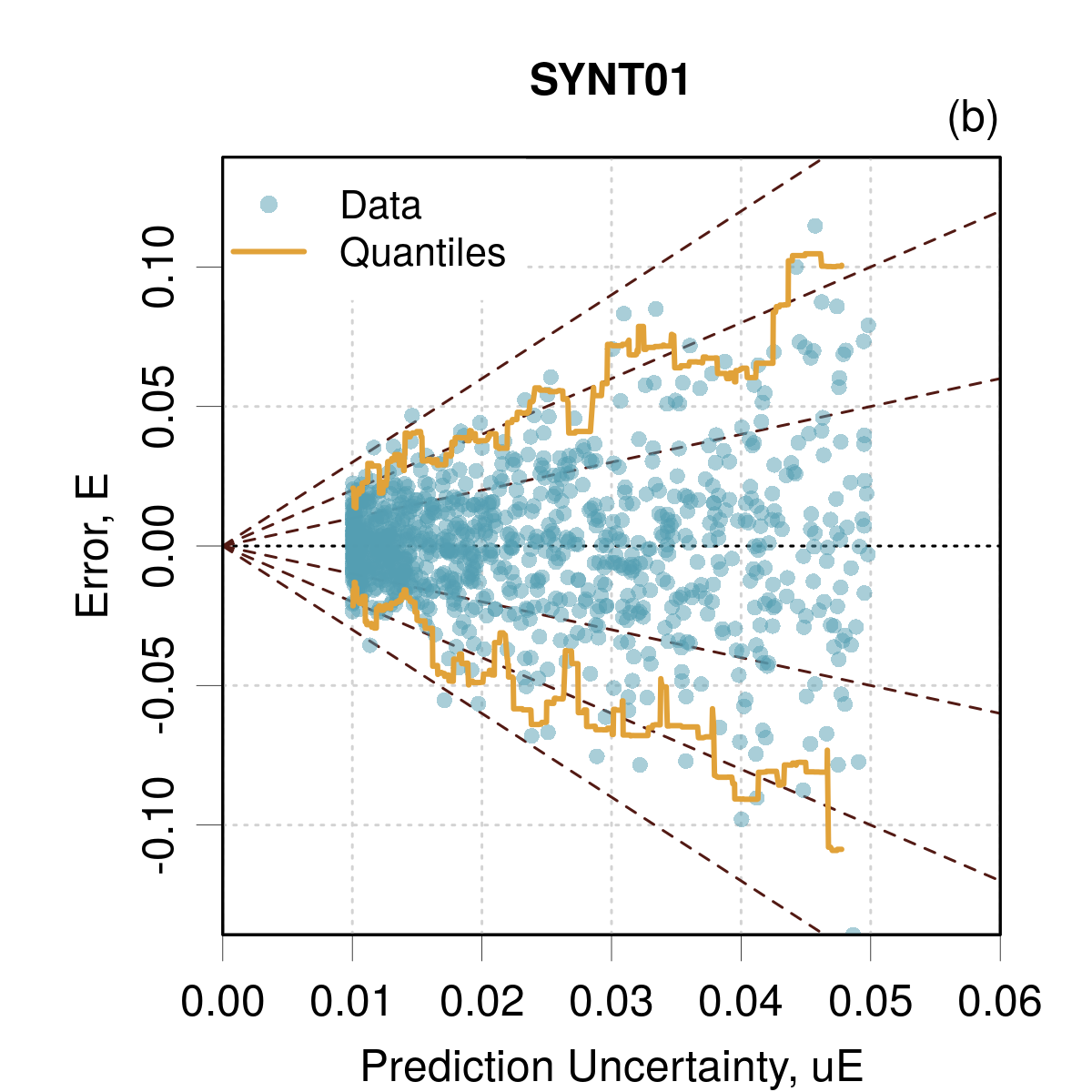}\includegraphics[height=6cm]{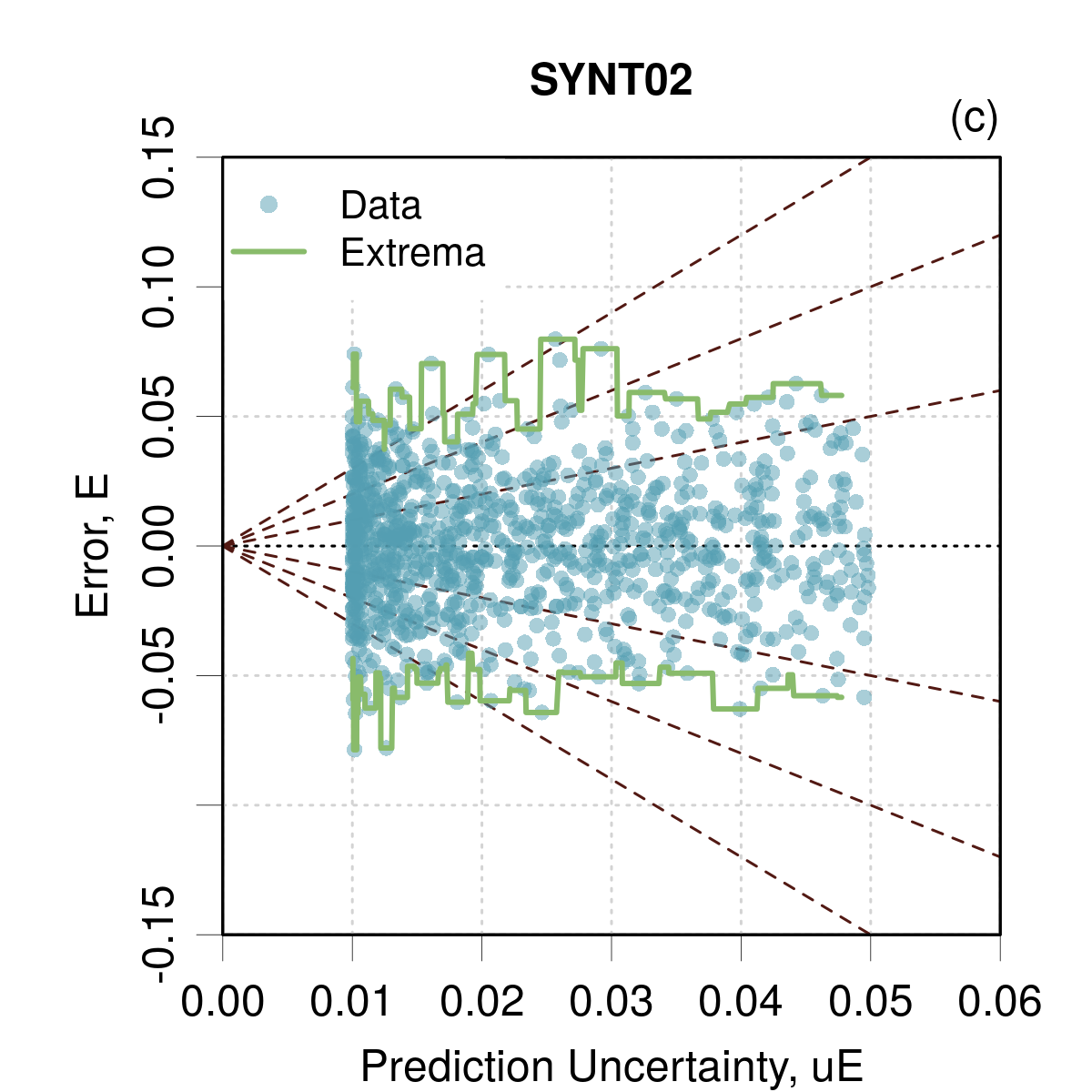}
\par\end{centering}
\caption{\label{fig:01}(a) Simple $(u_{E},E)$ plot for dataset SYNT01; (b)
same plot augmented with running quantiles; (c) idem for dataset SYNT02
with running extrema.}
\end{figure}

When $u_{E}$ depends notably on the validation point, one can simply
plot $E$ vs $u_{E}$ to check how the dispersion of $E$ scales with
$u_{E}$.\citep{Janet2019} An example is shown in Fig.\,\ref{fig:01}(a),
where guiding lines $y=\pm kx;\,k=1-3$ have been added to facilitate
the appraisal of the expected linear scaling. One sees indeed for
the consistent dataset SYNT01 that larger errors are associated with
larger uncertainty values, giving a typical fan-like structure to
the data cloud. The symmetry of the cloud with respect to the $y=0$
axis is furthermore a good indication that the errors have no noticeable
bias. 

\paragraph{Improvements.}

The proper scaling of $E$ with $u_{E}$ can be difficult to appreciate
visually for small datasets. A little more sophisticated approach
consists in adding an estimator of the local range of $E$ on the
graph. This might be done by using a sliding window to estimate either
the \emph{extrema} or the limits of a 95\,\% probability interval.
The latter method is called \emph{running quantiles }and is depicted
in Fig.\,\ref{fig:01}(b). In this implementation, the sliding window
contains a fixed number of points $n$ (not a fixed width of $u_{E}$),
which is automatically estimated by using the Rice formula for histograms
$n=2M^{1/3}$. One sees in Fig.\,\ref{fig:01}(b) that the quantile
lines oscillate around the $y=\pm2x$ lines, which can be expected
from the properties of the normal distribution used to generate the
SYNT01 dataset. In the case of the non-consistent SYNT02 dataset,
Fig.\,\ref{fig:01}(c) shows clearly the absence of scaling between
$E$ and $u_{E}$ (the larger errors occur anywhere along the $u_{E}$
axis). This trend is underlined in this plot by \emph{running extrema}
lines, which are easier to compute than quantiles, but oscillate more
strongly (strong dependence to outliers) and might be more difficult
to interpret. 

An alternative representation, plotting $\log(|E|)$ vs. $\log(u_{E})$,
is used in the literature.\citep{Musil2019} It is motivated by the
fact that, for a normal error distribution with mean 0 and variance
$\sigma^{2}$, the probability density function of the logarithm of
absolute errors has its mode at $\sigma$. In these conditions, one
should observe a strong concentration of points along the identity
line for statistically consistent validation sets and a \emph{running
mode} line should lie close to it. It is important to understand the
logic behind this type of plot, but I did not develop it further here
because (i) it is less intuitive than the $(u_{E},E)$ plot, (ii)
it requires the estimation of the mode (or some high density levels
of the data cloud) which limits the application to large datasets,
and (iii) it is sensitive to deviations from the zero-centered normal
error distribution which complicates the interpretation of a negative
diagnostic.

\subsection{Homoscedastic validation sets}

The problem when $u_{E}$ is constant is that the $(u_{E},E)$ plot
proposed above cannot be used. In such cases, the expected scaling
can be appreciated by using $z$-scores $Z=E/u_{E}$ and plotting
them against a relevant feature of the dataset, for instance the points
index or the QoI $V$. The latter is good to appreciate systematic
trends in scaling and to relate them to a range of predicted values.
Note that it might be difficult or impossible to spot \emph{z}-score
problems on such plots if they are not localized in $V$ space. 

\begin{figure}[t]
\noindent \begin{centering}
\includegraphics[height=6cm]{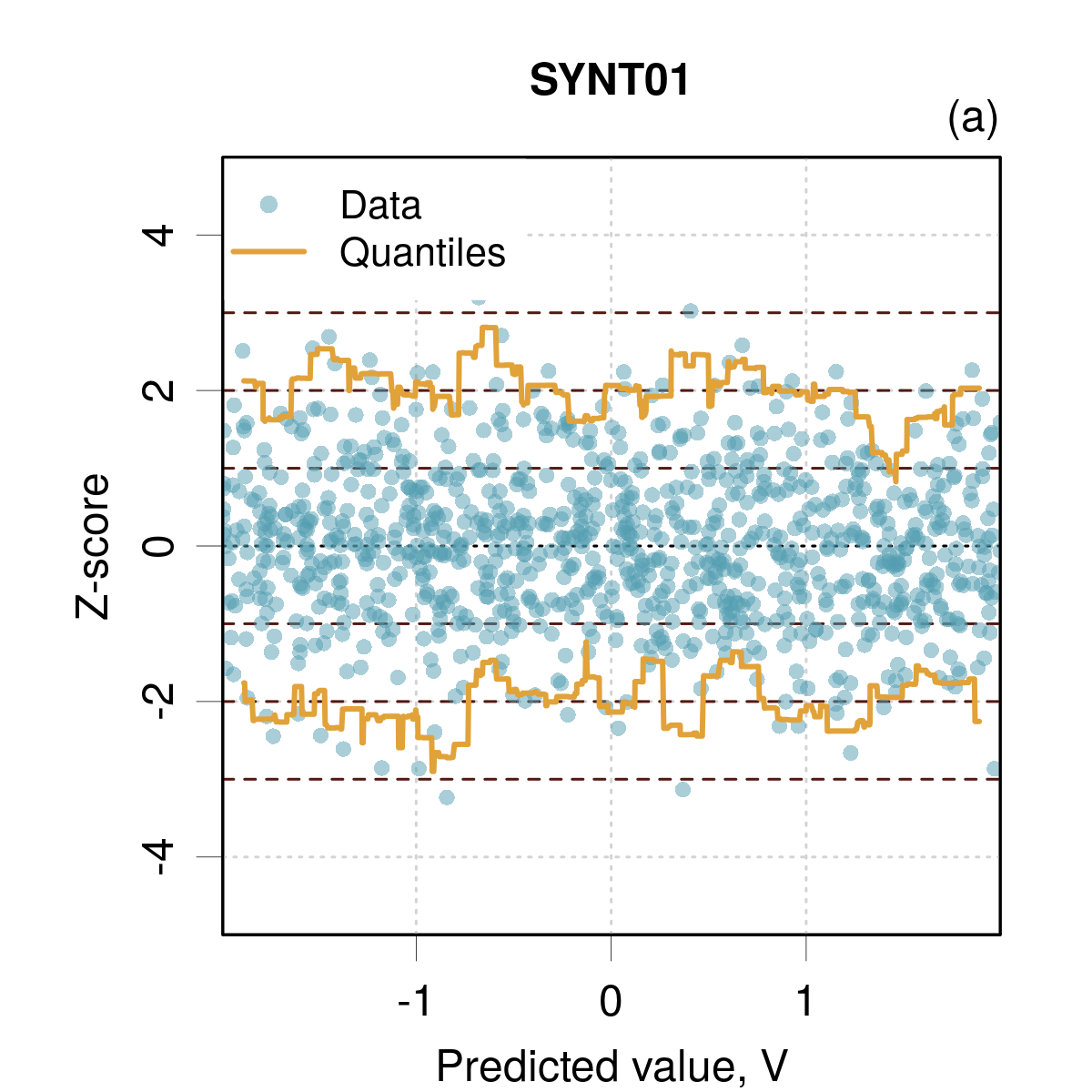}\includegraphics[height=6cm]{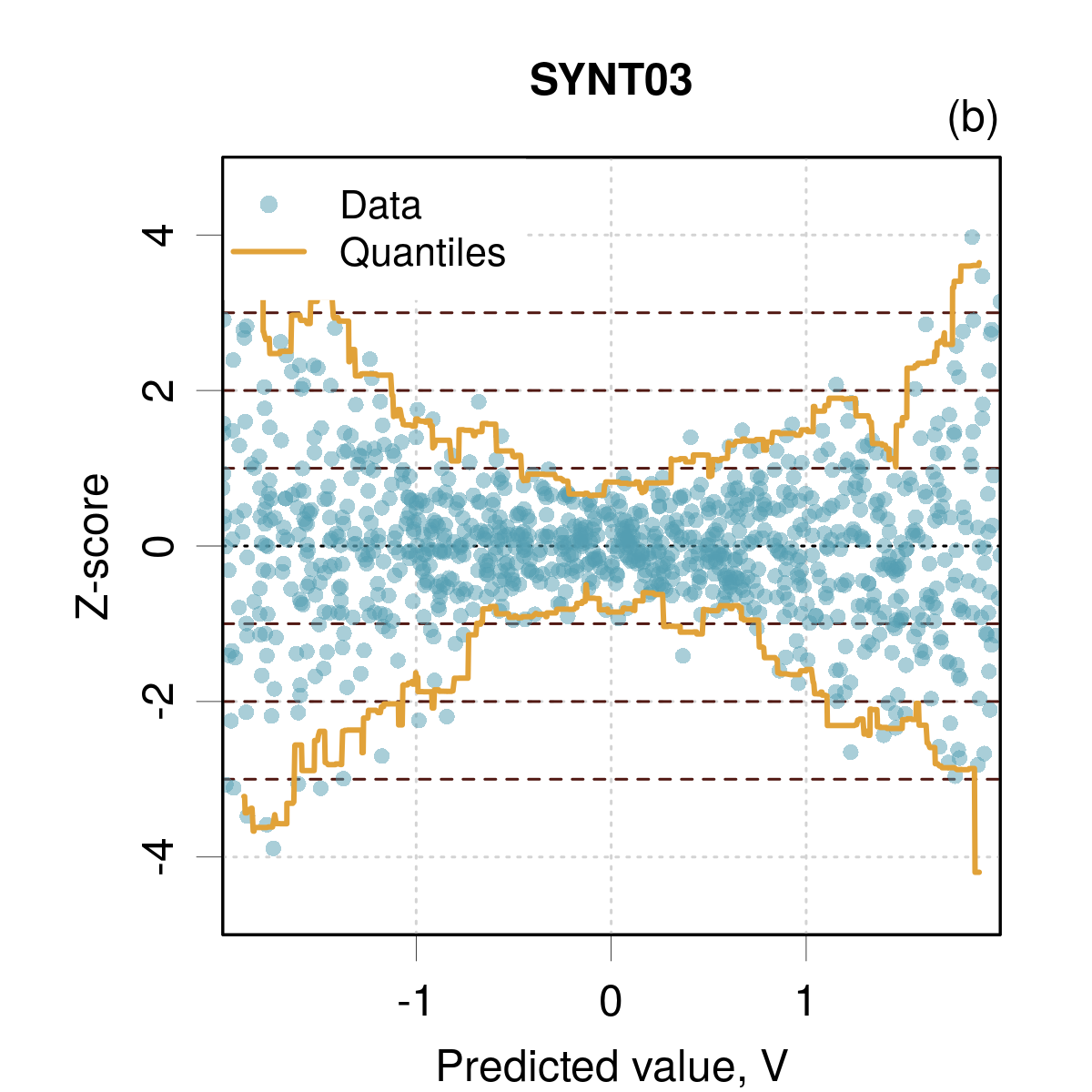}\includegraphics[height=6cm]{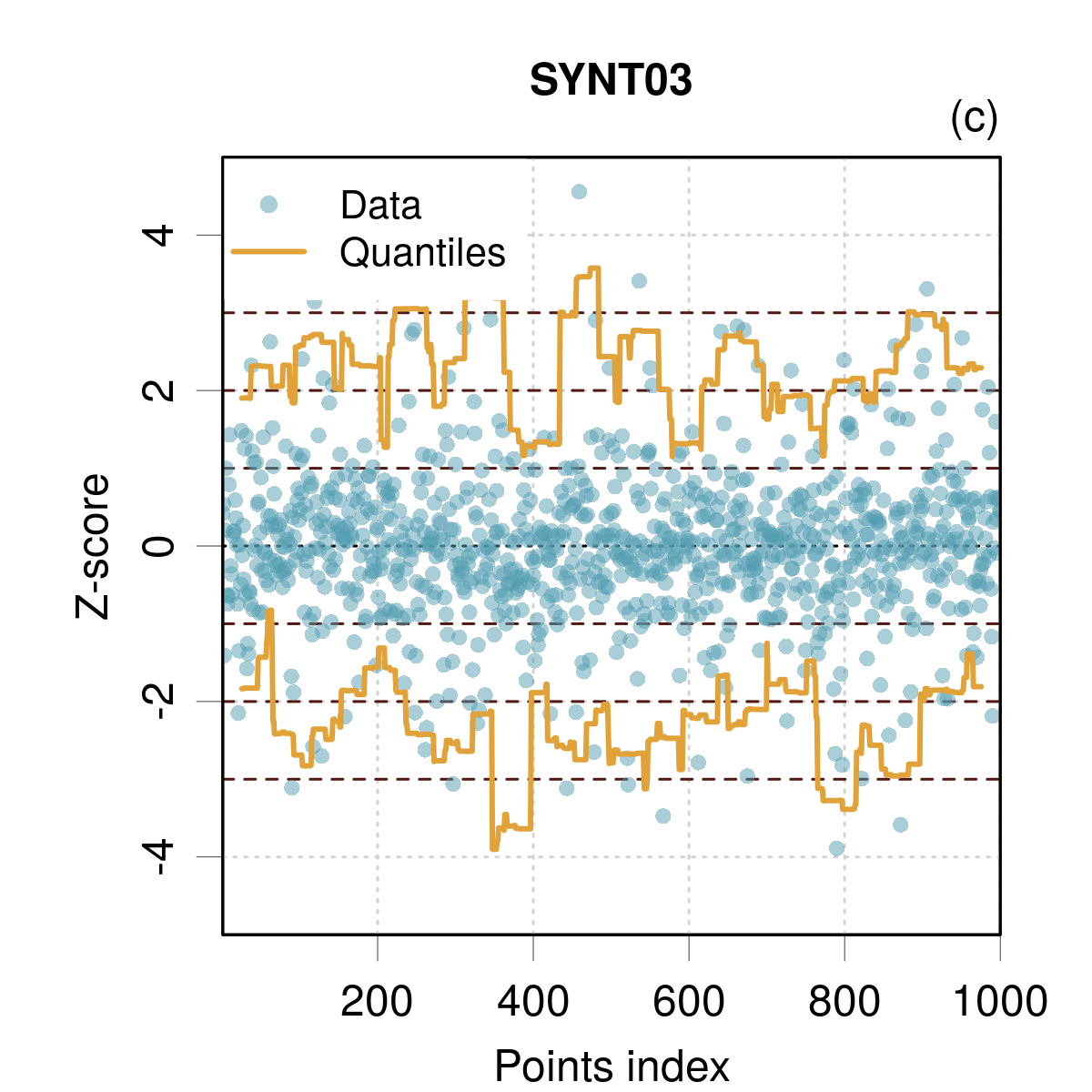}
\par\end{centering}
\caption{\label{fig:02}(a,b) $(V,Z)$ plots for dataset SYNT01 and SYNT03
with added \emph{running quantiles}; (c) $(index,Z)$ plot for SYNT03.}
\end{figure}

The guiding lines are now horizontal ($y=\pm k;\,k=1-3$), and, as
above, a simple $(V,Z)$ plot can be improved by running statistics.\textcolor{orange}{{}
}The\emph{ running quantiles} lines should run parallel to the guiding
lines. An example is shown in Fig.\,\ref{fig:02}(a) for an heteroscedastic
consistent dataset (SYNT01). In Fig.\,\ref{fig:02}(b) for a homoscedastic
non-consistent dataset (SYNT03), the envelope of the data clearly
deviates from the guiding lines. Although calibration is difficult
to assess, one might safely conclude to a lack of tightness. Note
that the diagnostic depends on the choice of a plotting ordinate.
Fig.\,\ref{fig:02}(c) presents the same dataset as a function of
the point index. The non-reliability of the uncertainties is difficult
to appreciate on this plot, as the running quantiles follow more or
less the guiding lines, although with large oscillations when compared
to the SYNT01 case. 

\subsection{Remarks}
\begin{itemize}
\item These simple graphical methods should help to detect frank departures
from calibration/tightness. They cannot be used to validate these
properties. In cases where one cannot easily reject the consistency
between errors and uncertainties, calibration and tightness have to
be assessed by quantitative methods described below.
\item They are applicable to both standard ($u_{E}$) and expanded ($U_{E,p}$)
uncertainties, albeit with different interpretations of the guiding
lines. 
\item The $z$-score based plot $(u_{E},Z)$ could also be used for dataset
with non-constant $u_{E}$ {[}e.g. Fig.\,\ref{fig:02}(a){]}, but
in such cases, I think the $(u_{E},E)$ plot enables easier diagnostics
{[}e.g. Fig.\,\ref{fig:01}(b){]}.
\item To lessen the dependence of the running statistics on a specific validation
set, one might think of a bootstrapping\citep{Efron1979,Efron1991}
approach to estimate \emph{mean running statistics} and their uncertainty.
However, this might be a little far fetched for this simple qualitative
visualization. 
\end{itemize}

\section{Quantitative methods\label{sec:Quantitative-methods}}

\subsection{Statistical framework}

\subsubsection{Average calibration\label{subsec:Calibration}}

\paragraph{Intervals-based testing.}

In the CS framework, a method is considered to be calibrated if the
confidence of its predictions matches the probability of being correct
for all confidence levels,\citep{Tran2020,Tomani2021} which can be
reformulated as ``prediction intervals should have the correct coverage''.\citep{Kuleshov2018} 

It is convenient here to deal with prediction errors instead of predicted
values, and one defines the \emph{prediction interval coverage probability}
(PICP) as \footnote{Note that this definition differs from the one in the calibration/sharpness
literature (see e.g. Kuleshov \emph{et al.}\citep{Kuleshov2018})
by my consideration of reference data uncertainty in the prediction
interval.}
\begin{equation}
\nu_{p,M}=\frac{1}{M}\sum_{i=1}^{M}\boldsymbol{1}\left(E_{i}\in I_{E_{i},P}\right)
\end{equation}
where $\boldsymbol{1}(x)$ is the \emph{indicator function} for proposition
$x$, taking values 1 when $x$ is true and 0 when $x$ is false,
and $I_{E_{i},P}$ is a $P=100p$\,\% prediction interval for $E_{i}$.
Hence, estimating a PICP simply amounts to count the number of times
a validation error falls within the corresponding prediction interval. 

Using PICPs, a method is calibrated if\citep{Kuleshov2018}\emph{
\begin{equation}
\lim_{M\rightarrow\infty}\nu_{p,M}=p,\,\forall p\in[0,1]\label{eq:calibration-1-1}
\end{equation}
}In practice, one has a limited amount of validation data to test
the equality, and a standard procedure to validate $\nu_{p,M}$ is
to estimate a 95\,\% confidence interval on the statistic, $I_{95}(\nu_{p,M})$,
and to test if it contains $p$:
\begin{equation}
p\stackrel{?}{\in}I_{95}(\nu_{p,M}),\,\forall p\in[0,1]
\end{equation}
The stacked notation $p\stackrel{?}{\in}I$ is used as a shorthand
for ``does $p$ belong to $I$ ?''. Note that in the CC-UQ literature
one often has to accept a weaker form of calibration, based on a single
$p$ value (0.95).

$\nu_{p,M}$ is the bounded ratio of two integers ($0\le\nu_{p,M}\le1$)
and is known in the literature as a \emph{binomial proportion}.\citep{Vollset1993}
The finite set of realizable values for $\nu_{p,M}$ depends on $M$,
and it might not contain a given value of $p$. There are many methods
to estimate $I_{95}(\nu_{p,M})$, with competing features such as
optimal coverage or minimal range, and the choice of the best one
is debated among experts. The main difficulty is that some properties
of the confidence interval are sharply oscillating with $M$, so that
the best choice might depend on $M$. However, all the experts agree
that the textbook method (known as the \emph{Wald method}), based
on a normality hypothesis, has to be avoided. 

An exploratory comparison \citep{Pernot2022a} over a set of methods
available in the R language\citep{RTeam2019} showed that it is reasonable
in the present setup to choose between the Agresti-Coull \citep{Agresti1998},
Clopper-Pearson \citep{Clopper1934} and continuity-corrected Wilson
\citep{Newcombe1998} methods. The latter is my standard choice in
this study. 

A limitation of PICP testing is the saturation of the coverage at
the upper limit: if a prediction interval for $p<1$ achieves a coverage
probability $\nu_{p,M}=1$, one gets no information on the amplitude
of the mismatch with the target probability. As a complementary diagnostic,
I find useful to consider the ranges ratio (RR), i.e. the ratio of
the mean range of prediction intervals over the range of the empirical
interval at probability $p$:
\begin{equation}
R_{p}=\frac{\frac{1}{M}\sum_{i=1}^{M}\left(I_{E_{i},P}^{+}-I_{E_{i},P}^{-}\right)}{\tilde{Q}_{E}\left((1+p)/2\right)-\tilde{Q}_{E}\left((1-p)/2\right)}
\end{equation}
where $I_{X}^{+/-}$ is the upper/lower limit of a prediction interval
$I_{X}$ and $\tilde{Q}_{E}$ is the empirical quantile function of
errors, estimated over the validation set (not to be confounded with
$\tilde{q}_{E}$ which is defined for individual predictions). Deviations
of $R_{p}$ from unity quantify the mismatch amplitude. The effect
of the validation set size on the value of $R_{p}$ can be estimated
by bootstrapping.\citep{Efron1979,Efron1991}

A second limitation of PICP testing appears when one has no sufficient
information to design a reliable prediction interval. This occurs,
frequently, when only standard uncertainties are available, in absence
of information on the underlying error distribution. In such cases,
one should turn to variance-based validation.

\paragraph{Variance-based testing.}

The underlying probabilistic model for variance-based testing is given
by Eq.\,\ref{eq:probmod-1}. Hence, for homoscedastic data, the consistency
between errors and uncertainty can readily be checked by comparing
the error variance to the squared uncertainty
\begin{equation}
\mathrm{Var}(E)\overset{?}{=}u_{E}^{2}\label{eq:VarE}
\end{equation}
To extend this equation to heteroscedastic data, let us assume that
the errors are drawn from a distribution $D(0,\sigma)$ (Eq.\,\ref{eq:probmod-1})
with a scale parameter $\sigma$ distributed according to $G(\sigma)$.
The distribution of errors is then a \emph{compound distribution},
more specifically a \emph{scale mixture distribution}. The variance
of the compound distribution is obtained by the \emph{law of total
variance}, i.e. 
\begin{equation}
\mathrm{Var}(E)=\left\langle \mathrm{Var}_{D}(E|\sigma)\right\rangle _{G}+\mathrm{Var}_{G}\left(\left\langle E|\sigma\right\rangle _{D}\right)\label{eq:totalVar}
\end{equation}
The first term of the RHS can be estimated as the mean squared uncertainty
$<u_{E}^{2}>$. For unbiased errors, the second term of the RHS should
be small to negligible, but in a general case, its estimation requires
binning of the errors according to the corresponding uncertainties,
estimating the mean error in each bin and taking the variance of the
mean errors over the bins. Accuracy of this procedure depends on the
sample size and binning strategy, and the main limitations of this
technique are the same as advanced by Scalia \emph{et al.}\citep{Scalia2020}
for the application of reliability diagrams (see Sect.\,\ref{par:Reliability-diagram.}).
Besides these technical complications, the test for unbiased errors
would thus be
\begin{equation}
\mathrm{Var}(E)\overset{?}{\simeq}\left\langle u_{E}^{2}\right\rangle \label{eq:varE}
\end{equation}
which does not account for the essential pairing between errors and
uncertainties, and could enable fortuitous agreements, i.e. an equality
does not guarantee that the probabilistic model (Eq.\,\ref{eq:probmod-1})
is respected by the data.

For heteroscedastic data, it seems thus more reliable to use \emph{scaled}
uncertainties, or $z$-scores $Z_{i}=E_{i}/u_{E,i}$, which account
for the pairing between errors and uncertainties, and for which Eq.\,\ref{eq:varE}
becomes
\begin{equation}
\mathrm{Var}(Z)\overset{?}{=}1
\end{equation}
Note that this test is valid for both homoscedastic and heteroscedastic
data.\citep{Pernot2022a} In the hypothesis of unbiased errors, one
should also have $<Z>=0$. Formally, $\mathrm{Var}(Z)$ can be linked
to the Birge ratio used in metrology to test statistical consistency.\textcolor{red}{\citep{Birge1932,Kacker2010,Bodnar2014}
}See Appendix\,\ref{sec:Var(Z)-vs.-Birge} for more details. 

Following the same logic as for PICPs, practical validation of $\mathrm{Var}(Z)$
relies on the test
\begin{equation}
1\stackrel{?}{\in}I_{95}\left(\mathrm{Var}(Z),M\right)
\end{equation}
where $I_{95}\left(\mathrm{Var}(Z),M\right)$ can be estimated by
an adapted bootstrapping method (BC$_{a}$, ABC...) to avoid the normality-based
textbook method.\citep{DiCiccio1996} A faster, but slightly less
accurate method to estimate $I_{95}\left(\mathrm{Var}(Z),M\right)$
is based on the estimation of $\mathrm{Var}(\mathrm{Var}(Z))$ introduced
by Cho \emph{et al.} \citep{Cho2005}, using the central moments of
the $Z$ sample
\begin{equation}
W_{z}=\mathrm{Var}(\mathrm{Var}(Z))=\frac{1}{M}\left(\mu_{4}-\frac{M-3}{M-1}\mu_{2}^{2}\right)
\end{equation}
where $\mu_{k}=1/M\sum_{i=1}^{M}(Z_{i}-\mu)^{k}$ and $\mu$ is the
arithmetic mean of $Z$. In absence of further information on the
distribution of $\mathrm{Var}(Z)$, a normality hypothesis leads to
the test
\begin{equation}
1\stackrel{?}{\in}\mathrm{Var}(Z)\pm t_{97.5,M-1}\sqrt{W_{z}}\label{eq:ChoCI}
\end{equation}
where $x\pm y$ denotes the upper and lower bounds of the interval,
and $t_{P,\nu}$ is the $P$\,\% quantile of the Student's-$t$ distribution
with $\nu$ degrees of freedom. The symmetry of the testing interval
might be problematic for small $M$ values, but in most scenarios
tested in PER2022, this method performed nearly as well as the best
bootstrapping methods and better than the worst ones.\citep{Pernot2022a}
Moreover, it is much faster.

\paragraph{Statistical power.}

The efficiency of the tests described above depends on the size of
the validation set. The \emph{power} of the test is the probability
to correctly reject the hypothesis that a PICP or Var(\emph{Z}) value
is compatible with its target value. A power threshold (typically
0.8) is defined to determine a minimal sample size. 

\begin{figure}[t]
\noindent \begin{centering}
\includegraphics[height=8cm]{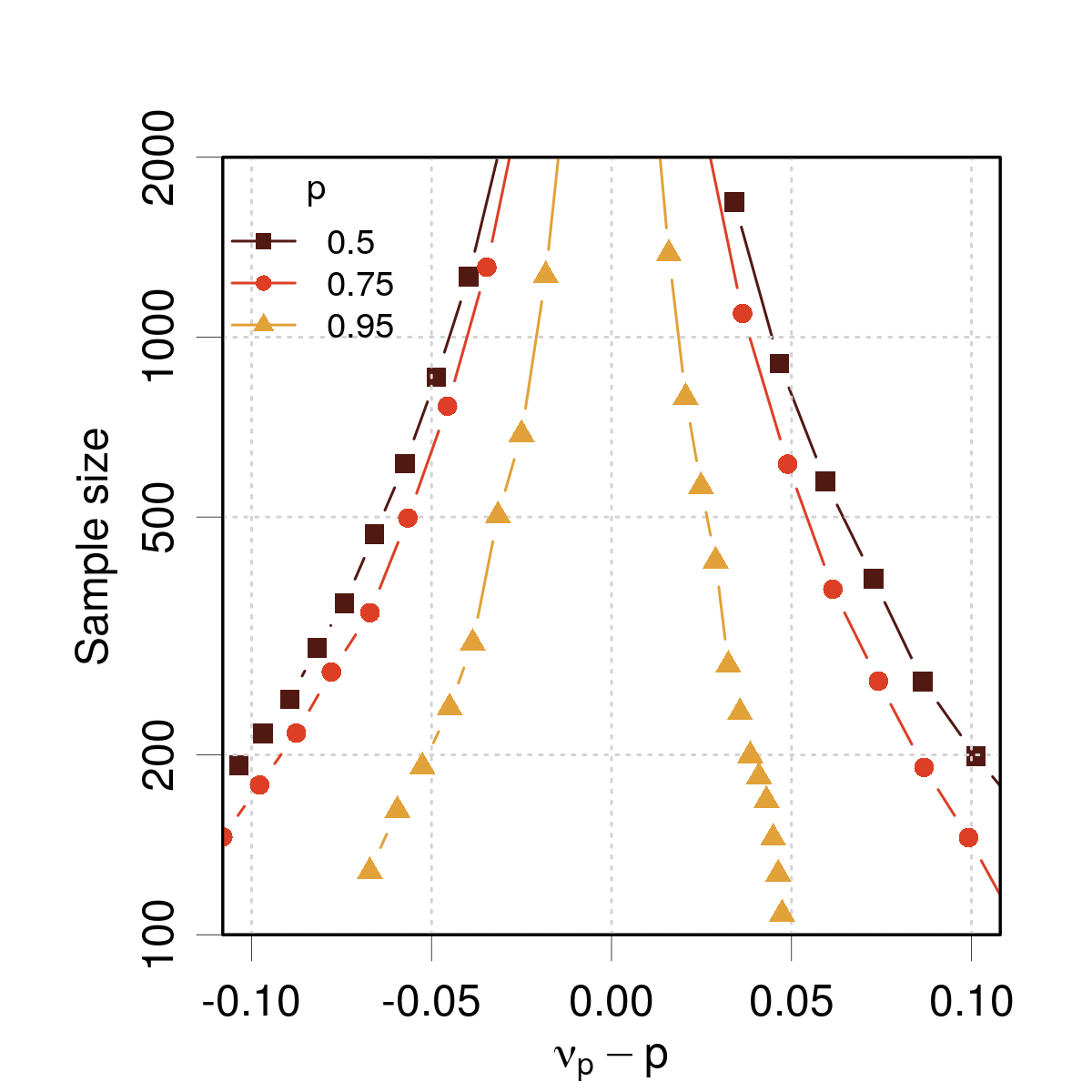}
\par\end{centering}
\caption{\label{fig:03}Minimal sample size required to achieve a power of
0.8 when testing a PICP value $\nu_{p}$ against its target $p$.
The continuity-corrected Wilson method is used to estimate the confidence
interval on $\nu_{p}$. }
\end{figure}

Fig.\,\ref{fig:03} reports the minimal sample sizes necessary to
reach a power of 0.8 for differences between a PICP value $\nu_{p,M}$
and its target value $p$ (see also PER2022\citep{Pernot2022a} (Fig.
S2) for an alternative representation). For instance, a sample size
of $M\simeq200$ is necessary to achieve a power of 0.8 in differentiating
a PICP value of $\nu_{0.95,M}=0.90$ from its $p=0.95$ target. Rejecting
safely a difference $|\nu_{p,M}-p|=0.01$ would take more than 2000
points for the same target. The situation worsens for smaller target
values (above 0.5, which is a symmetry point): for $\nu_{0.5,M}=0.45$,
one needs about 800 points to reject the compatibility between $\nu_{p,M}$
and $p$. 

As a guiding rule, similar sample sizes are required to test $\mathrm{Var}(Z)$. 

\subsubsection{Tightness\label{subsec:Tightness}}

As evoked previously, using the tests presented above on a validation
set provides only an \emph{average} \emph{calibration} diagnostic,
which is not sufficient to guarantee the desired \emph{tightness}
of prediction intervals, i.e. the small-scale reliability of the probabilistic
predictions. 

\paragraph{Local calibration.}

A simple way to assess tightness is to split the dataset into $N_{g}$
groups of sizes $\left\{ m_{j},\thinspace j=1,N_{g}\right\} $ and
test the average calibration for each group. For PICPs, one should
therefore test
\begin{equation}
p\stackrel{?}{\in}I_{95}(\nu_{p,m_{j},j}),\,\forall p\in[0,1],\thinspace j=1,N_{g}
\end{equation}
with a similar formula for $\mathrm{Var}(Z)$. 

The focus is here on the design of contiguous or overlapping groups
partitioning some relevant feature. In this case, tightness is similar
to \emph{local calibration}. In contrast to randomly generated groups
used for adversarial group calibration, local calibration enables
a diagnostic of tightness problems in specific areas of the grouping
feature. For continuous grouping features, several designs can be
considered: contiguous groups, overlapping groups or a sliding window.
For the kind of datasets we are considering in this study, the features
of choice to design groups are typically the predicted value $V$
and the prediction uncertainty ($u_{E}$ or $U_{E,p}$) for heteroscedastic
validation sets. This approach leads to the local coverage probability
(LCP),\citep{Pernot2022a} local $Z$ variance (LZV)\citep{Pernot2022a}
and local range ratio (LRR) analyses used in the graphical representations
described below. 

\paragraph{Reliability diagram.\label{par:Reliability-diagram.}}

To compare the uncertainty to the error, Scalia \emph{et al.} \citep{Scalia2020}
considered the proposition of Levi \emph{at al.}\citep{Levi2020}
to use a generalization of Eq.\,\ref{eq:VarE} in order ascertain
the conformity of the empirical error variance with the predicted
one, i.e.
\begin{equation}
\mathrm{Var}\left(E|u_{E}^{2}=\sigma^{2}\right)=\sigma^{2},\,\forall\sigma^{2}\label{eq:Levi}
\end{equation}
where, for each value of $\sigma^{2}$, the variance is estimated
on those points of the validation set having $\sigma^{2}$ as predicted
variance. The practical implementation of this scheme, resulting in
a so-called \emph{reliability diagram}, requires binning of $u_{E}$
values into intervals of $\sigma$. For each bin one plots the standard
deviation of the corresponding errors, noted $\mathrm{SD}(E)$, vs
the root of the mean squared value of the selected $u_{E}$ data,
noted $\mathrm{RMS}(u_{E})$. Its applicability, as for Eq.\,\ref{eq:totalVar}\emph{,}
is limited by low bin counts, notably for small validation sets or
those with highly skewed uncertainty distributions.\citep{Scalia2020}
Note that this formulation is closely related to the LZV analysis,
but instead of estimating $\mathrm{Var}(Z)$ for binned values of
$u_{E}$, one estimates $\mathrm{SD}(E)$, with the same caveat about
the neglect of pairing between $E$ and $u_{E}$ values as for Eq.\,\ref{eq:varE}. 

Levi \emph{et al.\citep{Levi2020}} demonstrated the advantage of
their method over the intervals-based approach of Kuleshov \emph{et
al.}\citep{Kuleshov2018}. I want to emphasize here that both methods
do not test for the same ``calibration''. The former one tests for
\emph{tightness} (Levi \emph{et al.} speak of \emph{perfect} calibration),
where the latter one tests for \emph{average calibration}. I think
that one interest of the tightness concept is to make such a distinction
more legible.

\paragraph{Statistical power.}

The sizes of the groups should ideally be large enough to retain sufficient
testing power, which in some cases might limit the number of groups
and the resolution of the tightness analysis. For small validation
sets, the use of overlapping groups, and notably a sliding window
design, enables to preserve diagnostic resolution without loosing
too much testing power. 

Smaller groups mean wider confidence intervals for local statistics,
and one might find situations where the average calibration is rejected,
while it seems locally acceptable for all or most of the groups. In
such cases, it is unlikely that the power of local tests is high enough
to reach conclusions. As mentioned above, \emph{predictions which
are not average calibrated cannot be accepted as tight}. Nevertheless,
even in absence of enough power, the presence of trends in the local
statistics remains of diagnostic interest.

\subsubsection{Ranking-based validation\label{subsec:ranking-based}}

These methods evaluate how the amplitude of errors is associated with
different $u_{E}$ values. They are mostly used in applications such
as active learning, where uncertainty is used to select predictions
with potentially large errors.\citep{Tynes2021,Zheng2022} They are
not applicable to homoscedastic validation sets. 

\paragraph{Correlation coefficients.}

The rank correlation coefficient (RCC) between $uE$ and $|E|$ has
been advocated by Tynes \emph{et al.}\citep{Tynes2021} over the linear
correlation coefficient (LCC) as a validation statistic. The LCC and
RCC are intuitively expected to be positive if the larger absolute
errors are associated with larger uncertainties and null if there
is no correlation between both properties. For instance, a consistent
dataset such as SYNT01 gives a RCC of $0.49$, while SYNT02 gives
a null value. Tynes \emph{et al.}\citep{Tynes2021} report values
between 0.2 and 0.65 for various ML-UQ datasets in computational chemistry.
These are rather weak correlation coefficients, but a perfect correlation
coefficient (RCC=1) would result from an unlikely perfect predictor
(an \emph{oracle}) such as $u_{E}\propto|E|$. However, such a validation
set with perfect ranking might still fail calibration tests, as the
scaling between $u_{E}$ and $E$ is not accounted for in the RCC
value. One might therefore conclude on the absence of tightness from
a null RCC, but nothing can be inferred about calibration or tightness
from a non-null value.

Note that the $R^{2}$ score from a linear regression \emph{with}
intercept\footnote{Not to be confounded with the Birge ratio, also noted $R^{2}$ (Appendix\,\ref{sec:Var(Z)-vs.-Birge}).}
might be used to the same effect. In this case the $R^{2}$ score
is the square of the LCC. The user should however be warned that there
are several definitions of the $R^{2}$ score, one of them using a
linear regression \emph{without} intercept. This one does not relate
to the correlation coefficient. Unfortunately, this is the only version
of the $R^{2}$ score statistic implemented in a popular machine learning
package,\citep{scikit-learn} with a notable risk to be misused.

\paragraph{Confidence curves.}

A confidence curve is established by estimating a statistic of error
sets pruned from those points with uncertainties larger than a threshold.\citep{Scalia2020}
Technically, this is a ranking-based method, as the ordering of the
data plays a determinant role. 

For instance, if one defines a threshold $u_{k}$ by removing the
$k$\,\% largest uncertainties (this applies also to expanded uncertainties),
on gets a normalized confidence statistic as
\begin{equation}
c_{S}(k)=S\left(E\,|\,u_{E}<u_{k}\right)/S(E)\label{eq:Levi-1}
\end{equation}
where $S$ is an error statistic (typically the Mean Absolute Error).
A continuously decreasing confidence curve reveals a desirable association
between the larger errors and the larger uncertainties. 

Usually, an \emph{oracle curve} is plotted as reference,\citep{Scalia2020}
generated from an hypothetical dataset with perfect correlation between
$u_{E}$ and $|E|$ {[}see Fig.\,\ref{fig:05}(c){]}. This reference
is not realistic for the type of error distributions expected here.
As an alternative, I proposed\citep{Pernot2022c} to generate a reference
curve $c_{S}(k;\widetilde{E},u_{E})$ from a pseudo-error set $\widetilde{E}$
sampled from a distribution with mean 0 and standard deviation $u_{E}$
\begin{equation}
\widetilde{E}_{i}\sim D(0,u_{E_{i}})
\end{equation}
The sampling is repeated to provide a stable mean reference curve
and a confidence band (at the 95\,\% level). To avoid any ambiguity
with the \emph{oracle}, I refer to this curve as a \emph{probabilistic}
reference. The difference between \emph{oracle} and \emph{probabilistic}
reference curves can be seen in Fig.\,\ref{fig:05}(c). The effect
of the choice of $D$ has been studied elsewhere.\citep{Pernot2022c}
To summarize, it does not practically affect the reference curve itself,
but mostly the width of the confidence band. A normal distribution
is a reasonable choice in absence of specific information about $D$.

An essential point is that comparing the confidence curve $c_{S}$
to the oracle reference does not provide information about calibration
nor tightness. In fact, any transformation of the uncertainties that
does not affect their rank would result into exactly the same confidence
curve. By contrast, pairing the confidence curve with the probabilistic
reference can be considered as a proper variance-based tightness validation
method. Interestingly, it provides two kinds of diagnostics: (1) a
continuously decreasing confidence curve validates the use of the
predictive uncertainties for active learning, regardless of calibration;
and (2) a confidence curve in agreement with the probabilistic reference
validates the tightness of the uncertainties. Its main weakness when
compared to a local calibration method or to a reliability diagram
is to depend explicitly (but weakly, as discussed above) on the choice
of a probabilistic model. For the same reasons as discussed earlier,
this tightness diagnostic has to be conditioned on a positive average
calibration test. 

\subsection{Graphical representations\label{subsec:Graphical-representations}}

In practice, it is often more informative to plot the statistics and
their confidence intervals than to perform the validation tests. Several
plots have been proposed in the literature to check average calibration
(e.g. calibration curves, PIT histograms) \citep{Gneiting2014,Tran2020,Chung2021}.
They were tested in PER2022 and I found that they were of limited
interest to the typical scenarios of computational chemistry UQ. They
might nevertheless become handy in the cases where one has full probabilistic
predictions,\citep{Tran2020} but are not presented here as they do
not provide tightness diagnostics. 

\subsubsection{Local coverage probability (LCP) and local range ratio (LRR) analyses\label{subsec:Local-coverage-probability}}

The local values of $\nu_{p,m_{j},j}$ and $I_{95}(\nu_{p,m_{j},j})$
can be plotted against the location of the group centers and compared
to $p$. The same plot can represent a series of $p$ values of interest
(for instance, 0.5, 0.75 and 0.95). For a self-contained calibration/tightness
diagnostic, the values for average calibration can also be displayed
in the right margin of the plot. 

\begin{figure}[t]
\noindent \begin{centering}
\includegraphics[height=6cm]{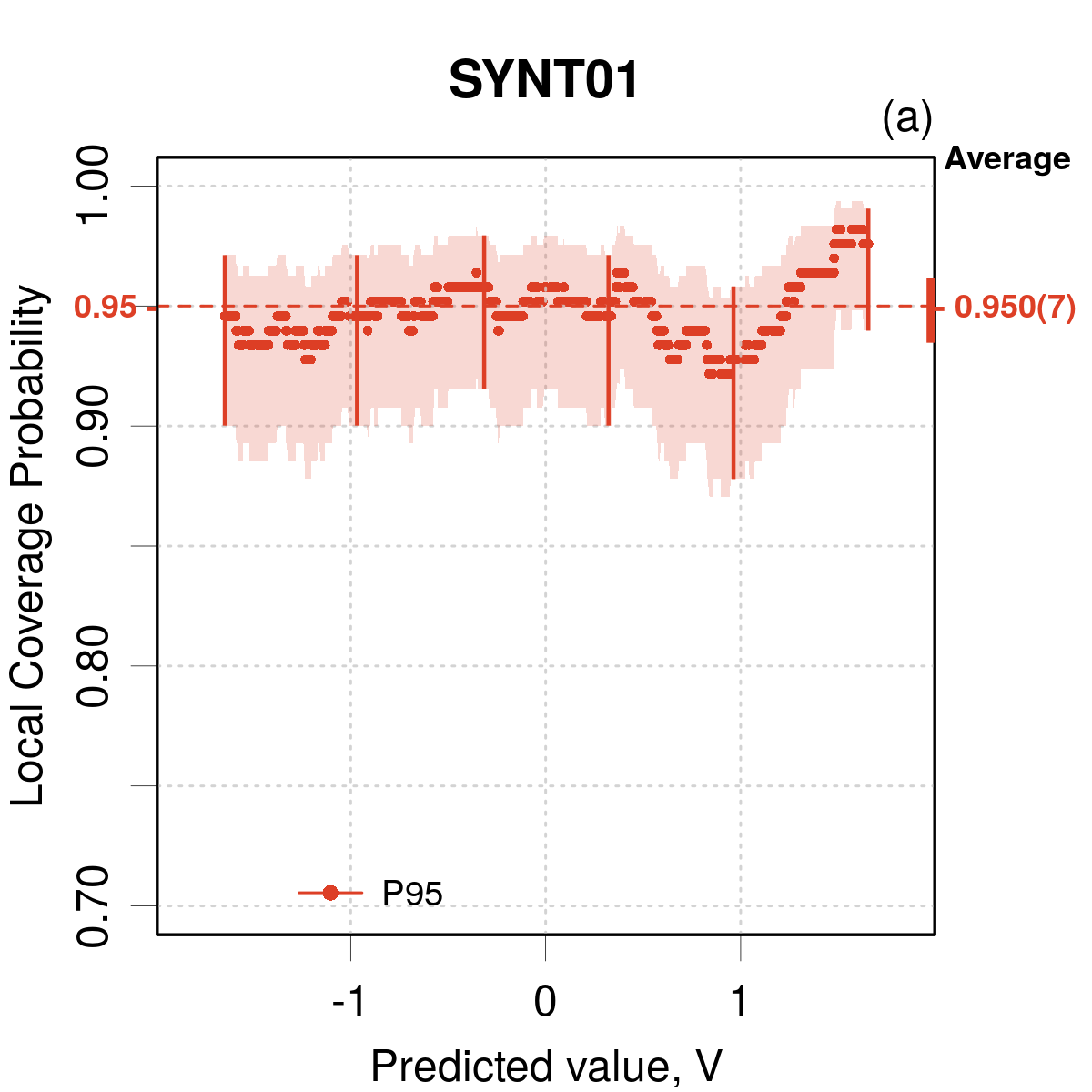}\includegraphics[height=6cm]{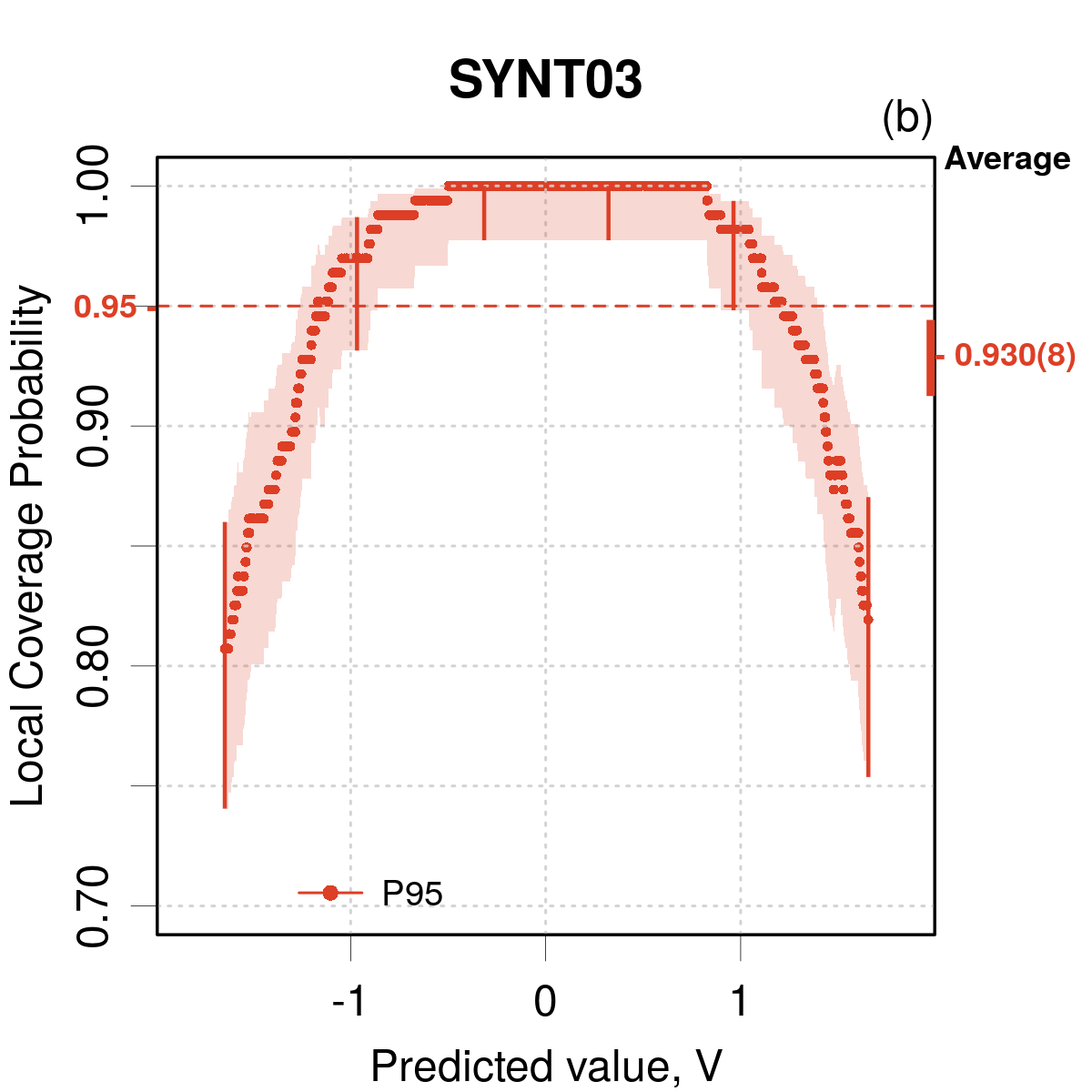}
\par\end{centering}
\noindent \begin{centering}
\includegraphics[height=6cm]{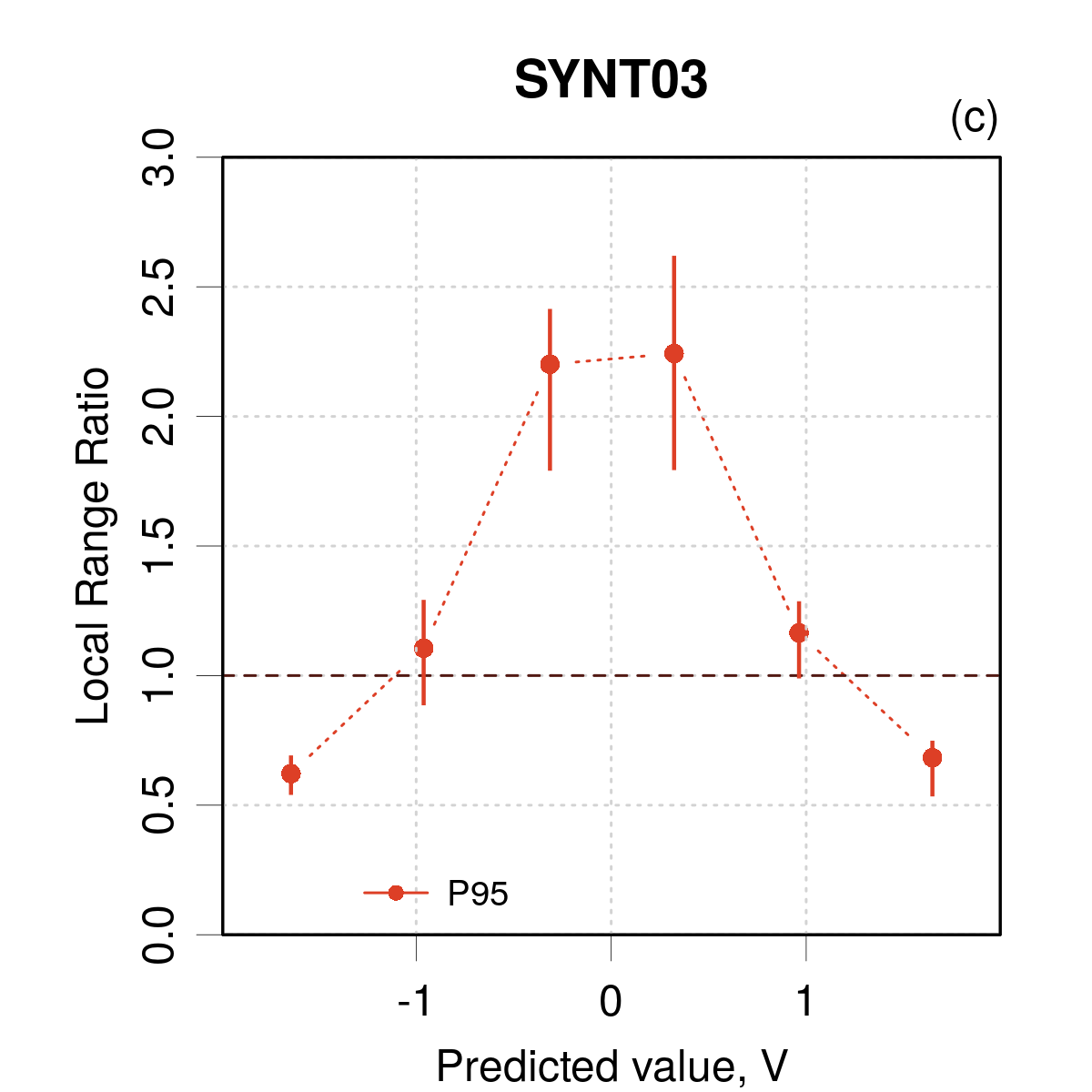}\includegraphics[height=6cm]{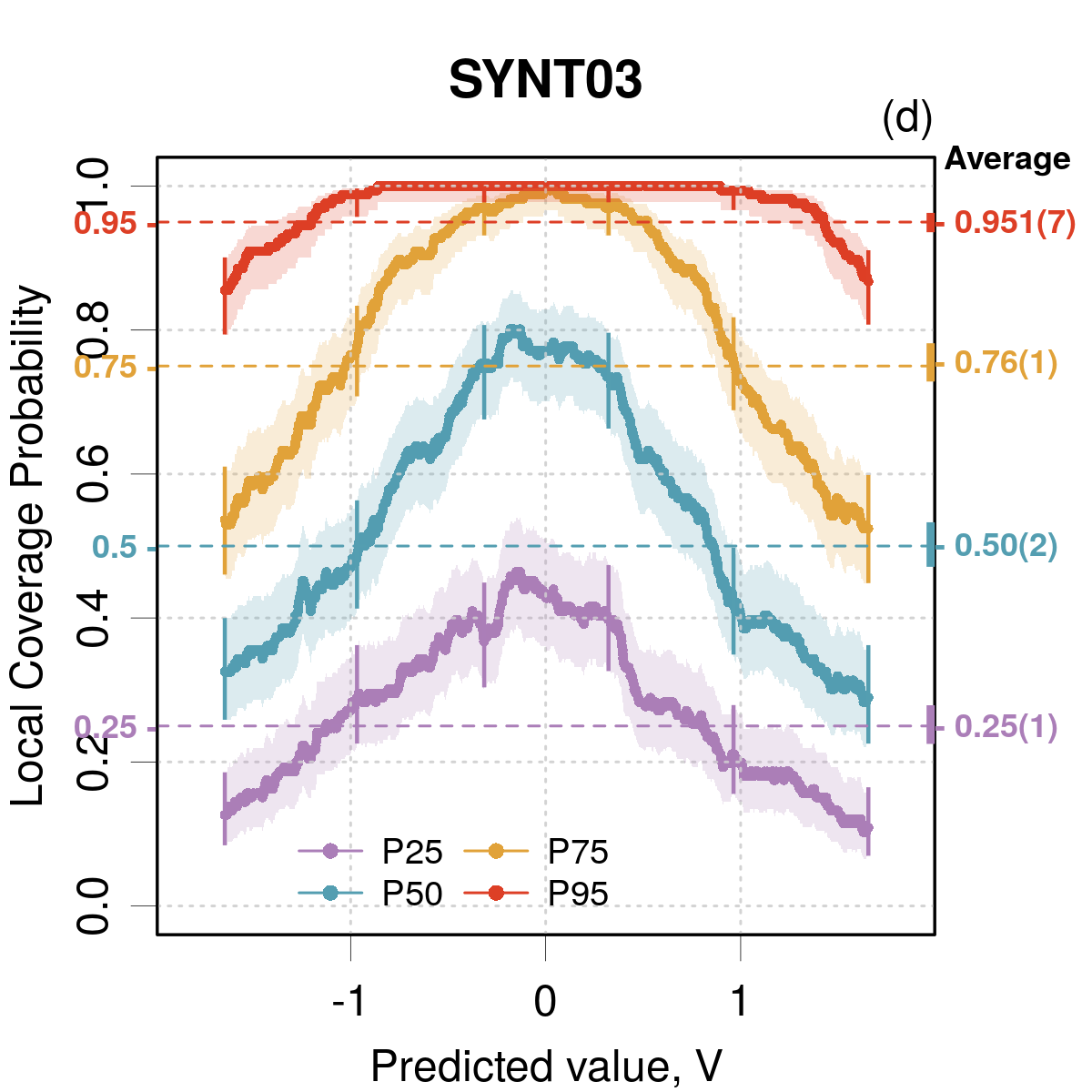}
\par\end{centering}
\caption{\label{fig:04}(a,b) Local coverage probability (LCP) analysis of
the SYNT01 and SYNT03 datasets for 95\,\% prediction intervals based
on the normality of the error generation process ($U_{E_{i},95}=1.96u_{E_{i}}$);
(c) Local range ratio (LRR) analysis for the SYNT03 dataset; (d) LCP
analysis of SYNT03 with recalibrated uniform prediction intervals
estimated from the errors.}
\end{figure}

Application to a 95\% prediction interval (based on $U_{E,95}=1.96u_{E}$)
for the SYNT01 set is shown in Fig.\,\ref{fig:04}(a), where one
can see that the error bars based on $I_{95}(\nu_{p,m_{j},j})$ for
all groups along $u_{E}$ overlap the target probability, indicating
a good tightness of prediction intervals. In the right margin, average
calibration is also attested by the PICP value for the full dataset.
These prediction intervals are therefore calibrated and tight. 

In contrast, the same analysis for the SYNT03 dataset {[}Fig.\,\ref{fig:04}(b){]}
shows unambiguously a lack of tightness (average calibration is not
optimal either considering the PICP value of 0.930(8)).\textcolor{orange}{{}
}It is clear from this graph that the constant uncertainty used for
these data is only adapted for a few groups along $V$. Moreover,
the PICP values for the overestimated uncertainties saturate to 1.0,
and we get no idea of the amplitude of the miscalibration from the
LCP analysis. Plotting the local range ratio (LRR) statistic $R_{p}$
provides us with this information {[}Fig.\,\ref{fig:04}(c){]}. For
the small uncertainties, underestimation is by a factor about 2.0,
while for the large ones, the prediction interval's width is overestimated
by a factor about 2.3. Note that the LRR analysis did not use a sliding
window because the excess computing time due to repeated bootstrapping
does not contribute to the diagnostic. The computation overload is
much less stringent for the LCP analysis, which does not use bootstrapping
for confidence intervals estimation. 

The same dataset can be used to illustrate how the statistics from
a validation set enable to make calibrated predictions without ensuring
tightness.\citep{Levi2020} Expanded uncertainties $U_{E,p}$ are
estimated from the empirical quantiles of the errors set, for $p=0.25,\,0.5,\thinspace0.75,\thinspace0.95$
and used to build uniform prediction intervals for all the points.
The LCP analysis in Fig.\,\ref{fig:04}(d) shows that calibration
is good at all the levels (the PICP values in the right margin are
consistent with their probability targets), but that tightness is
not ensured at any level, as most LCP intervals do not overlap their
target probability. 

\subsubsection{Local Z variance (LZV) analysis and reliability diagram\label{subsec:Local-Z-variance}}

\begin{figure}[t]
\noindent \begin{centering}
\includegraphics[height=6cm]{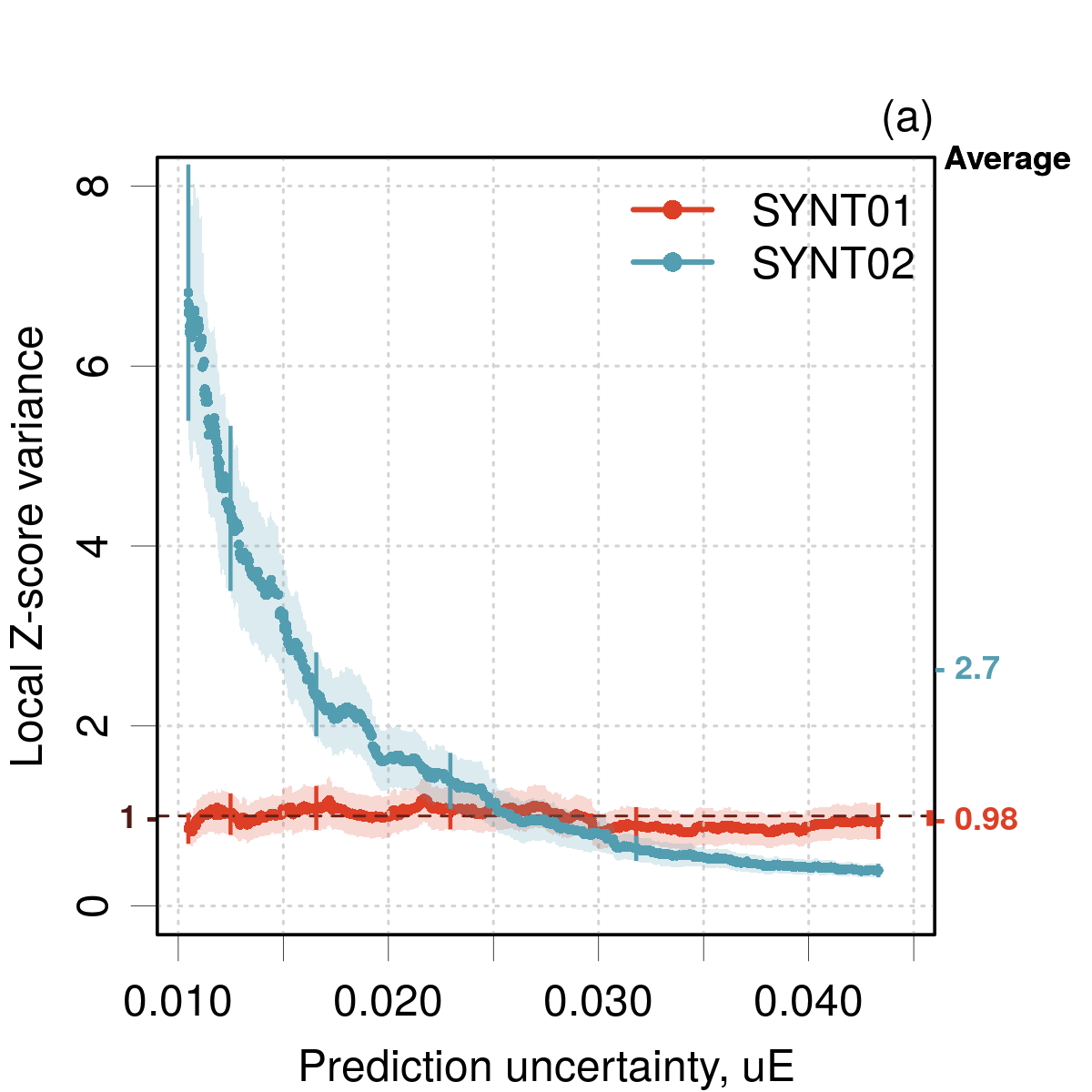}\includegraphics[height=6cm]{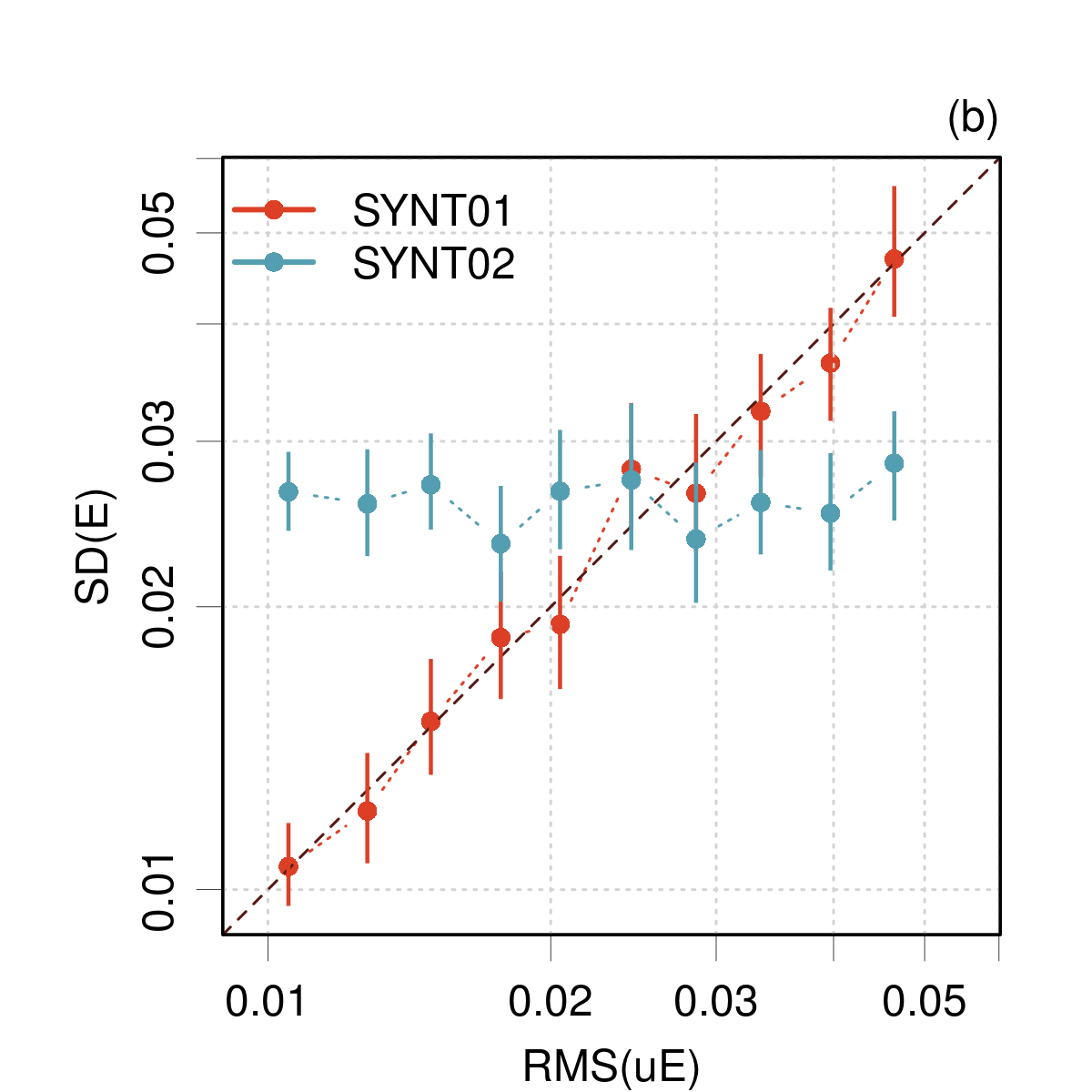}\includegraphics[height=6cm]{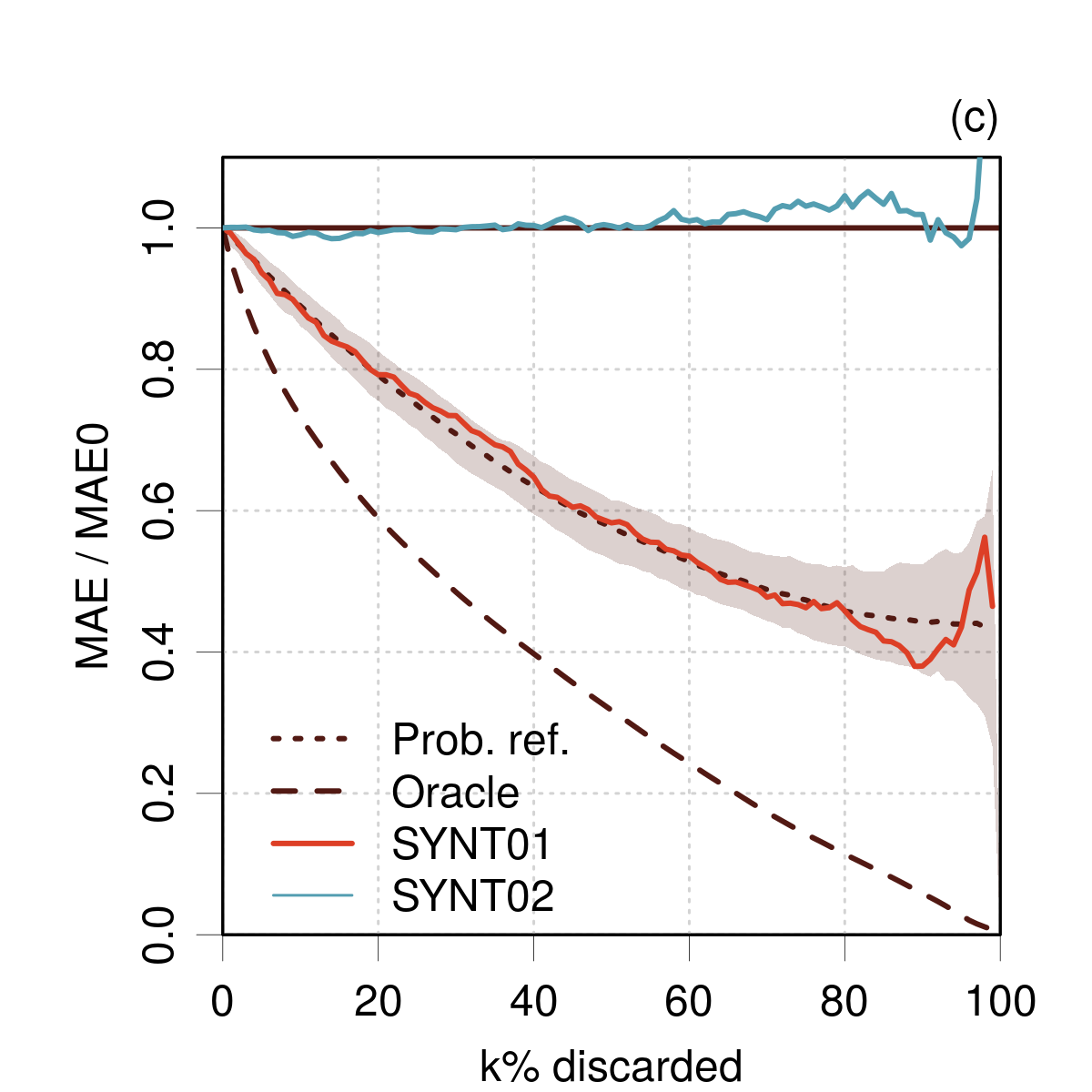}
\par\end{centering}
\caption{\label{fig:05}(a) Local \emph{z}-score variance (LZV) analysis of
the SYNT01 and SYNT02 datasets; (b) reliability diagrams; and (c)
confidence curves with oracle and probabilistic references. }
\end{figure}

A similar representation can be used for the local validation of $\mathrm{Var}(Z)$
(LZV analysis). For the SYNT01 dataset {[}Fig.\,\ref{fig:05}(a){]},
the test is fully consistent with $\mathrm{Var}(Z)=1$, which is not
the case for the SYNT02 dataset, for which $\mathrm{Var}(Z)$ varies
between about 0.5 and 7, with an average value of 2.7. Note that for
large validation sets, the use of a sliding window might present the
same computation overload as for the LRR analysis, unless replacing
the bootstrapping methods by the Cho method to estimate confidence
intervals. 

For comparison, the reliability diagram for SYNT01 and SYNT02 is presented
in Fig.\,\ref{fig:05}(b). The curve for SYNT01 follows closely the
identity line, meaning that all levels of uncertainty describe correctly
the dispersion of the corresponding errors (tightness). A contrario,
the flat line for SYNT02 reveals the lack of consistency between errors
and uncertainties. For the LZV analysis, the mismatch factor of the
prediction uncertainty can be estimated locally by the square root
of $\mathrm{Var}(Z)$. With the reliability diagram, the same information
can be obtained by taking the ratio between $\mathrm{SD}(E)$ and
$\mathrm{RMS}(u_{E})$.

For the same datasets, I plotted also the confidence curves {[}Fig.\,\ref{fig:05}(c){]}.
As the curve for SYNT02 is non-decreasing, one can conclude readily
to an absence of tightness. Comparing the confidence curve for SYNT01
to the oracle does not bring any information about calibration nor
tightness. It seems to be far from the oracle, but still, the continuously
decreasing curve is a positive feature. Comparison with the probabilistic
reference let us unambiguously conclude that the errors match the
probabilistic model relating them to uncertainties. Considering the
good value of $\mathrm{Var}(Z)$ for this dataset, one might also
conclude to a good tightness.

\subsection{The problem of small probabilistic ensembles\label{subsec:Small-ensembles}}

It was assumed in Sect.\,\ref{subsec:Notations} that probabilistic
predictions were made through distributions or prediction ensembles
that were implicitly large enough to enable an accurate estimation
of statistical summaries or empirical quantile functions used for
validation. However, it is not uncommon to find applications where
uncertainties are obtained as the standard deviation of small ensembles,
typically with less than 10 values (see examples in Sect.\,\ref{sec:Examples}).
Estimation of quantiles from such small ensembles is not possible,
barring recourse to intervals-based validation, and I would like to
consider here how ranking- and variance-based validation methods perform
in this context. 

To illustrate the problem, let us consider a normal error distribution
$N(0,\sigma)$ from which $n$ samples are drawn to estimate $\sigma$.
Let us note $s_{n}$ the standard deviation of the ensembles. The
distribution of $s_{n}$ for repeated sampling follows a scaled \emph{chi}
distribution with $n-1$ degrees of freedom
\begin{equation}
\sqrt{n-1}s_{n}/\sigma\sim\chi_{n-1}
\end{equation}
When considering a validation set one has therefore a variance source
for $u_{E}$ entangled with the variance of $E$, which makes the
validation equation $\mathrm{Var}(E/u_{E})=1$ irrelevant. Kacker
\emph{et al.}\citep{Kacker2010} formulated this in other terms by
showing that the Birge ratio should not be used to estimate the statistical
consistency of GUM\citep{GUM} type A uncertainties. In fact, when
uncertainty is estimated by the standard deviation of a small ensembles,
the ratio of the mean error to the standard error is a \emph{t}-score
(or\emph{ t-}statistic)
\begin{equation}
T=<E>/(s_{n}/\sqrt{n})
\end{equation}
In the case of a normal error distribution, $T$ has a Student's-\emph{t
}distribution with $n-1$ degrees of freedom, and its variance is
\begin{equation}
\mathrm{Var}(T)=(n-1)/(n-3)\label{eq:varT}
\end{equation}
When $n$ increases, the Student's-\emph{t} distribution converges
to the standard normal, and one recovers $\mathrm{Var}(T)=1$. 

We are thus left with two questions:
\begin{enumerate}
\item \emph{For average calibration, how does Eq.\,\ref{eq:varT} hold
for non-normal distributions ?} This point is studied in Appendix\,\ref{sec:Calibration-of-t-scores}
and summarized here. Deviations from Eq.\,\ref{eq:varT} for a large
range of distribution shapes can mostly be neglected for $n\ge10$.
For smaller ensembles, one should allow for a wider range of $\mathrm{Var}(T)$
values, that can be extracted from Fig.\,\ref{fig:A-02}. For instance,
for $n=5$, $\mathrm{Var}(T)$ values around 2, between 1.7 and 2.4,
could be accepted.
\item \emph{Which diagnostics can be used for tightness assessment ?} This
point is explored in Appendix\,\ref{sec:Validation-of-t-scores}.
The main conclusion is that all plots against $u_{E}$ (e.g. $(u_{E},E)$
plots, LZV analysis vs. $u_{E}$ or reliability diagrams) are strongly
perturbed by statistical noise and should not be used. A LZV analysis
vs. $V$ is more useful in this context. 
\end{enumerate}
My recommendation for probabilistic predictions based on small ensembles
($n<30$) would thus be (1) to check average calibration through Eq.\,\ref{eq:varT},
possibly adapted for $n<10$, and (2), conditional to average calibration,
to check tightness through a LZV analysis vs. $V$.

\section{Examples\label{sec:Examples}}

I present below several case studies based on datasets extracted from
the computational chemistry literature. The first one, PRO2022,\citep{Proppe2022}
was already presented in PER2022 (under the PRO2021 tag). It is reconsidered
here to show the interest and limits of the ($u_{E}$,$E$), LRR plots
and confidence curves. In the same spirit, two other examples treated
in PER2022 are also briefly treated together (PAN2015 and PAR2019).
A recent dataset extracted from the ATOMIC-2$_{um}$ protocol\citep{Bakowies2022}
is introduced, along with two cases dealing with uncertainties extracted
from small ensembles of predictions: LIN2021\citep{Lin2021} from
five repeats of a Free Energy Perturbation protocol and ZHE2022\citep{Zheng2022}
from an ensemble of eight neural networks (NN) predictions in a query
by committee (QbC)\citep{Smith2018} protocol.

\subsection{PRO2022\protect\footnote{The data provided by Jonny Proppe were initially associated with a
2021 preprint by Proppe and Kircher \citep{Proppe2021} and labeled
PRO2021. For consistency, I now refer to the published version of
the paper \citep{Proppe2022} and label the data PRO2022.}}

\begin{figure}[t]
\noindent \begin{centering}
\includegraphics[height=6cm]{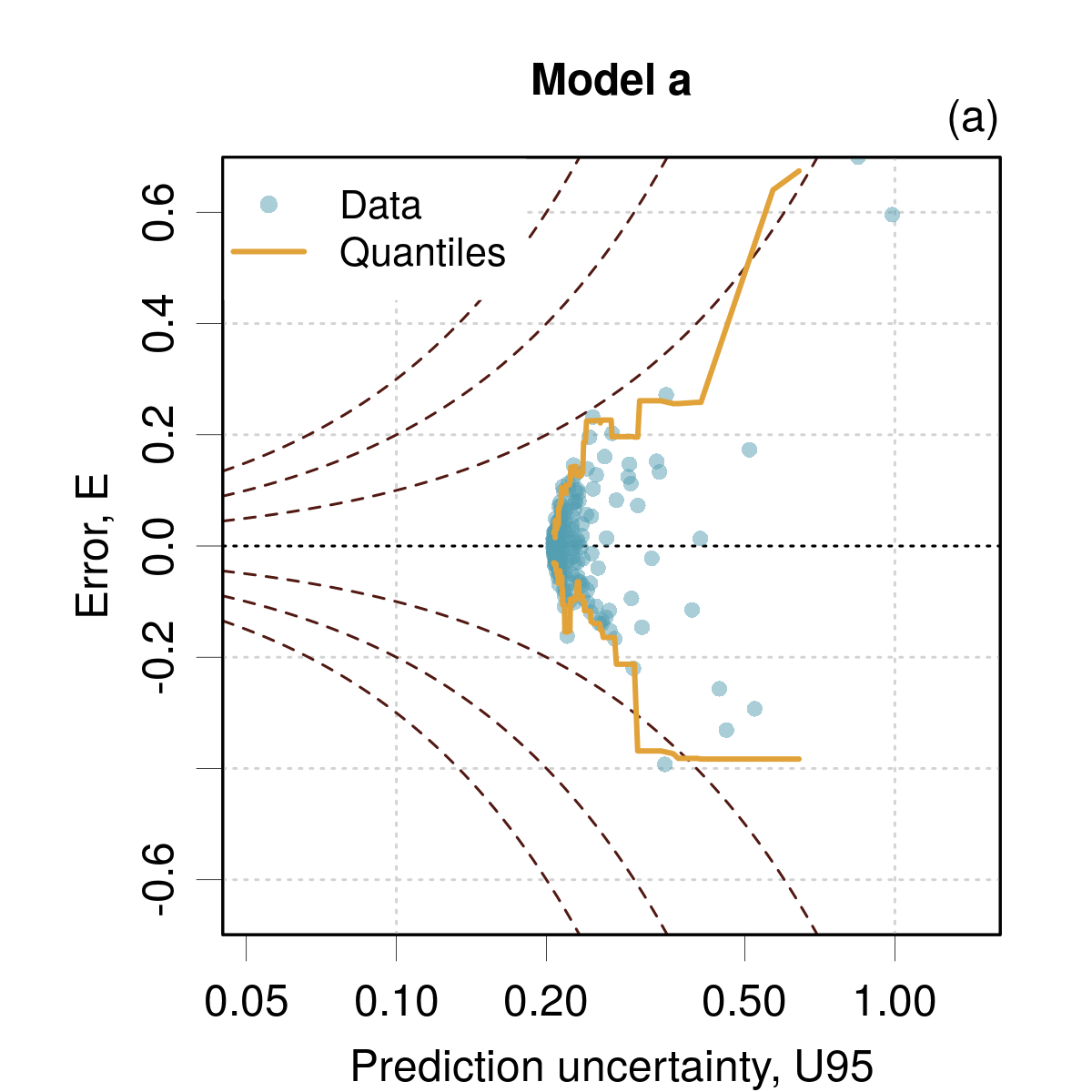}\includegraphics[height=6cm]{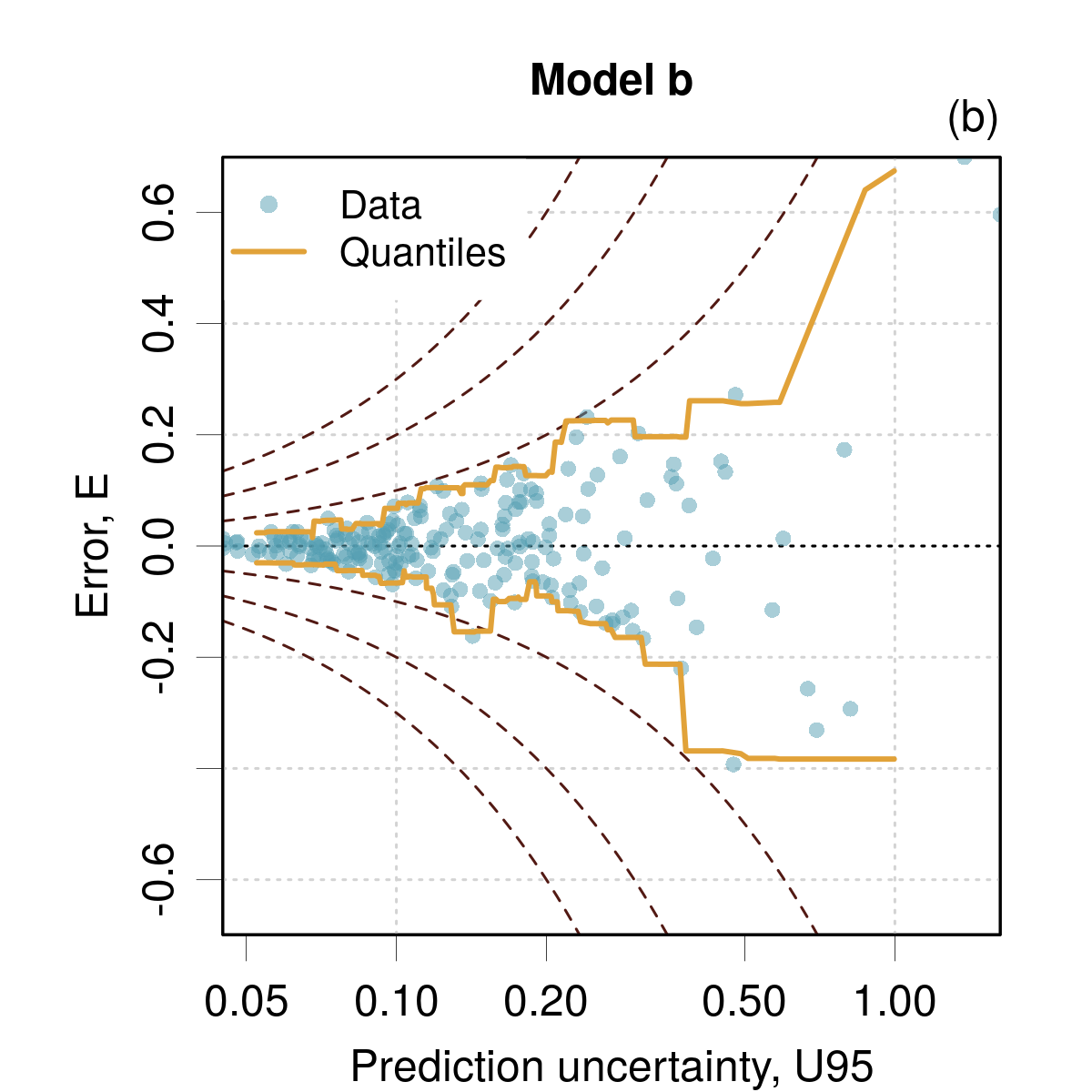}\includegraphics[height=6cm]{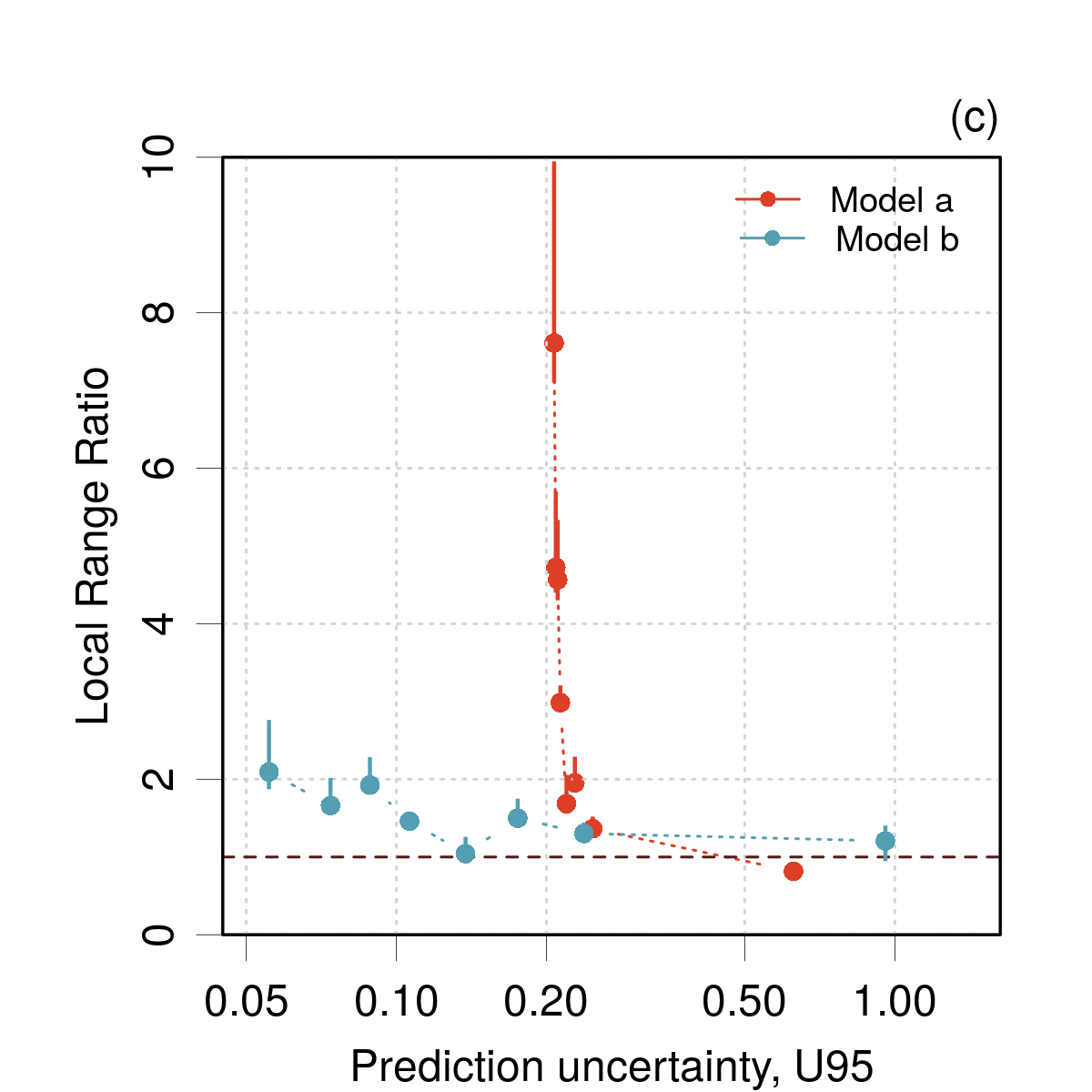}
\par\end{centering}
\noindent \begin{centering}
\includegraphics[height=6cm]{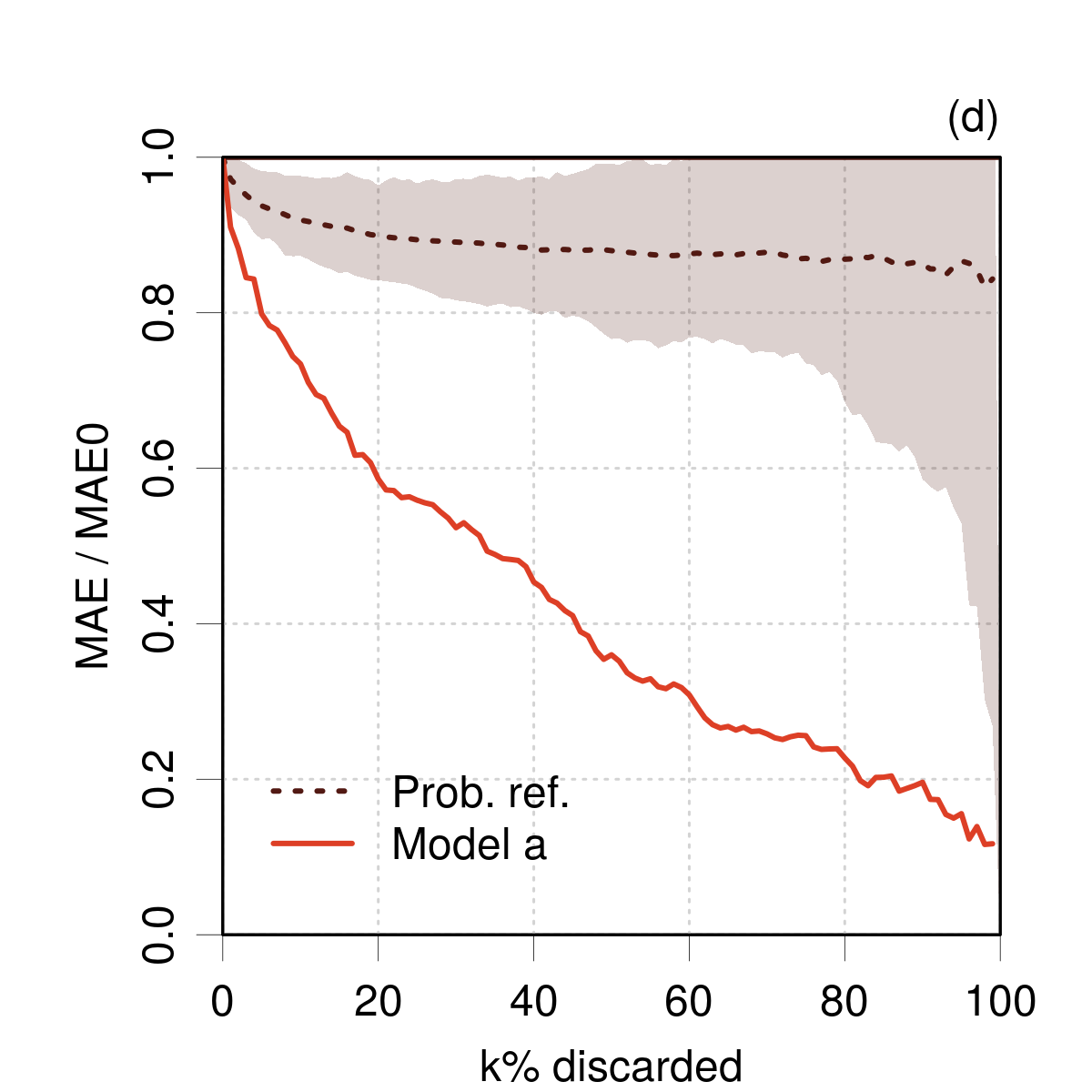}\includegraphics[height=6cm]{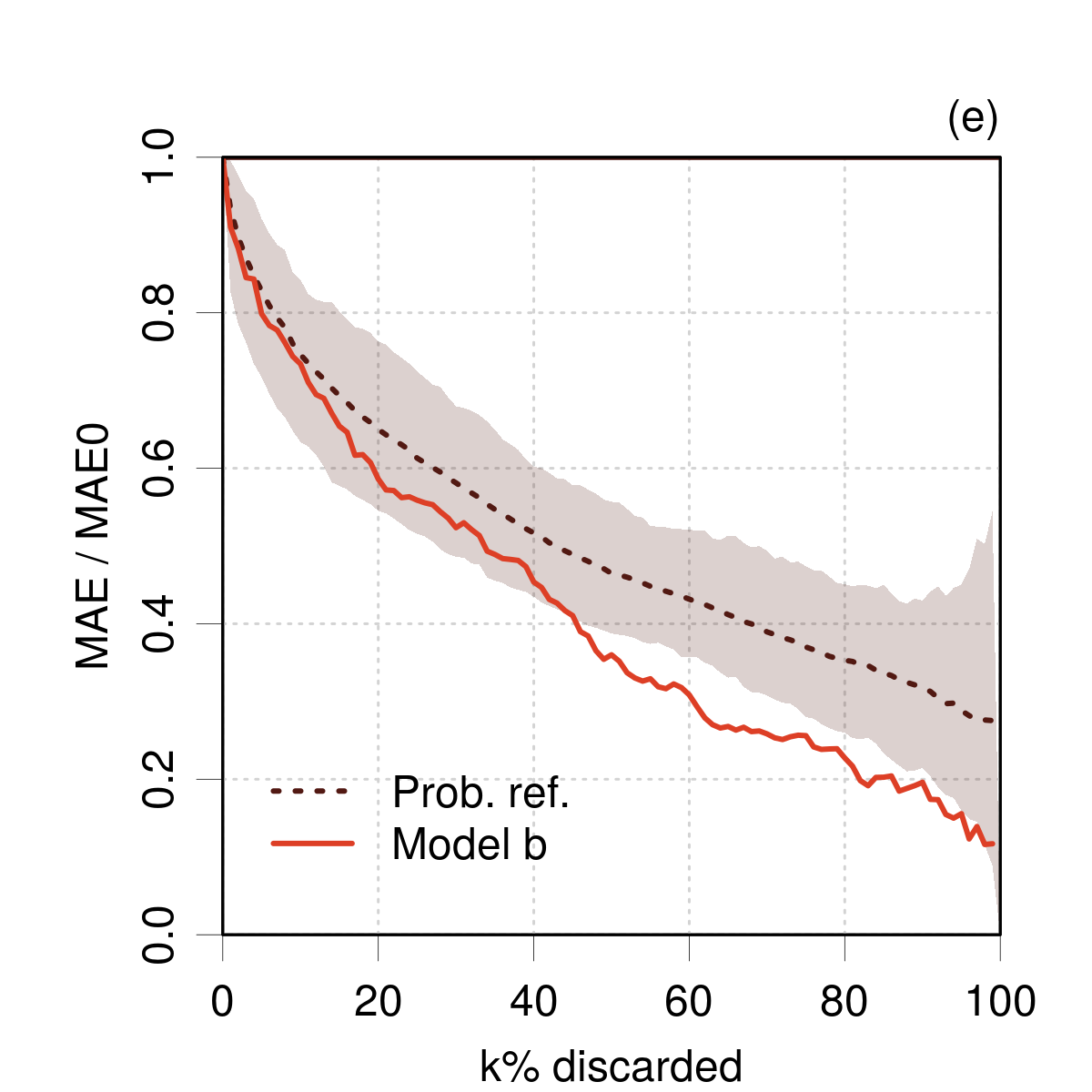}
\par\end{centering}
\caption{\label{fig:06}Calibration/tightness study for error Models $a$ and
$b$ of the PRO2022 dataset: (a,b) $(U_{E,95},E)$ plots for Model
\emph{a }and Model \emph{b}; (c) LRR analysis (8 groups); (d) and
(e) confidence curves for Models $a$ and $b$. All data are unitless.}
\end{figure}

I revisit here the treatment I made of these data in PER2022\citep{Pernot2022a}.
Proppe and Kircher\citep{Proppe2022} compared two models to estimate
a prediction uncertainty for the logarithm of reaction rates and provided
expanded uncertainties $U_{E,95}$ for both models. The data are dimensionless.
($U_{E,95}$,E) plots {[}Fig.\,\ref{fig:06}(a,b){]} show unambiguously
that model $b$ is much better than model $a$, in the sense that
there is a better match between the scale of errors and uncertainties,
notably for smaller uncertainties (below 0.2). However, one notices
(as already done by Proppe and Kircher) that in both cases a single
point is located outside of the $y=\pm x$ interval, which suggests
an overestimation of $U_{E,95}$. 

In fact, average calibration provides a PICP value of $\nu_{0.95}=0.995(5)$
for both methods. The statistical uncertainty is too small for the
confidence interval $I_{95}(\nu_{0.95})$ to include the target value
(0.95).\textcolor{orange}{{} }It is striking that, despite the difference
observed in the ($U_{E,95}$,E) plots, both models are identically
calibrated, illustrating the shortcomings of considering average calibration
without considering tightness. 

When PICP values are close to their upper limit, a LRR plot should
be used to get a quantitative appreciation of the overestimation of
$U_{E,95}$ (in PER2022, in the absence of the LRR analysis, a LZV
analysis using $u_{E}=U_{E,95}/2$ was done). One can see on Fig.\,\ref{fig:06}(c)
that Model $a$ provides prediction intervals that can be up to eight
times too wide, while this does not exceed a factor two for model
$b$. 

Despite their considerable difference in calibration, both models
have identical confidence curves {[}generated using $u_{E}=U_{E,95}/2$,
Fig.\,\ref{fig:06}(d,e){]}, a reflection of the fact that both uncertainty
sets have the same ranking (the Spearman (rank) correlation coefficient
between both uncertainty sets is 1). However, the probabilistic reference
curves enable to confirm the diagnostic that Model \emph{b} is much
closer to a good tightness than Model \emph{a}. For Model \emph{b},
it appears that the problem lies essentially in the small uncertainties.
Despite the non-perfect calibration, one sees that both models provide
uncertainties that would be suitable for active learning.

\subsection{BAK2022}

Like its predecessor (ATOMIC\citep{Bakowies2019,Bakowies2020}), the
ATOMIC-2$_{um}$ method\citep{Bakowies2022} provides uncertainties
on its predictions by a composite protocol. A set of 184 predictions
has been compared to ATcT\citep{Ruscic2014} reference values. The
corresponding data ($R$, $U_{R,95}$, $V$ and $U_{V,95}$) have
been collected from Table S20 of the reference article. 

\begin{figure}[t]
\noindent \begin{centering}
\includegraphics[height=6cm]{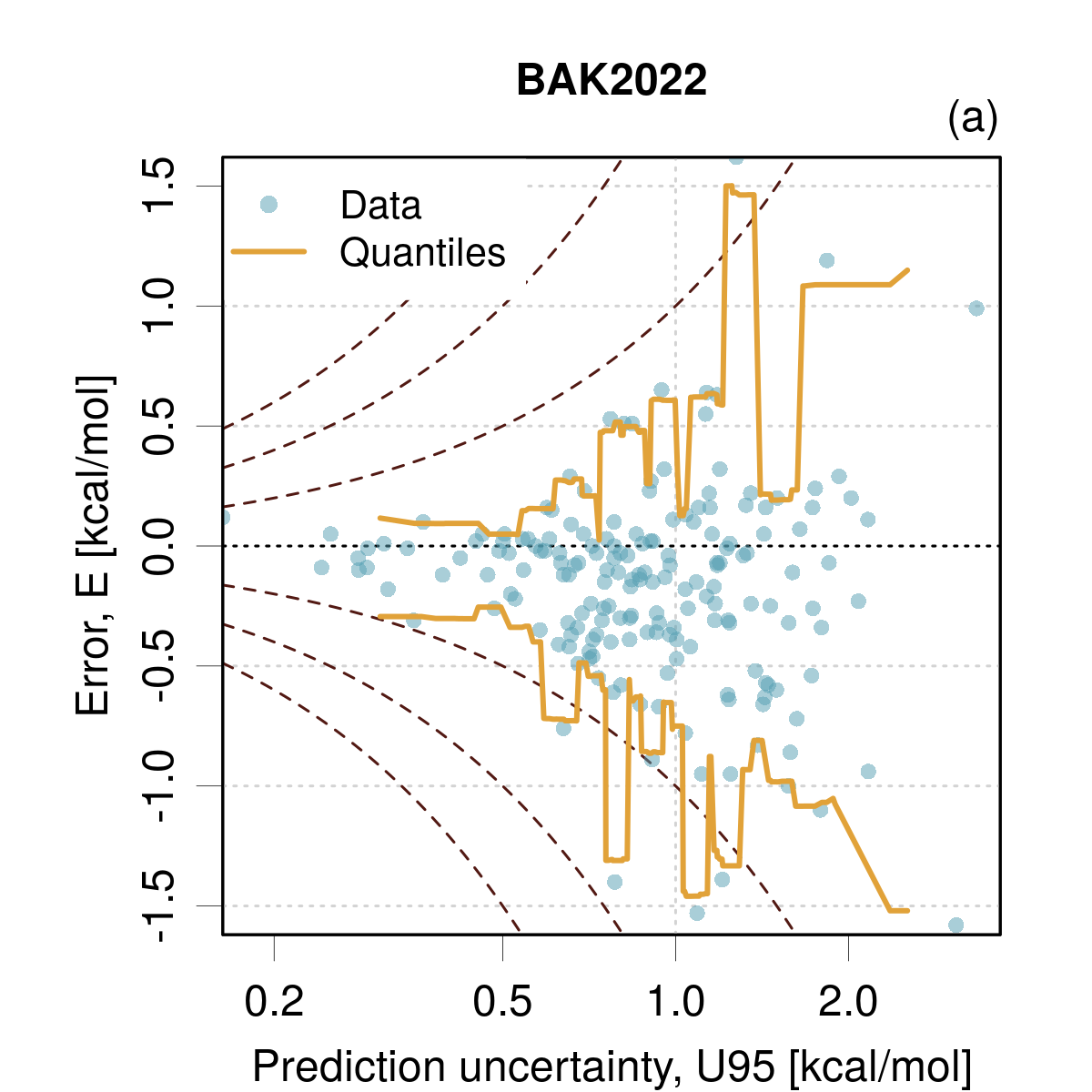}\includegraphics[height=6cm]{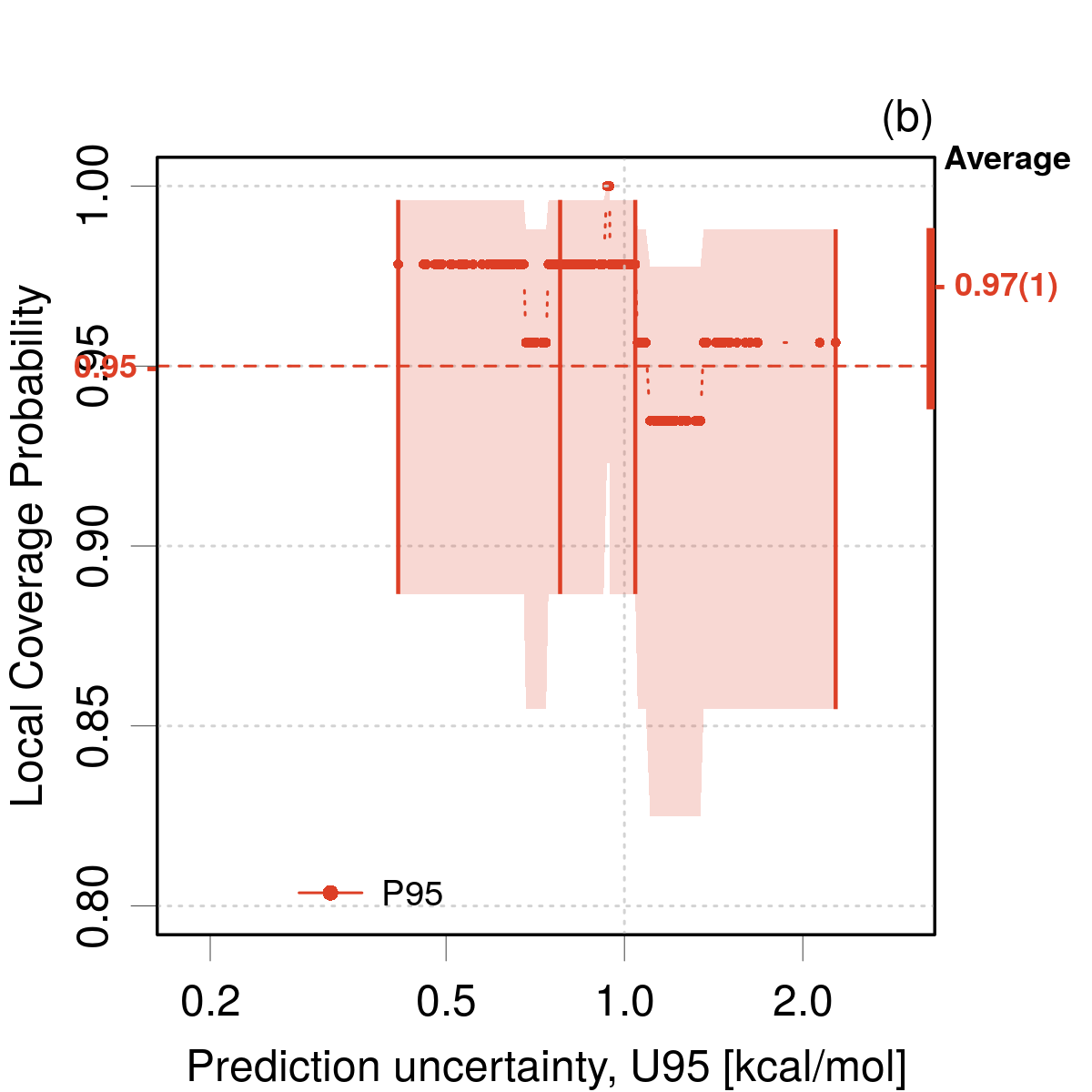}\includegraphics[height=6cm]{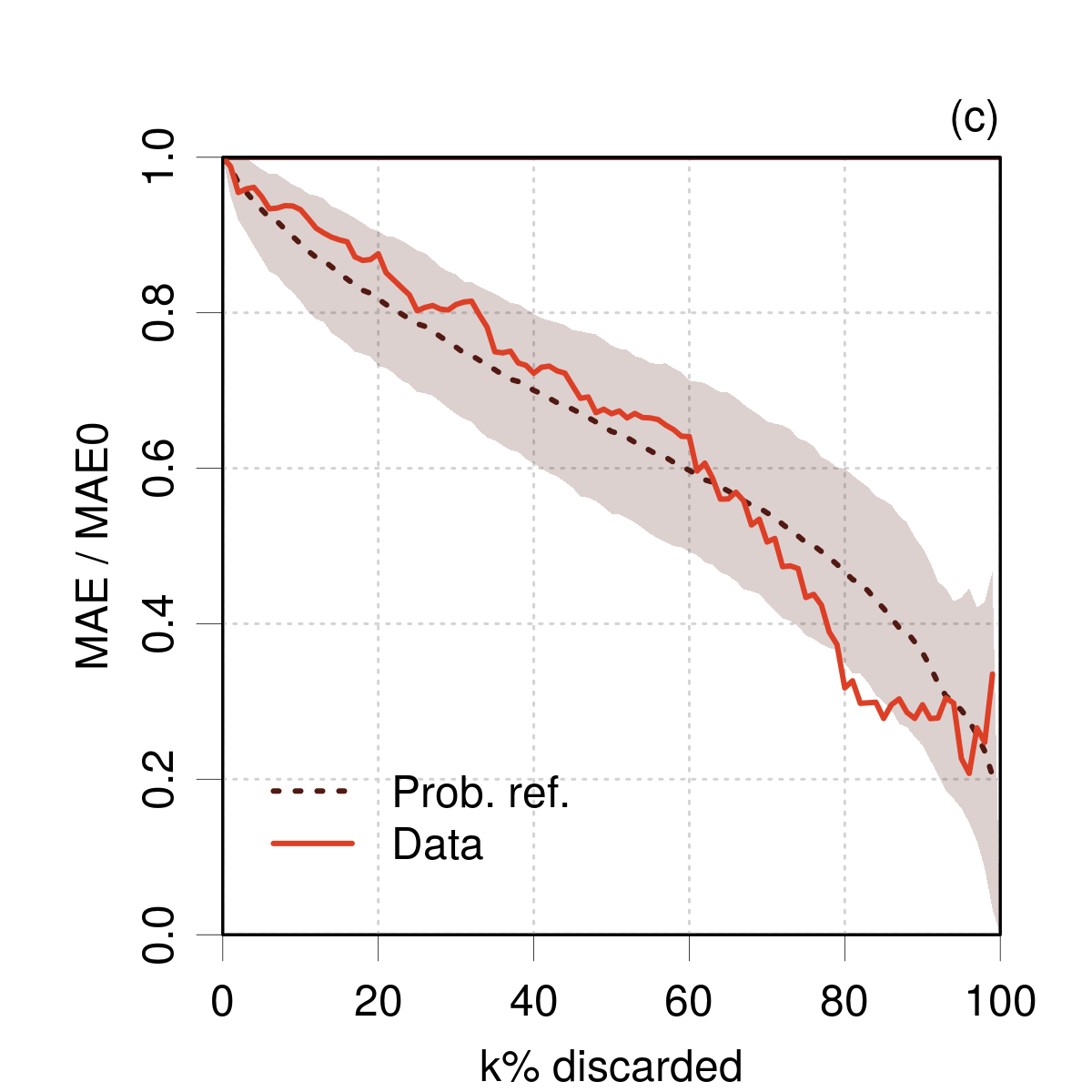}
\par\end{centering}
\caption{\label{fig:07}Calibration/tightness study for the BAK2022 dataset:
(a) $(U_{E,95},E)$ plot; (b) LCP analysis (4 groups); (c) confidence
curve.}
\end{figure}

On the $(U_{E,95},E)$ plot {[}Fig.\,\ref{fig:07}(a){]}, the errors
are well contained between the $y=\pm x$ lines with a few outlying
points. The seems however to be a slight bias of the errors towards
the negative values that might compromise the hypothesis of symmetric
prediction intervals. I checked that the correction of this bias does
not improve the calibration/tightness results, so I worked with the
original data.

The uncertainties have been calibrated by Bakowies to target a 95\%
coverage in subsets of a large dataset (more than 1100 values), with
an overall coverage of 97.2\%.\citep{Bakowies2022} From the more
limited dataset used here, I get a compatible value of 0.97(1), which
does not exclude the 0.95 target {[}Fig.\,\ref{fig:07}(b){]}, although
the LCP analysis shows a trend for overestimation of the small uncertainties.
A confidence curve, built using $u_{E}=U_{E,95}/2$ confirms this
diagnostic {[}Fig.\,\ref{fig:07}(c){]}. It shows a very good tightness,
except for the bottom 25\% of the uncertainties, where the curve drops
and makes an excursion out of the probabilistic reference band.

Globally, these diagnostics confirm that the uncertainties estimated
by the ATOMIC-2$_{um}$ protocol are globally and locally reliable,
with a small trend to be conservative, notably for the smaller uncertainties
(below ca. 0.7\,kcal/mol). Note that in this range, the calculated
uncertainties are in average four times larger than the reference
uncertainties. It is therefore unlikely that the overestimation problem
comes from the reference uncertainties.

\subsection{PAN2015 and PAR2019}

BEEF-based CC-UQ methods are calibrated through a \emph{parameters
uncertainty inflation} (PUI) scheme \citep{Pernot2017,Pernot2017b}
that implies strong functional constraints which play against their
tightness. \citep{Simm2016,Reiher2022} As the calibration is quantified
by the mean prediction variance, there is no guarantee that the prediction
uncertainty is reliable for any single prediction. 

\paragraph{PAN2015.}

\begin{figure}[t]
\noindent \begin{centering}
\includegraphics[height=6cm]{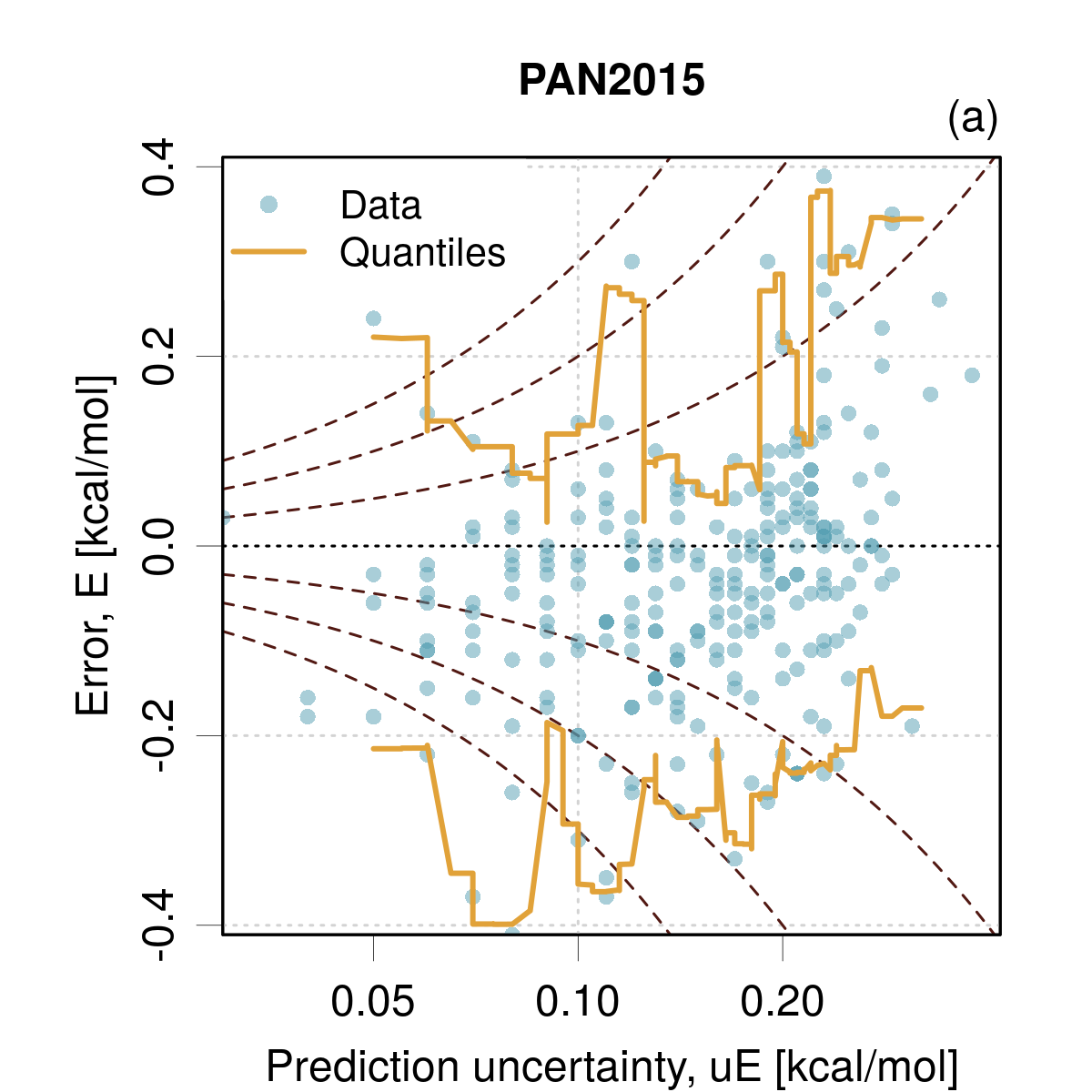}\includegraphics[height=6cm]{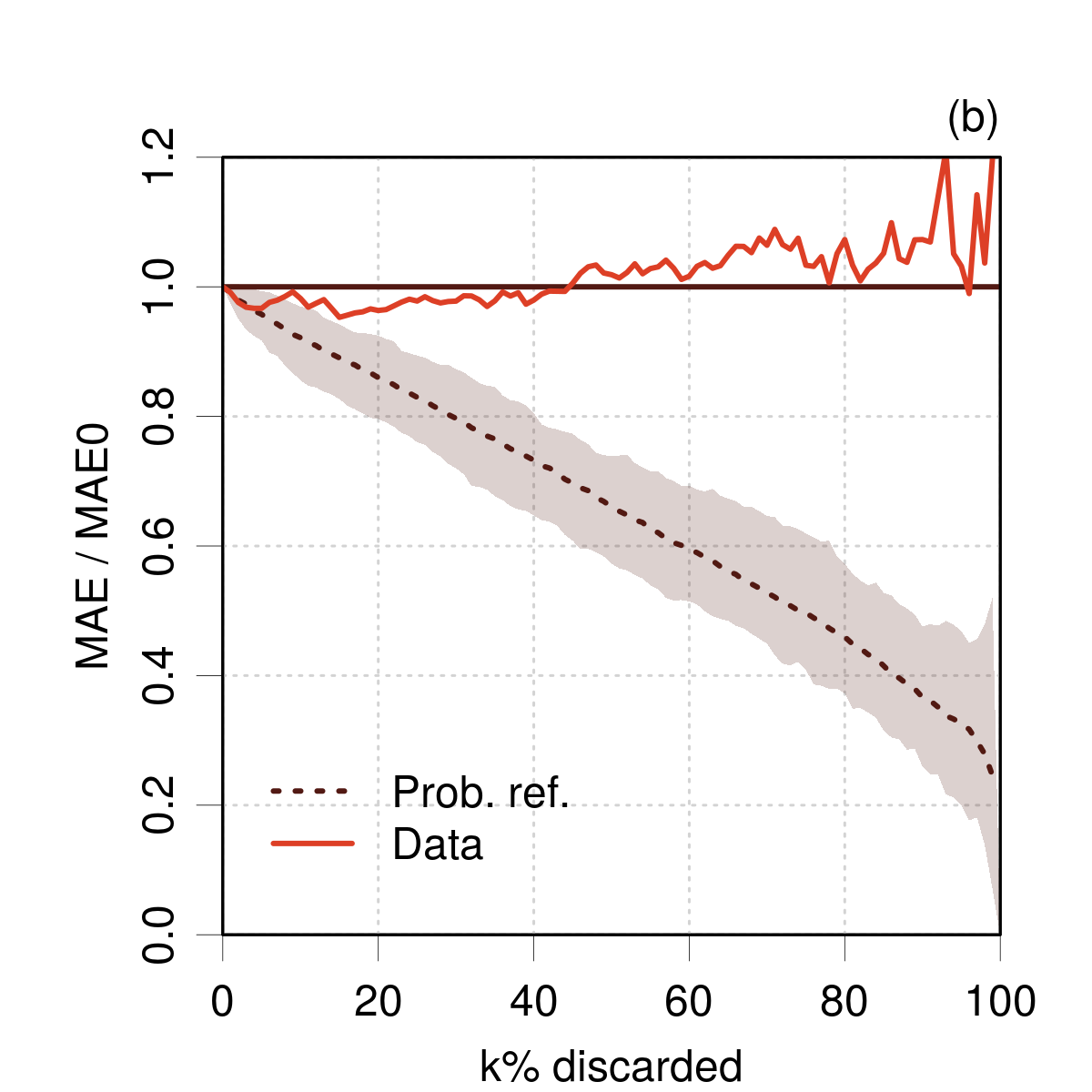}
\par\end{centering}
\noindent \begin{centering}
\includegraphics[height=6cm]{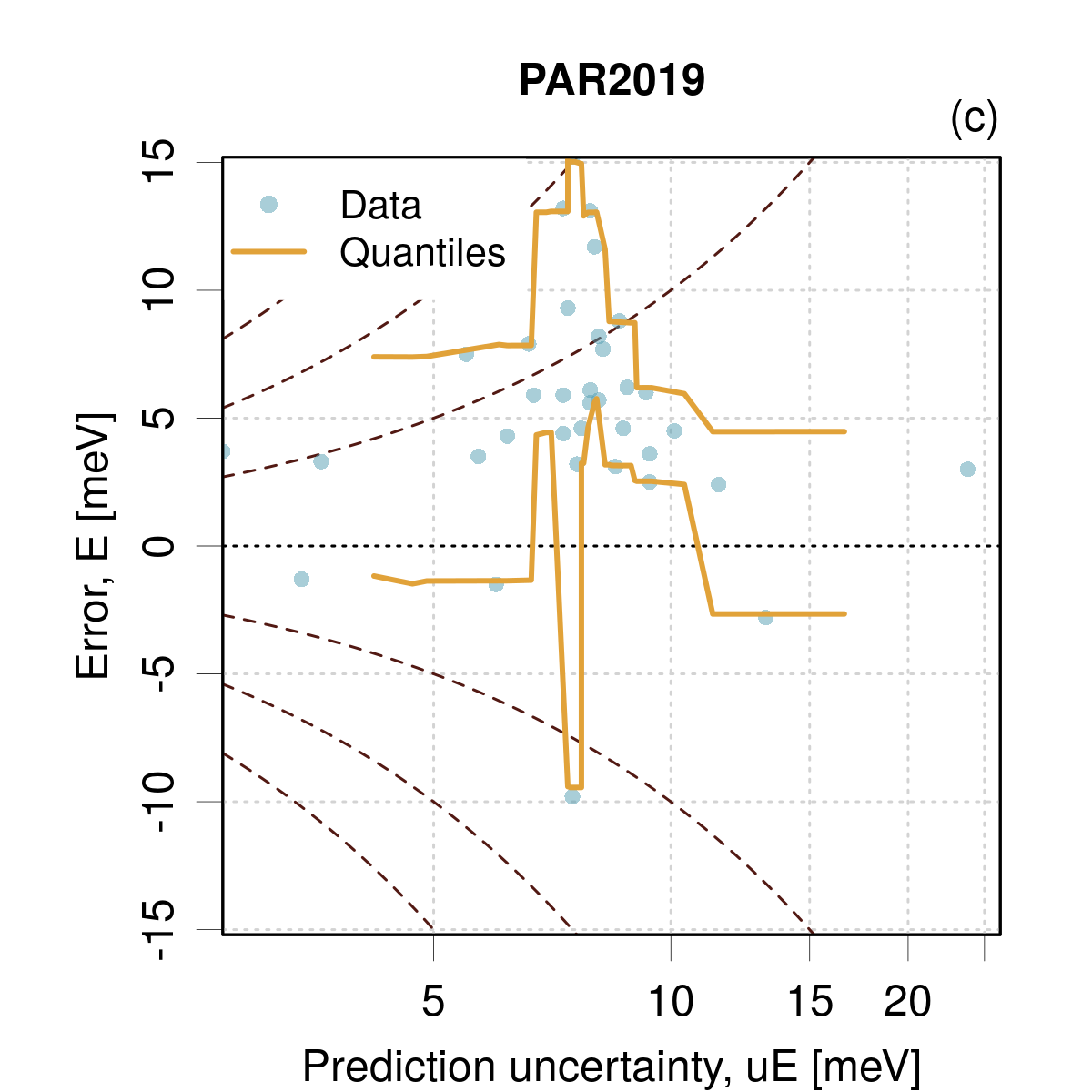}\includegraphics[height=6cm]{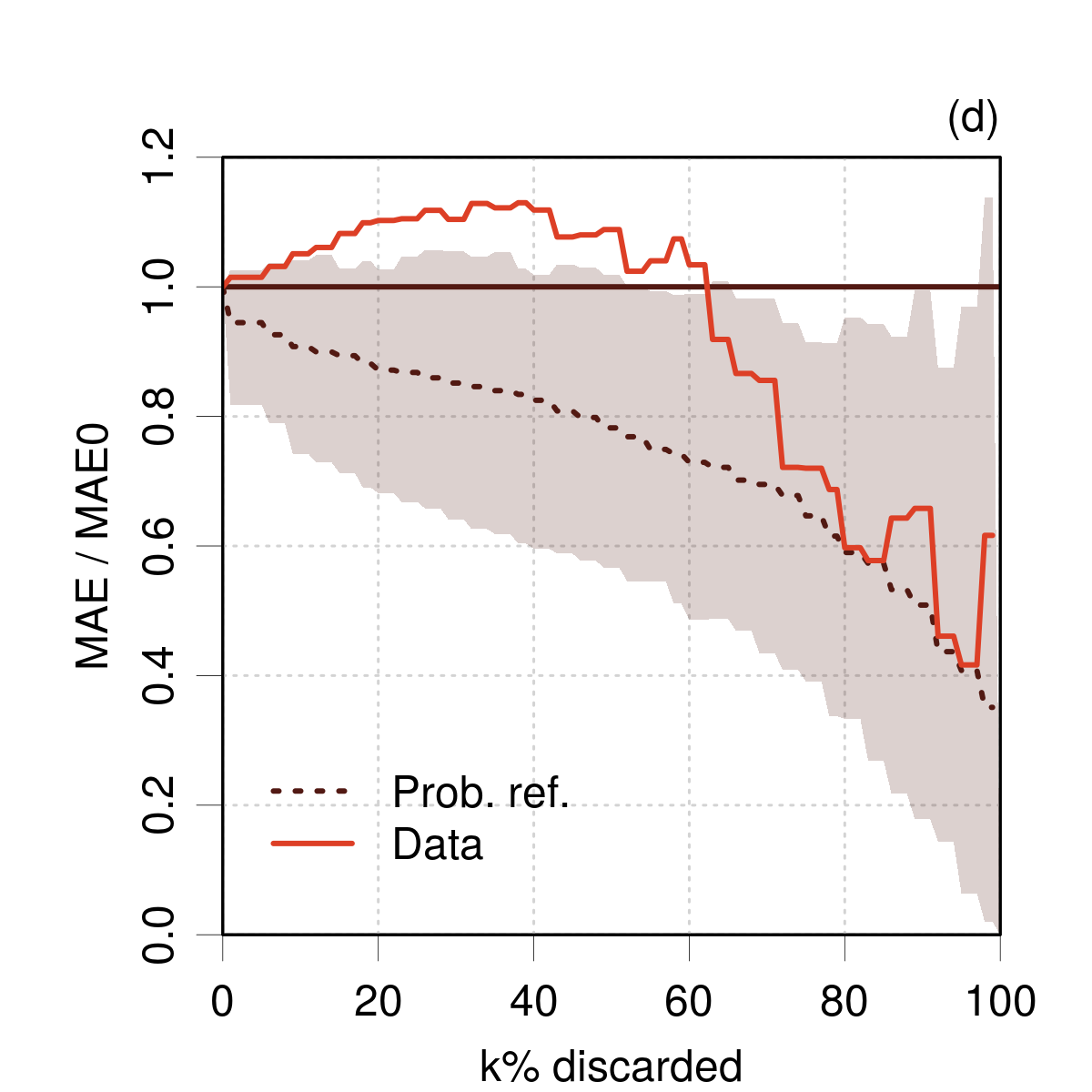}
\par\end{centering}
\noindent \centering{}\caption{\label{fig:08}(a,c)$(u_{E},E)$ curves for cases PAN2015 and PAR2019;
(b,d) corresponding confidence curves.}
\end{figure}

This validation set of 257 formation heats and their standard uncertainties
predicted by the mBEEF DFT has been extracted from a 2015 article
by Pandey \emph{et al}.\citep{Pandey2015} I previously analyzed this
dataset \citep{Pernot2017b,Pernot2022a}, showing an inconsistency
between the prediction uncertainties and the errors amplitudes. A
variance-based analysis has been performed in PER2022, showing a correct
calibration with $\mathrm{Var}(Z)=1.28(20)$. However, the LZV analysis
with respect to the prediction uncertainty revealed an absence of
tightness, with $\mathrm{Var}(Z)$ values varying between 3 and 0.5.
To complement this analysis, an $(u_{E},E)$ plot and a confidence
curve are reported in Fig.\,\ref{fig:08}(a,b). The $(u_{E},E)$
plot shows that uncertainties do not quantify correctly the dispersion
of errors, but the most striking plot is certainly the confidence
curve. As it is non-decreasing, it clearly reveals that errors and
uncertainties are not statistically consistent. I find this representation
to be the most revealing when compared to those presented in my earlier
studies of this dataset.\citep{Pernot2017b,Pernot2022a}

\paragraph{PAR2019.}

As another example of BEEF-generated uncertainties, I considered in
PER2022 a small set of 35 harmonic vibrational frequencies issued
from an article by Parks \emph{et al. }\citep{Parks2019}. For this
set, one has $\mathrm{Var}(Z)=0.42(13)$, a negative calibration test.
The data are too sparse to attempt a LZV analysis. Fig.\,\ref{fig:08}(c,d)
reports the $(u_{E},E)$ plot and confidence curve. Both plots enable
to conclude to an absence of tightness, the confidence curve showing
again an inconsistent ranking between absolute errors and uncertainties.

These examples clearly confirm that a method designed for average
calibration on a learning set should not be expected to produce reliable
prediction uncertainties. 

\subsection{Small-ensemble predictions}

\subsubsection{LIN2021}

In a recent study on the prediction of binding free energies by the
Free Energy Perturbation (FEP) protocol, Lin \emph{et al.} \citep{Lin2021}
provided a set of data including reference experimental values, FEP
values and FEP uncertainties for relative binding free energies (RBFE)
and absolute binding free energies (ABFE). The RBFE dataset contains
results for two versions of the FEP method, and I kept here the first
one (Full FEP protocol) for which more data are provided ($M=333$).
Predicted values and uncertainties on the FEP procedures were produced
by taking the mean and standard deviation of five repeats of the protocol
($n=5$). To check the statistical consistency of the errors, one
should therefore divide the reported standard deviations by $\sqrt{n}$. 

\begin{figure}[t]
\noindent \begin{centering}
\includegraphics[height=6cm]{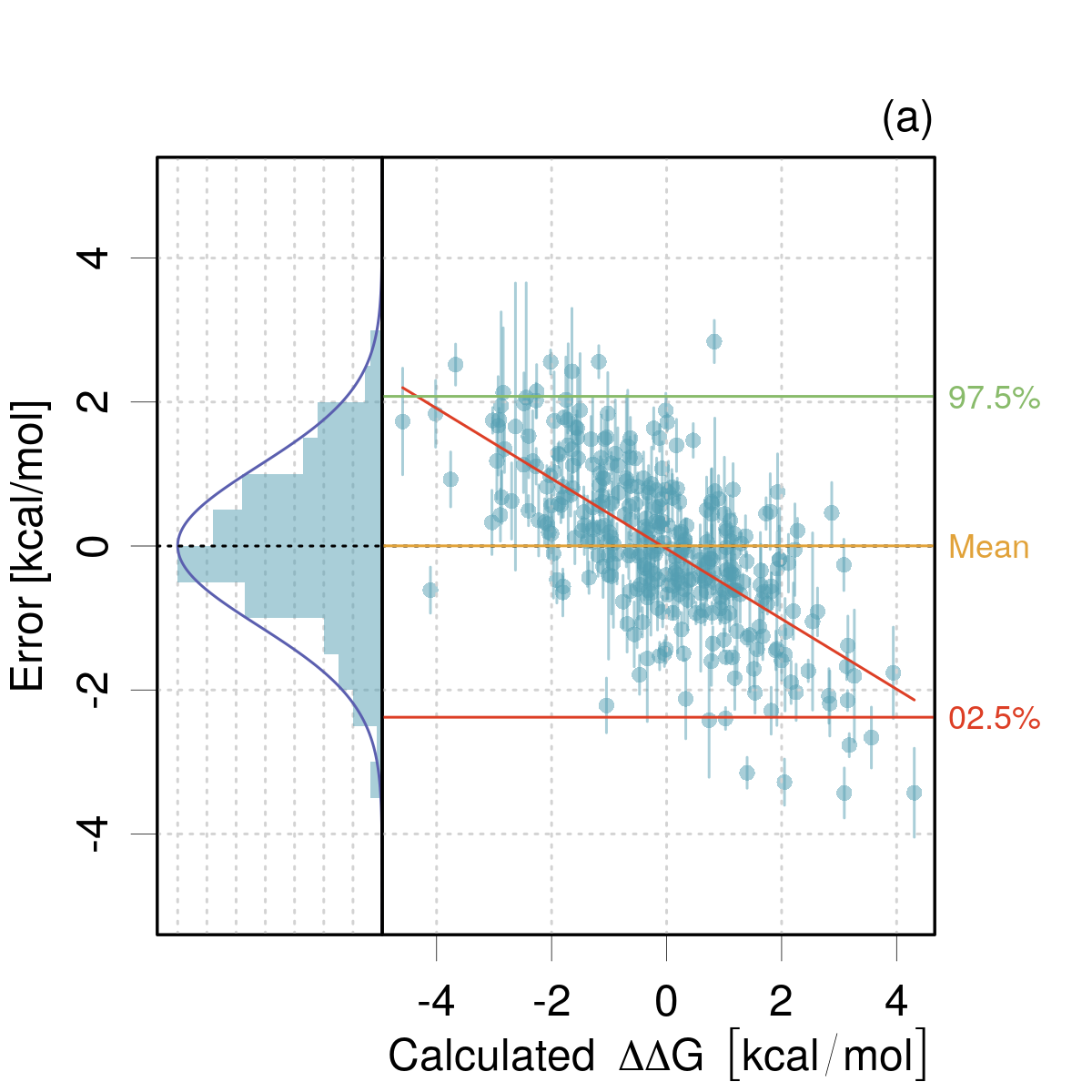}\includegraphics[height=6cm]{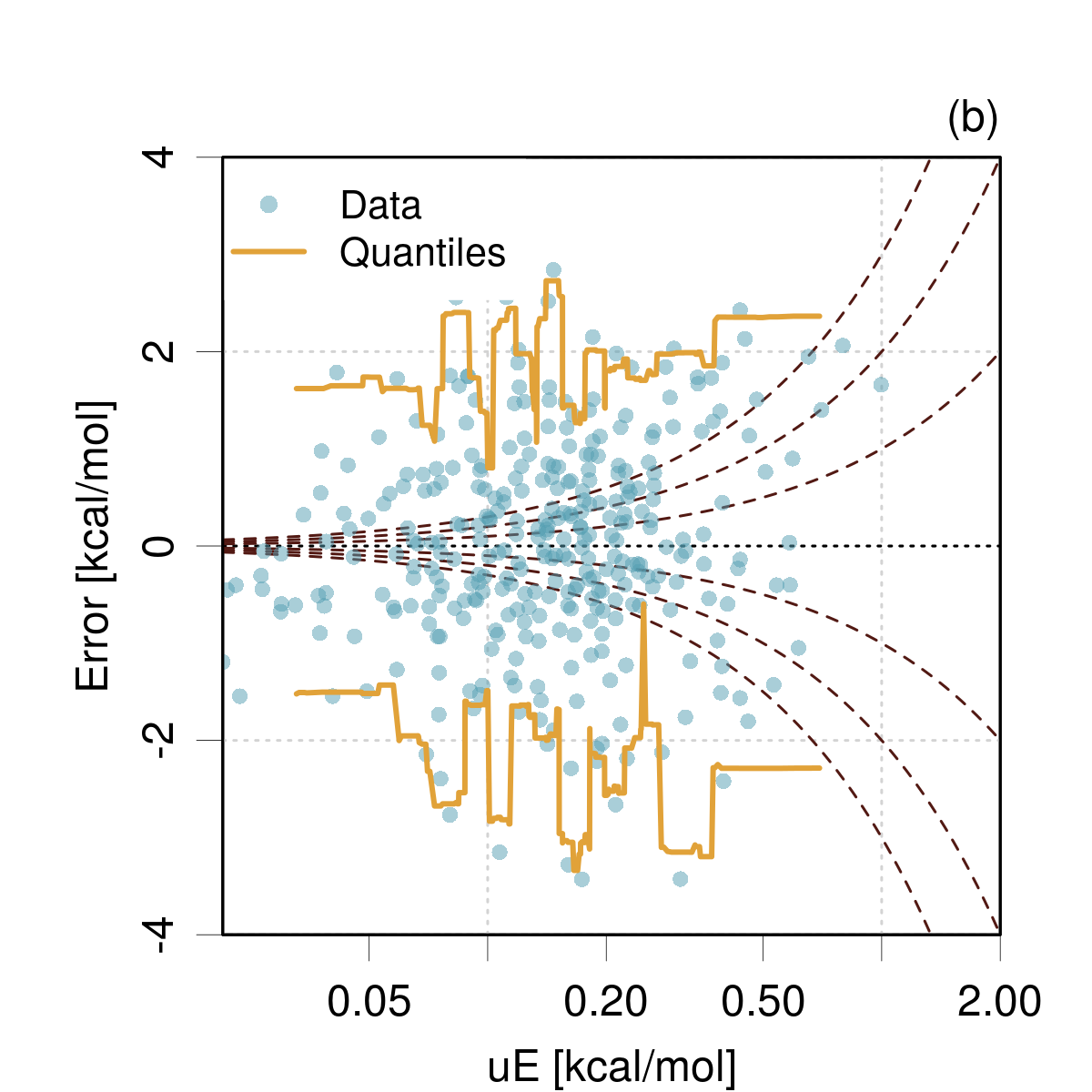}\includegraphics[height=6cm]{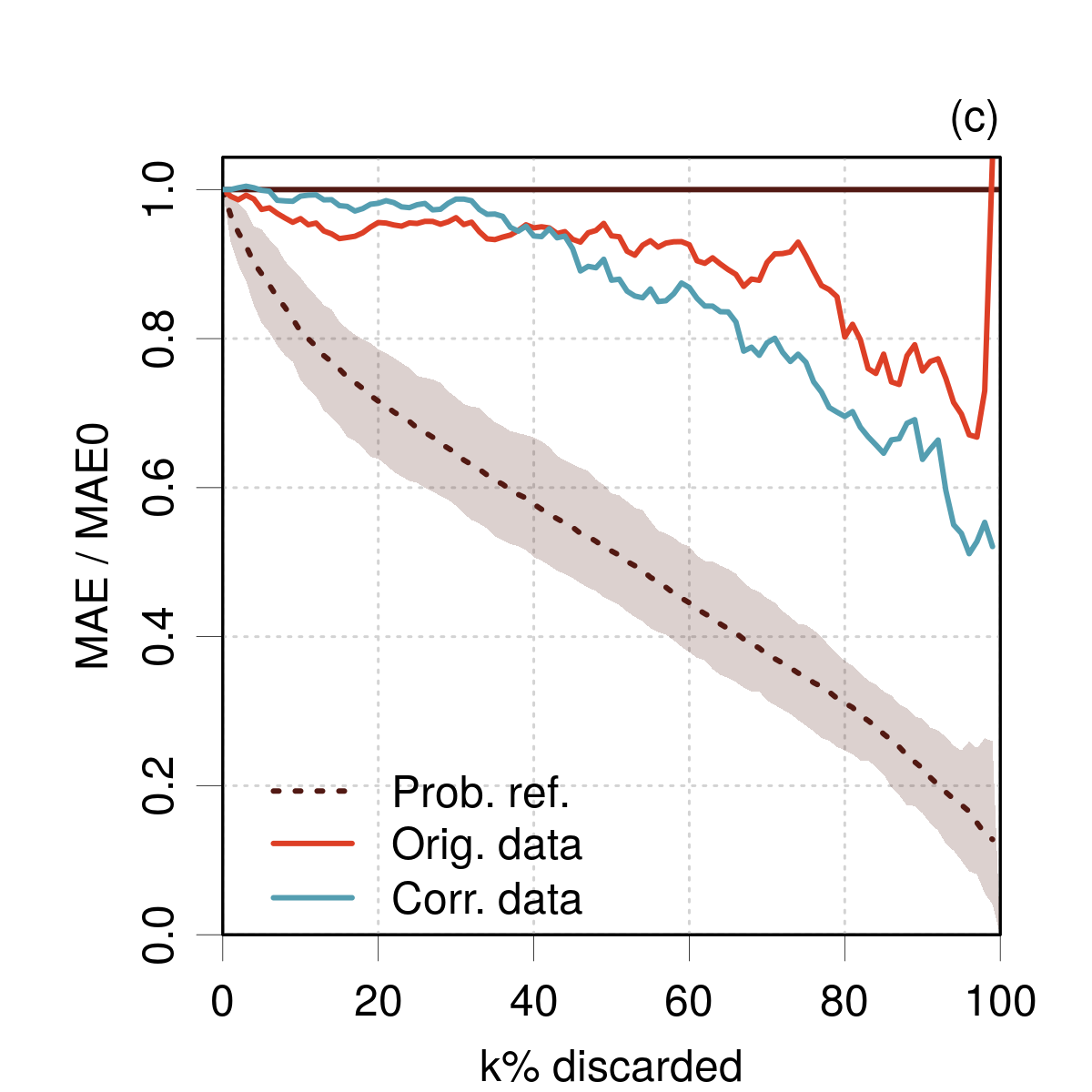}
\par\end{centering}
\noindent \begin{centering}
\includegraphics[height=6cm]{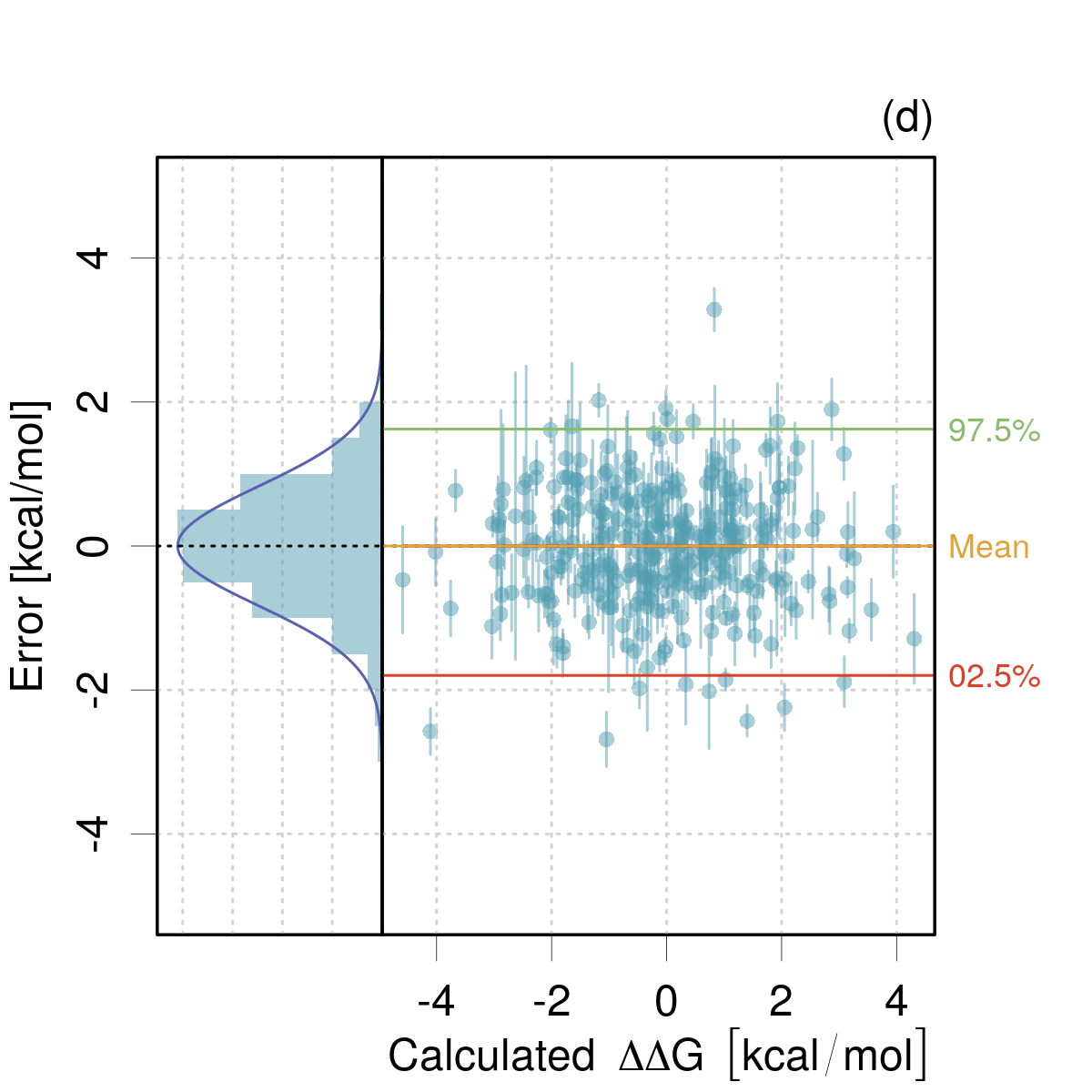}\includegraphics[height=6cm]{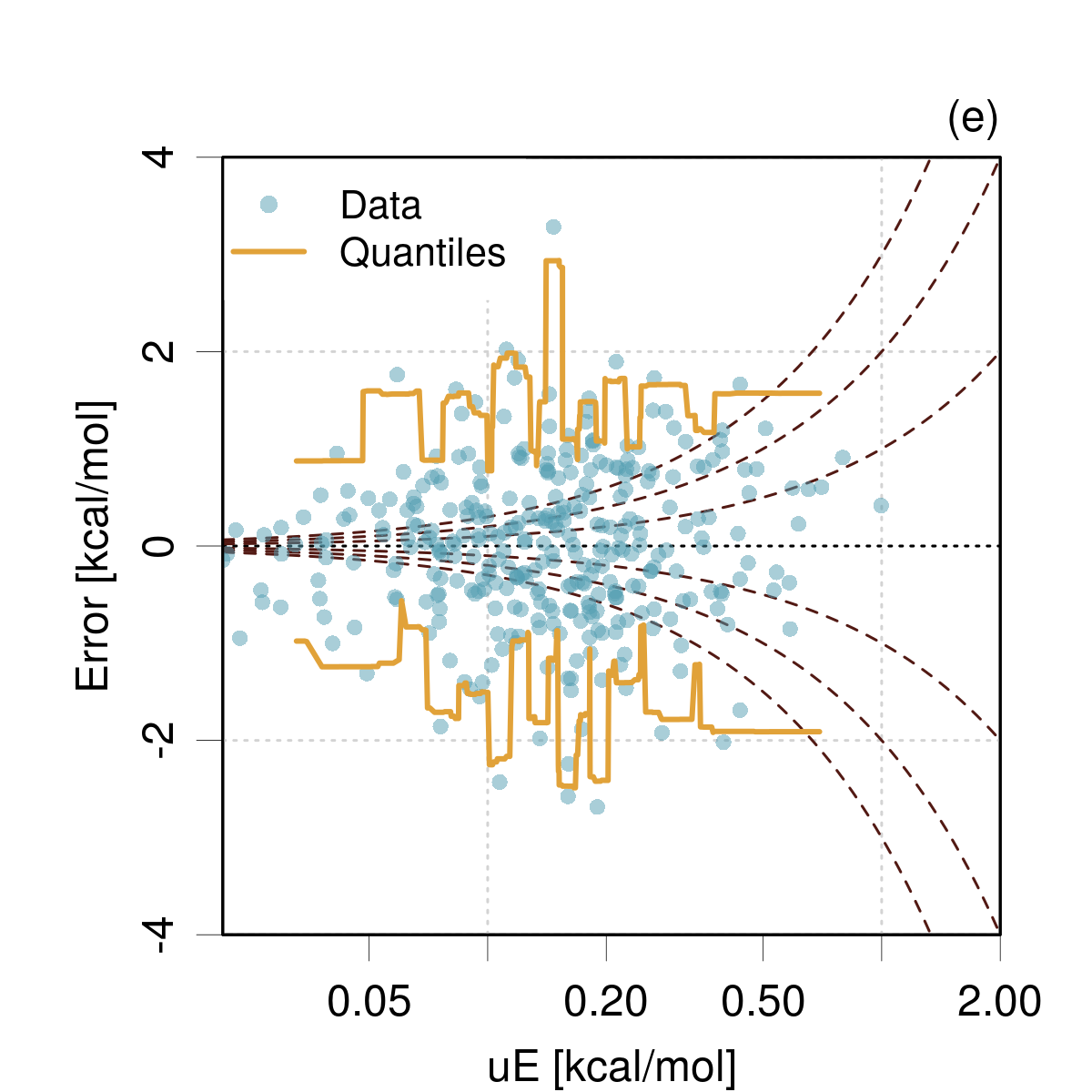}\includegraphics[height=6cm]{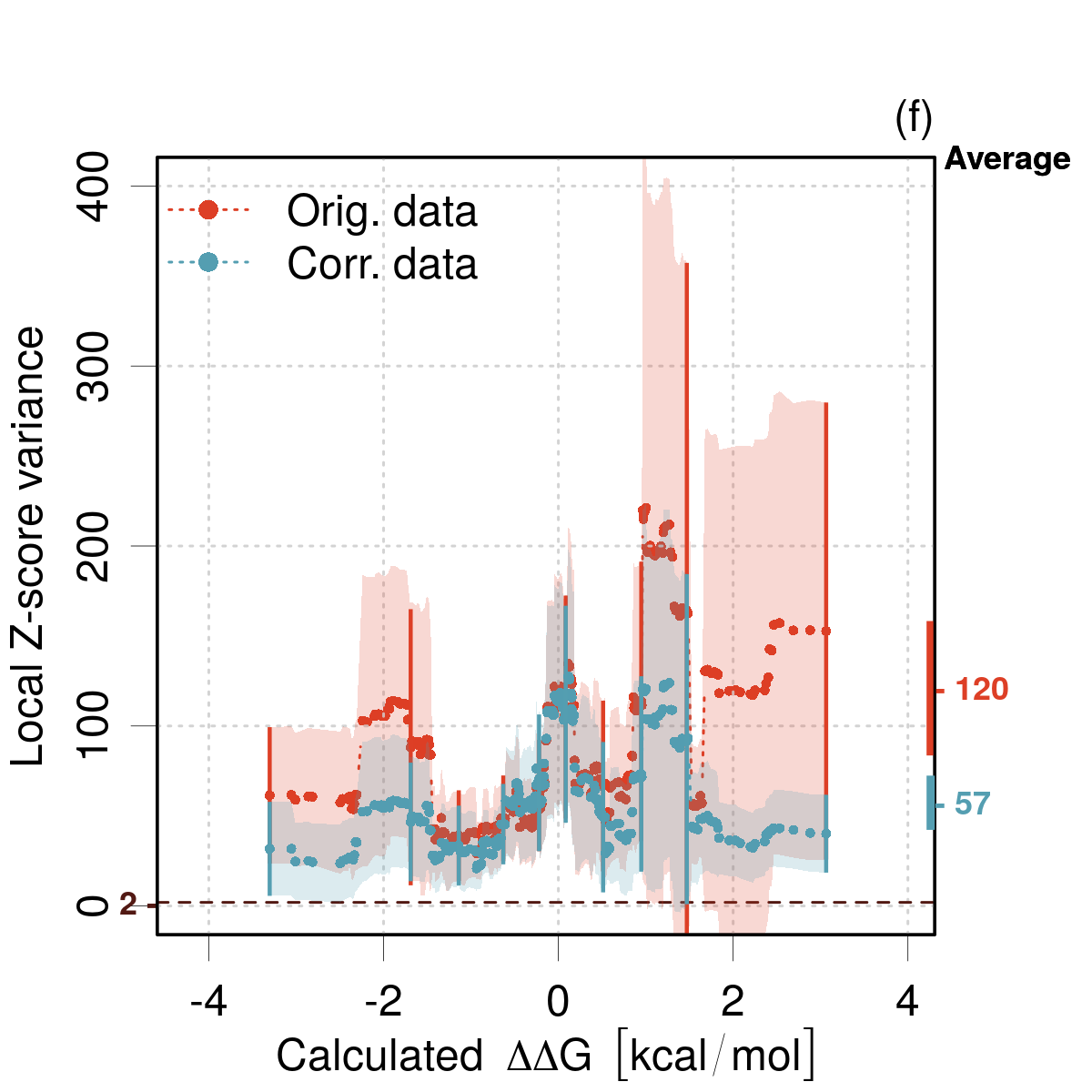}
\par\end{centering}
\noindent \raggedright{}\caption{\label{fig:09}Calibration analysis for dataset LIN2021. (a) Error
distribution vs $V$ (the error bars represent $2u_{E}$); (b) $(u_{E},E)$
plot with running quantiles; (c) confidence curves for the original
dataset and the corrected one; (d) same as (a) for the trend-corrected
error set; (e) same as (b) for the trend-corrected error set; (f)
LZV analysis for both sets. }
\end{figure}

The errors and their distribution are shown in Fig.\,\ref{fig:09}(a).
The errors have a quasi-normal distribution with a notable and unsuitable
trend. The $(u_{E},E)$ plot {[}Fig.\,\ref{fig:09}(b){]} shows that
the errors seem unrelated to the uncertainties and that the uncertainties
are too small to explain the dispersion of the errors. Confirming
this point, the variance of \emph{t}-scores is much too large ($\mathrm{Var}(T)=120$
vs. 2 for $n=5$). The confidence curve {[}Fig.\,\ref{fig:09}(c),
``Orig. data''{]} shows a very slow and shaky decrease, far above
the reference curve, confirming a weak link between errors and uncertainties.
The reported uncertainties for the FEP procedure should therefore
not be interpreted nor used as prediction uncertainties. They should
probably not be used either to identify predictions with large errors.

A linear trend correction, without modification of the uncertainties,
improves notably the error distribution in terms of bias {[}Fig.\,\ref{fig:09}(d){]},
but is insufficient to compensate for miscalibration {[}Fig.\,\ref{fig:09}(e){]}.
The variance of the t-scores is reduced to $\mathrm{Var}(T)=57$,
still far above the target value. The confidence curve is not improved
either {[}as they share the same uncertainty set, both curves share
the same probabilistic reference; Fig.\,\ref{fig:09}(e){]}, and
the LZV plots for the original and corrected data confirm the diagnostic.

The experimental uncertainty is reported to be about 0.4~kcal/mol
for the kind of experimental data used as reference.\citep{Wang2015c}
Combining quadratically this value with the original uncertainties
results in a significant but insufficient decrease of $\mathrm{Var}(T)$,
to 6.1 ($I_{95}=[5.1,7.1]$) for the original data and 3.5 ($I_{95}=[2.8,4.2]$)
for the corrected ones. 

Model errors are not accounted for in the FEP UQ procedure, and one
might conclude they have a non-negligible contribution to the error
budget.

\subsubsection{ZHE2022}

A recent article by Zheng \emph{et al.} \citep{Zheng2022} provides
formation enthalpies and uncertainties for two data-driven methods,
AIQM1 and ANI-1ccx. The uncertainties were obtained by a query by
committee (QbC) strategy,\citep{Smith2018} and taken as the standard
deviation (SD) of the results for an ensemble of $n=8$ neural networks
(NN). Zheng \emph{et al.} consider that NN SDs provide uncertainty
quantification on the methods predictions and used them to detect
unreliable simulations, outliers and suspicious reference data. Two
validation sets $\left\{ E_{i},u_{E_{i}}\right\} _{i=1}^{M}$ with
$M=472$ were gathered from the source article for AIQM1 and ANI-1ccx,
by aggregating data for $\Delta H_{f}$ and removing systems with
missing values. 

As in the QbC protocol the prediction value is taken as the mean of
the $n$ NN predictions,\citep{Smith2018} one should divide the reported
standard deviations by $\sqrt{n}$ for consistency. However, the authors
used the standard deviation (which would be the uncertainty estimate
for a single NN prediction) throughout their article, so I used it
also to define $u_{E}$. Furthermore, no information is provided about
the uncertainty on the experimental data used as reference, and I
ignore them in a first step.

\begin{figure}[t]
\noindent \begin{centering}
\includegraphics[height=6cm]{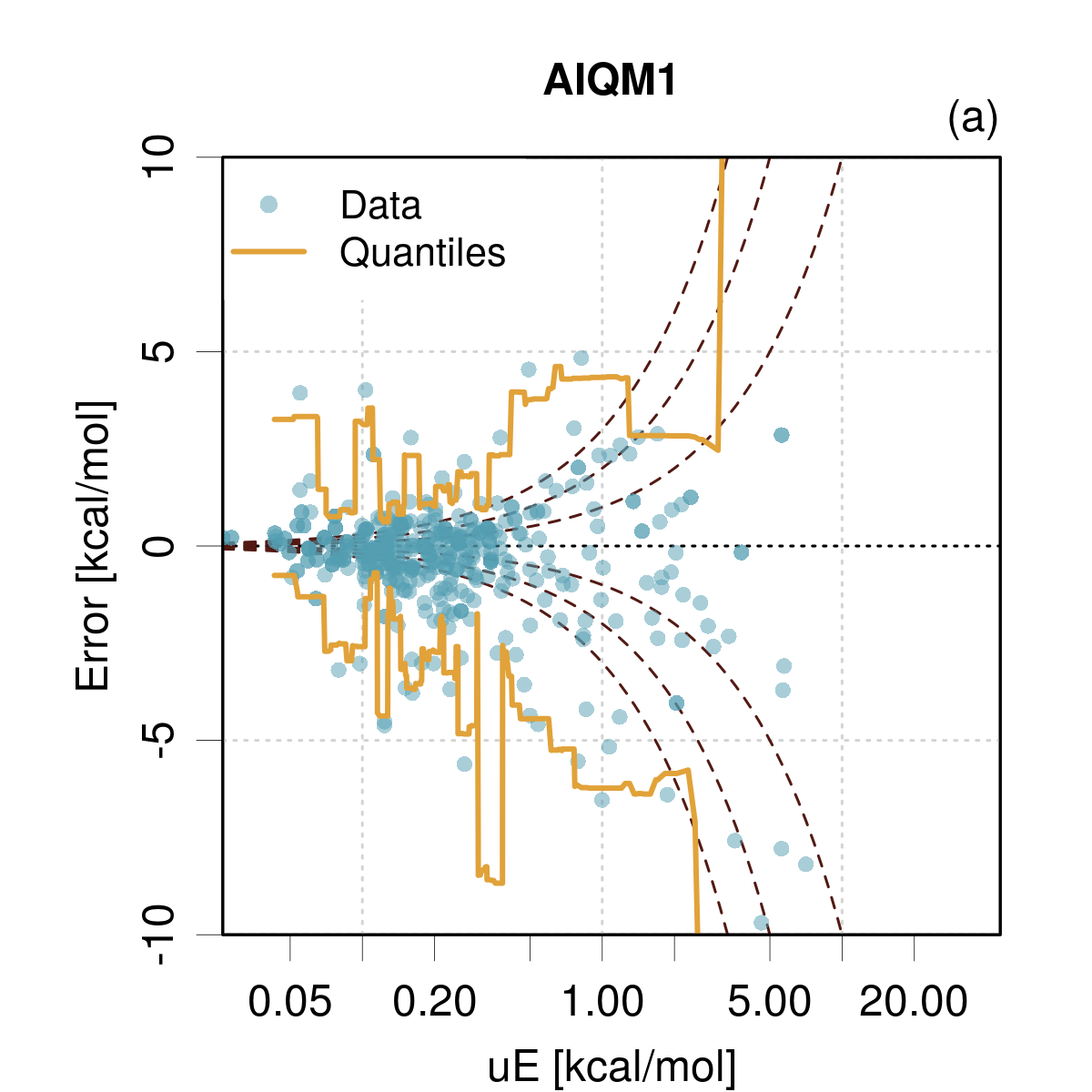}\includegraphics[height=6cm]{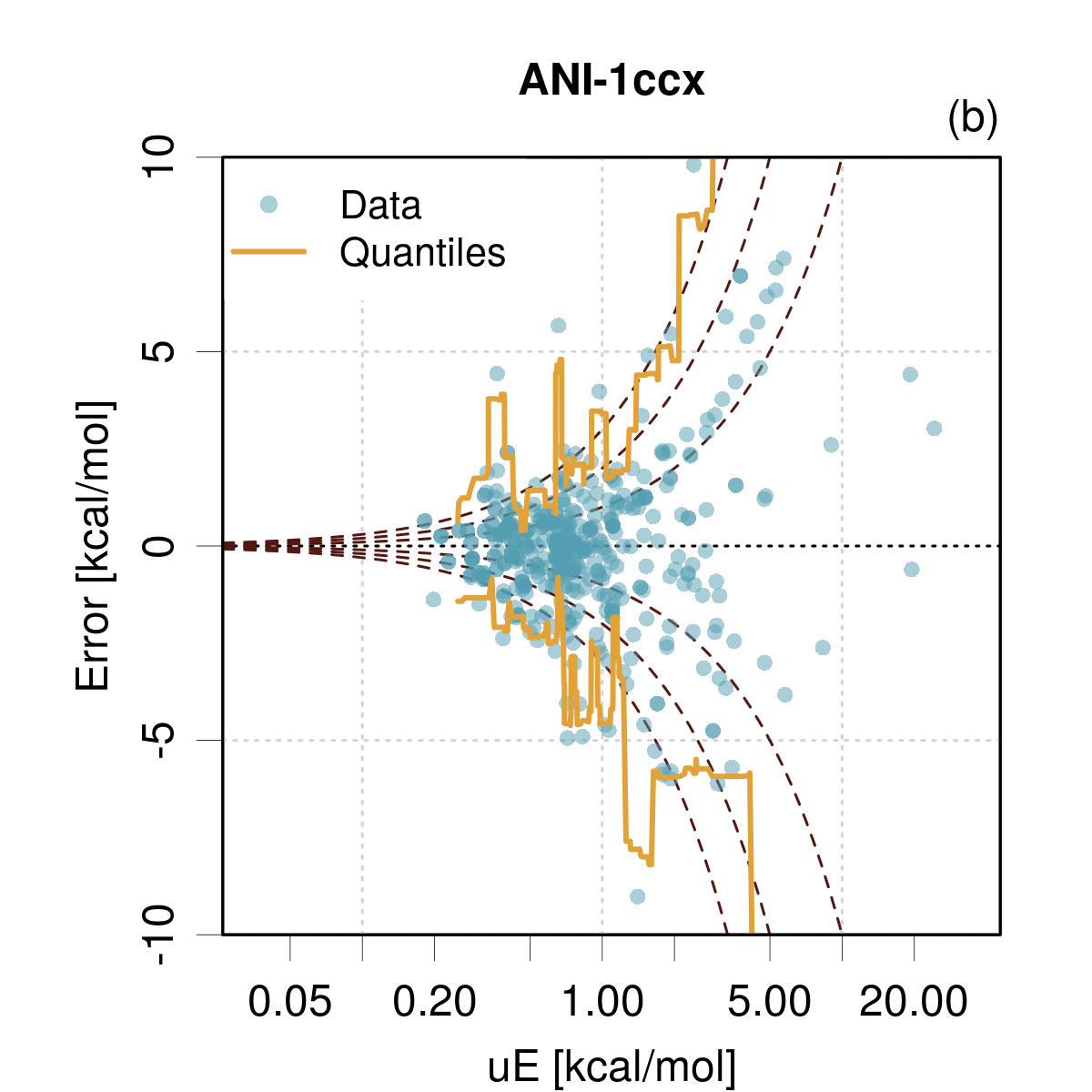}
\par\end{centering}
\noindent \begin{centering}
\includegraphics[height=6cm]{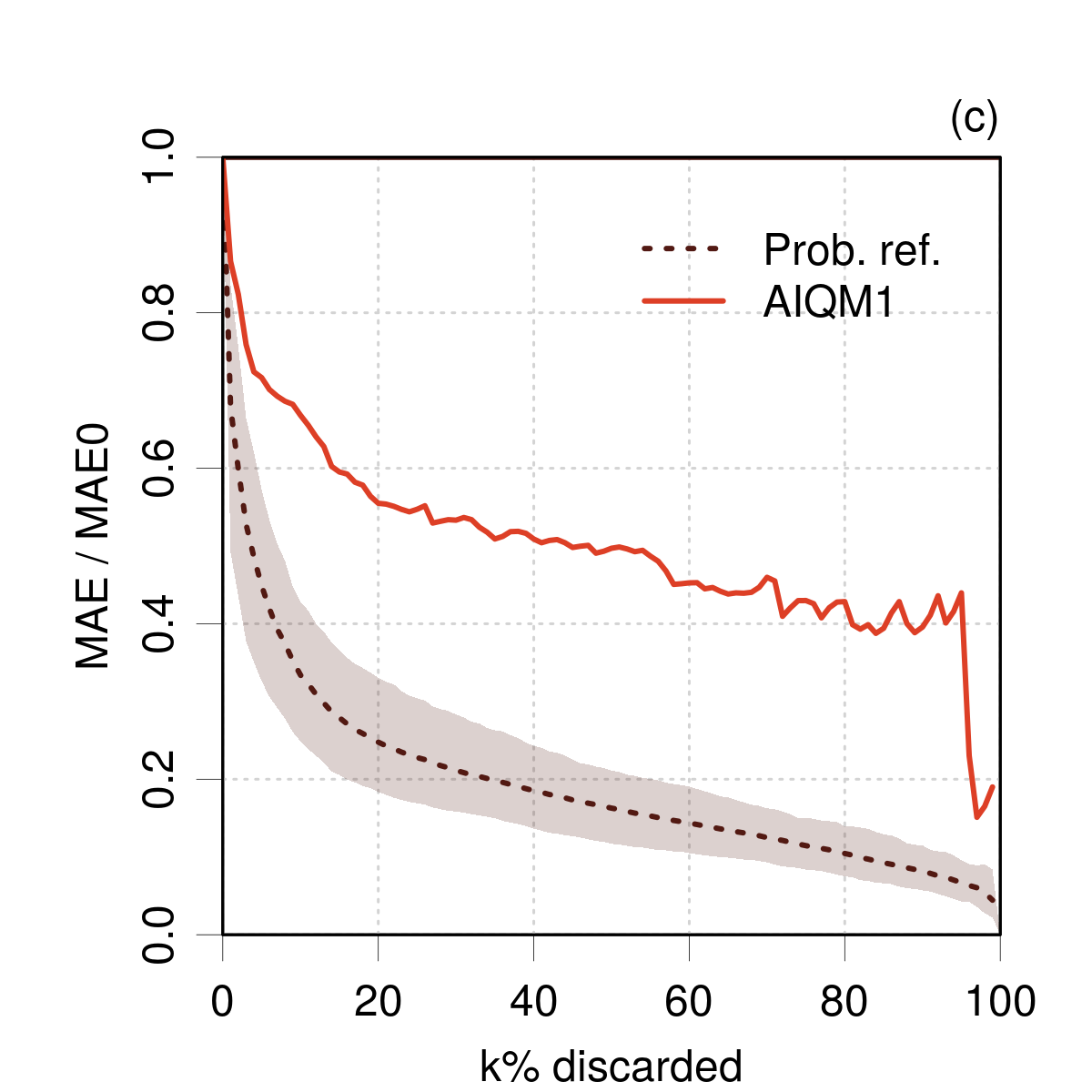}\includegraphics[height=6cm]{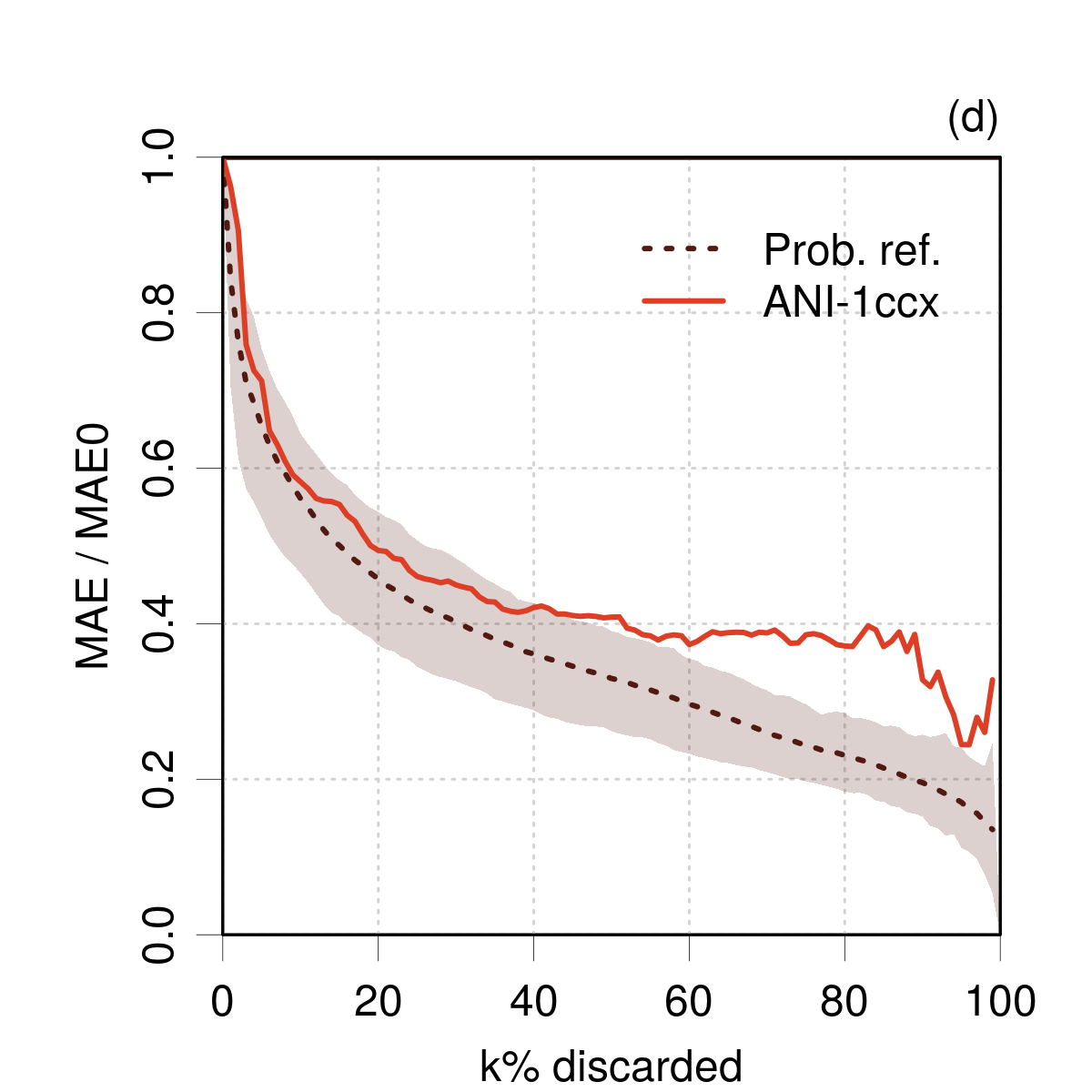}
\par\end{centering}
\noindent \begin{centering}
\includegraphics[height=6cm]{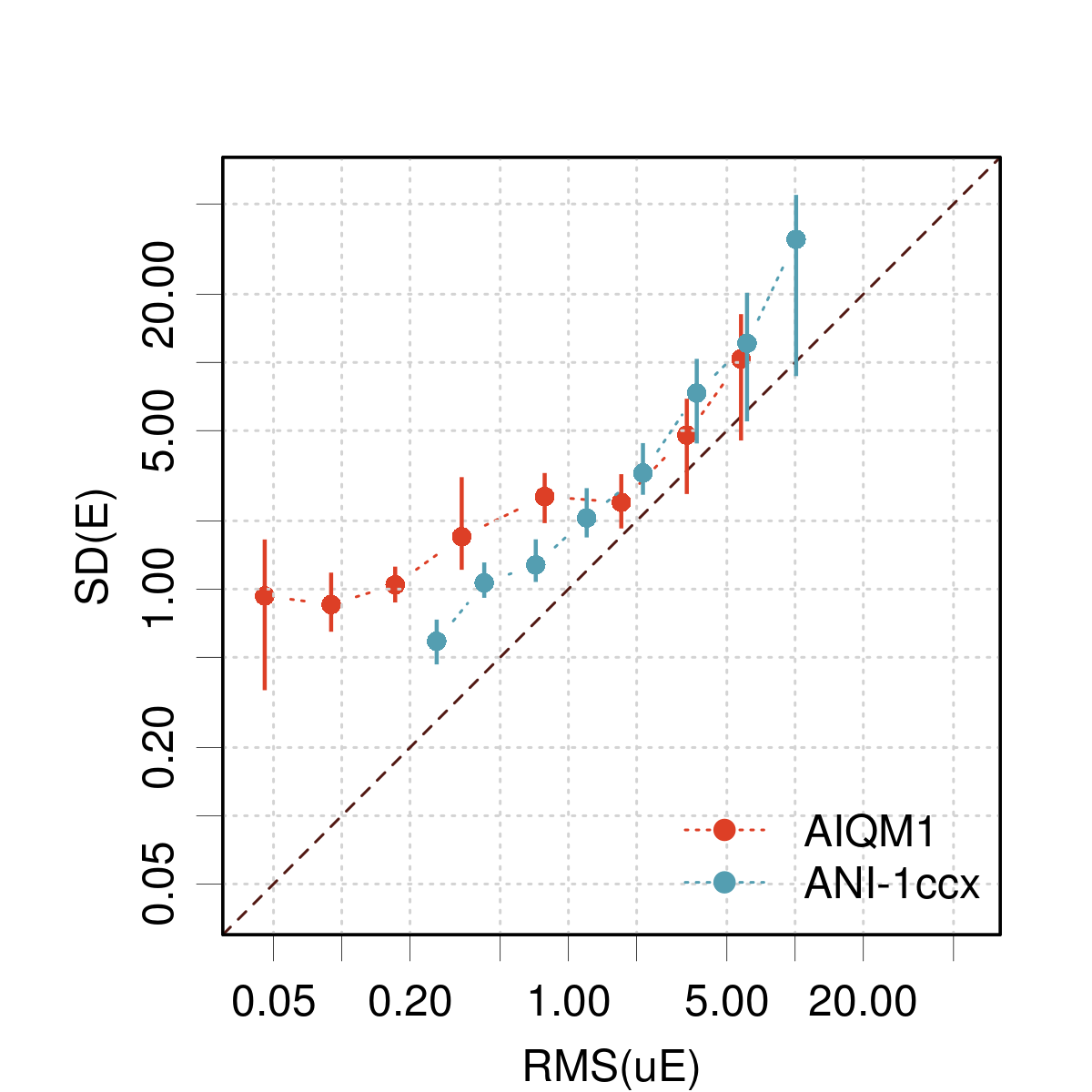}
\par\end{centering}
\caption{\label{fig:10}Calibration analysis for dataset ZHE2022: (a,b) $(uE,E)$
plot with running quantiles for methods AIQM1 and ANI-1ccx (the oracle
is derived from AIQM1); (c,d) confidence curves; (e) reliability diagrams. }
\end{figure}

As a first diagnostic, one plots $E$ vs $u_{E}$ for both methods
{[}Fig.\,\ref{fig:10}(a,b){]}. It is clear for the AIQM1 dataset
that a large part of the uncertainties are too small to explain the
amplitude of the errors. The consistency seems slightly better above
1\,kcal/mol. One can safely reject calibration for this dataset,
which is confirmed by the value of $\mathrm{Var}(Z)=59$, to compare
to the 1.4 target. Note that the scaling of the standard deviation
by the $1/\sqrt{8}$ factor would increase this value by a factor
8. The situation is somewhat better for the ANI-1ccx method {[}Fig.\,\ref{fig:10}(b){]},
where the cumulative quantile curves follow grossly the guidelines.
However, one has $\mathrm{Var}(Z)=4.3$ for this dataset, which leads
to reject calibration. 

As the $R^{2}$ score based on a linear regression without intercept
used by the authors does not inform us on the correlations between
$u_{E}$ and $|E|$, I estimated the rank correlation coefficients
for the CHNO subset and found 0.37 and 0.42 for AIQM1 and ANI-1ccx,
respectively. These values are within the range reported by Tynes
et al.\citep{Tynes2021} for uncertainty datasets used in active learning
(0.2 - 0.65). One might thus conclude that there is a rather strong
relation between the QbC uncertainties and the prediction errors.
This is assessed by the confidence curves {[}Fig.\,\ref{fig:10}(c,d){]}.
Let us however note that these curves show a sharp decrease for the
first fifth of the \emph{k} axis and progressively switch to a slower
decrease, or even a plateau for ANI-1ccx. This would mean that the
consistency between errors and uncertainties is visible only for the
20\% larger uncertainties. The departure of the confidence curves
from the the probabilistic reference confirms the absence of calibration
and tightness, with a better performance for ANI-1ccx.

Despite the absence of calibration, one might check the reliability
of both sets on a reliability diagram {[}Fig.\,\ref{fig:10}(d){]}.
Both sets have similar and slightly off reliability for the larger
uncertainties (above 1\,kcal/mol). Below this value, the ANI-1ccx
performs better than AIQM1, with a nearly constant offset from the
identity line. In contrast, the reliability curve for AIQM1 deviates
from the identity line to reach a plateau, indicating a poor reliability
of small uncertainties.

The median of the standard uncertainties derived\footnote{The ATcT provides expanded $U_{95}$ uncertainties.}
from the Active Thermochemical Tables (ATcT, ver. 1.112)\citep{Ruscic2004}
would be about 0.1~kcal/mol (the mean is about 0.17~kcal/mol). Adding
quadratically a uniform contribution of 0.1\,kcal/mol to the QbC
uncertainties reduces $\mathrm{Var}(Z)$ to 29 and 4.1 for AIQM1 and
ANI-1ccx, respectively. Experimental uncertainty alone is thus far
from explaining the missing uncertainty and there should be a significant
contribution of model errors. This is acknowledged by Zheng \emph{et
al.}, who observed that some predictions with large errors have small
QbC uncertainties.\textcolor{orange}{{} }

This analysis confirms the findings of recent studies about the overconfidence
of ensemble NN UQ protocols.\citep{Tran2020,Scalia2020} It is clear
from the present analysis that the QbC uncertainties cannot be considered
as prediction uncertainties, mostly because they are not integrating
model errors. However, as clearly demonstrated by Zheng \emph{et al.}
and observed on the confidence curves, they seem well fit for the
purposes of active learning and outliers detection. 

\section{Available software\label{sec:Available-software}}

Except for simple graphical diagnostics presented in Sect.\,\ref{sec:Graphical-methods},
extensive coding might be required to implement the CS/CT validation
methods. To my knowledge, three toolboxes are freely available that
implement some of these methods.
\begin{itemize}
\item \href{https://uncertainty-toolbox.github.io/}{Uncertainty Toolbox}.
``A python toolbox for predictive uncertainty quantification, calibration,
metrics, and visualizations''.\citep{Chung2021} The toolbox focuses
on regression tasks in ML-UQ. It implements, among other, calibration
and sharpness statistics, adversarial group calibration and some re-calibration
methods. 
\item \href{https://github.com/epiforecasts/scoringutils}{scoringutils}
``The \texttt{scoringutils} package provides a collection of metrics
and proper scoring rules and aims to make it simple to score probabilistic
forecasts against the true observed values.'' Issued in 2022, this
\texttt{R} package deals with predictive probability distributions
represented as sample or parametric distributions.\citep{Jordan2019,Bosse2022}
\item \href{https://github.com/ppernot/ErrViewLib}{ErrViewLib}. Coded in
\texttt{R},\citep{RTeam2019} the package implements functions for
simple graphical checks (\texttt{plotEvsPU}) and calibration/tightness
analysis \texttt{(plotLCP}, \texttt{plotLRR}, \texttt{plotLZV}, \texttt{plotRelDiag}
and \texttt{plotConfidence}). It is \emph{not} ML oriented and does
not presently treat prediction ensembles. All the plots of the present
study have been generated with \texttt{ErrViewLib-v1.5d} (\url{https://github.com/ppernot/ErrViewLib/releases/tag/v1.5d}),
also available at Zenodo (\url{https://doi.org/10.5281/zenodo.6783307}).
Alg.\,\ref{alg:Skeletal} presents a skeletal example to generate
a ($u_{E},E$) plot and a LCP analysis for an heteroscedastic synthetic
dataset. The \texttt{UncVal} graphical interface to explore the main
UQ validation methods provided by \texttt{ErrViewLib} is also available
on GitHub (\url{https://github.com/ppernot/UncVal}), either as source
code or as a Docker container. \textcolor{orange}{}
\begin{algorithm}
\textcolor{orange}{}
\begin{lstlisting}[language=R]
library(ErrViewLib) 
N   = 1000 
s2  = rchisq(N, df = 4)         # Random variance
uE  = 0.01 * sqrt(s2/mean(s2))  # Re-scale uncertainty
E   = rnorm(N, mean=0, sd=uE)   # Generate errors
ErrViewLib::plotEvsPU(uE, E)
U95 = 1.96*uE                   # U95 for normal law
ErrViewLib::plotLCP(E, U95, ordX = U95, prob = 0.95, ylim = c(0.5,1))
\end{lstlisting}

\textcolor{orange}{\caption{\label{alg:Skeletal}Example of \texttt{R} script using \texttt{ErrViewLib}.}
}
\end{algorithm}
\end{itemize}

\section{Discussion and conclusion\label{sec:Discussion-and-conclusion}}

This article presents a comprehensive panel of simple and more complex
graphical and statistical methods to test the calibration and tightness
of probabilistic predictions. Tightness has been introduced as a concept
to evaluate the small-scale reliability of probabilistic predictions.
As for sharpness, its use is conditional to average calibration. The
full validation of the reliability of probabilistic predictions requires
thus the estimation of (average) calibration \emph{and} tightness.

The tool set presented in PER2022 for intervals- and variance-based
validation has been completed by easy to implement graphical checks
and by ranking-based methods (correlation coefficients, confidence
curves) used in machine learning\citep{Scalia2020}. A summary of
the applicability and validation capacity of all the methods is presented
in Table~\ref{tab:Summary-table-of}. 
\begin{table}[t]
\noindent \begin{centering}
\begin{tabular}{lccccccccc}
\hline 
\multicolumn{1}{l}{Diagnostic} &  & \multicolumn{5}{c}{Applicability} &  & \multicolumn{2}{c}{Validation}\tabularnewline
\cline{3-7} \cline{9-10} 
 & ~~ & $\,\,\,\,q_{E}\,\,\,\,$ & $\,\,\,\,u_{E}\,\,\,\,$ & $\,\,\,\,U_{E,p}\,\,\,\,$ & Homosc. & Heterosc. & ~~ & Calibrat. & Tightness\tabularnewline
\cline{1-1} \cline{3-7} \cline{9-10} 
\emph{Average} &  &  &  &  &  &  &  &  & \tabularnewline
~~~PIT hist. &  & \textcolor{green}{$\checked$} & \textcolor{red}{$\mathsf{X}$} & \textcolor{red}{$\mathsf{X}$} & \textcolor{green}{$\checked$} & \textcolor{green}{$\checked$} &  & \textcolor{green}{$\checked$} & \textcolor{red}{$\mathsf{X}$}\tabularnewline
~~~Calib. curve &  & \textcolor{green}{$\checked$} & \textcolor{red}{$\mathsf{X}$} & \textcolor{red}{$\mathsf{X}$} & \textcolor{green}{$\checked$} & \textcolor{green}{$\checked$} &  & \textcolor{green}{$\checked$} & \textcolor{red}{$\mathsf{X}$}\tabularnewline
~~~PICP &  & \textcolor{green}{$\checked$} & \textcolor{red}{$\mathsf{X}$} & \textcolor{green}{$\checked$} & \textcolor{green}{$\checked$} & \textcolor{green}{$\checked$} &  & \textcolor{green}{$\checked$} & \textcolor{red}{$\mathsf{X}$}\tabularnewline
~~~Var($Z$) &  & \textcolor{green}{$\checked$} & \textcolor{green}{$\checked$} & \textcolor{red}{$\mathsf{X}$} & \textcolor{green}{$\checked$} & \textcolor{green}{$\checked$} &  & \textcolor{green}{$\checked$} & \textcolor{red}{$\mathsf{X}$}\tabularnewline
~~~Cor($u_{E}$,$|E|$) &  & \textcolor{green}{$\checked$} & \textcolor{green}{$\checked$} & \textcolor{green}{$\checked$} & \textcolor{red}{$\mathsf{X}$} & \textcolor{green}{$\checked$} &  & \textcolor{red}{$\mathsf{X}$} & \textcolor{red}{$\mathsf{X}$}$^{*}$\tabularnewline[\doublerulesep]
\emph{Local} &  &  &  &  &  &  &  &  & \tabularnewline
~~~LCP/LRR &  & \textcolor{green}{$\checked$} & \textcolor{red}{$\mathsf{X}$} & \textcolor{green}{$\checked$} & \textcolor{green}{$\checked$} & \textcolor{green}{$\checked$} &  & \textcolor{green}{$\checked$}$^{\dagger}$ & \textcolor{green}{$\checked$}\tabularnewline
~~~LZV &  & \textcolor{green}{$\checked$} & \textcolor{green}{$\checked$} & \textcolor{red}{$\mathsf{X}$} & \textcolor{green}{$\checked$} & \textcolor{green}{$\checked$} &  & \textcolor{green}{$\checked$}$^{\dagger}$ & \textcolor{green}{$\checked$}\tabularnewline
~~~Reliab. diag. &  & \textcolor{green}{$\checked$} & \textcolor{green}{$\checked$} & \textcolor{red}{$\mathsf{X}$} & \textcolor{red}{$\mathsf{X}$} & \textcolor{green}{$\checked$} &  & \textcolor{green}{$\checked$}$^{\dagger}$ & \textcolor{green}{$\checked$}\tabularnewline
~~~Confid. curve (oracle) &  & \textcolor{green}{$\checked$} & \textcolor{green}{$\checked$} & \textcolor{green}{$\checked$} & \textcolor{red}{$\mathsf{X}$} & \textcolor{green}{$\checked$} &  & \textcolor{red}{$\mathsf{X}$} & \textcolor{red}{$\mathsf{X}$}$^{*}$\tabularnewline
~~~Confid. curve (prob.) &  & \textcolor{green}{$\checked$} & \textcolor{green}{$\checked$} & \textcolor{red}{$\mathsf{X}$} & \textcolor{red}{$\mathsf{X}$} & \textcolor{green}{$\checked$} &  & \textcolor{green}{$\checked$}$^{\dagger}$ & \textcolor{green}{$\checked$}\tabularnewline
\hline 
\end{tabular}
\par\end{centering}
\caption{\label{tab:Summary-table-of}Summary of the applicability of uncertainty
validation methods for calibration and tightness. $^{*}$ A negative
diagnostic invalidates tightness. $^{\dagger}$The local validation
methods apply to calibration for very large validation sets only.
For small validation sets, both average calibration and tightness
have to be validated (see Sect.\,\ref{subsec:Concepts-and-definitions}).}
\end{table}

We have seen that the ranking-based methods are not able to give a
positive validation diagnostic, but they might be used to ascertain
a negative tightness diagnostic. Note that ranking-based methods find
their utility in active learning, where the main purpose of an uncertainty
is to identify cases susceptible of large errors. From a set of average-calibrated
methods, on should prefer the one with the best sharpness or confidence
curve, but we have no guarantee that it might have a good tightness.
Besides, ranking-based methods cannot be used for homoscedastic datasets
(i.e. validation sets for which all predictions have the same uncertainty).
We are thus left with intervals- and variance-based validation methods,
the choice of which is guided by available information.

When predictions are represented by analytical distributions or large
ensembles, all methods are available. Either for calibration or tightness
validation, testing for the adequacy of a set of intervals with different
coverage probabilities will be more demanding than testing for variance,
as the latter is less dependent on the shape of the distribution.
However, unless the distribution's shape is very far from normal,
e.g. with a strong asymmetry, variance-based methods should be adequate. 

In many instances, the literature about computational chemistry uncertainty
quantification reports only statistical summaries, i.e. standard uncertainties
or expanded uncertainties at the 95\,\% level. In such cases, the
choice of validation method is imposed: standard uncertainties should
be handled by variance-based methods and expanded uncertainties by
intervals-based methods. Of course, in cases where expanded uncertainties
were derived from standard uncertainties by a known expansion factor,
inverse transformation to standard uncertainties can give access to
variance-based methods.

In the proposed framework, calibration is validated on the full validation
set, using prediction intervals coverage probabilities (PICP) for
the intervals-based approach and the variance of scaled errors or
\emph{z}-scores ($\mathrm{Var}(Z)$) for the variance-based approach.
Tightness validation is based the same tools, but applied to subsets
or groups of the validation set to assess local or small-scale reliability,
leading to LCP analysis for the intervals-based approach and to LZV
analysis for the variance-based approach. The groups can be designed
according to any relevant criteria, but using the predicted value
and the prediction uncertainty are two interesting alternatives. I
have shown that the latter case is closely linked to the reliability
diagrams introduced by Levi \emph{et al.}\citep{Levi2020}. Note that
by using a new probabilistic reference, confidence curves have been
promoted from a ranking-based to a variance-based validation method
for tightness. Reliability diagrams and confidence curves can only
be used for heteroscedastic datasets. 

A special care has to be taken for those cases where uncertainty is
estimated as the standard deviation of a small ensemble ($n<30$).
In such cases, the scaled errors are not \emph{z}-scores, but \emph{t}-scores,
for which the theoretical variance used for validation is $(n-1)/(n-3)$
(for normal predictive distributions) instead of 1 for \emph{z}-scores.
With this caveat in mind, it is possible to validate calibration,
but we have seen that tightness is very sensitive to the statistical
noise characteristic of small ensembles. In particular, the LZV approach
with groups based on the prediction uncertainty, or the reliability
diagrams, will reject tightness. In this case, using the predicted
value $V$ as a grouping feature for the LZV analysis is a better
alternative.

The tools presented in this study are of interest primarily to CC-UQ
researchers in order to validate their methods to generate prediction
uncertainties, but the most simple of them, such as the $\left(u_{E},E\right)$
plot, can easily be applied by end users curious to evaluate and gain
confidence in uncertainties they might want to publish or reuse. UQ
outputs failing to satisfy these validation tests should be used with
caution and not over-interpreted. They should not be used to infer
probability intervals for the true value of a property, as would be
expected in the Virtual Measurements framework.\citep{Irikura2004}
All the examples taken from the literature, as well as those presented
in PER2022 show that designing reliable prediction uncertainties is
a very demanding process, which leaves ample room for future developments
in CC-UQ. 

\section*{Data availability statement }

The data and codes that enable to reproduce the figures of this study
are openly available at the following URL: \url{https://github.com/ppernot/2022_Tightness},
or in Zenodo at \url{https://doi.org/10.5281/zenodo.7059776}.

\section*{Acknowledgments}

I would like to thank Andreas Savin for enlightening and constructive
discussions, Jonny Proppe for providing the PRO2022 dataset, Pavlo
Dral for helpful comments on a former version of my analysis of the
ZHE2022 dataset, and Matthew Evans for pointing out the Uncertainty
Toolbox. 

\bibliographystyle{unsrturlPP}
\bibliography{NN}

\section*{Appendices}

\appendix

\section{Var(\emph{Z}) vs. Birge ratio\label{sec:Var(Z)-vs.-Birge}}

Variance-based validation is an essential tool in absence of prediction
intervals (see Sect.\,\ref{subsec:Calibration}). I implemented it
through the \emph{z}-scores variance statistic $\mathrm{Var}(Z)$,
where $z_{i}=E_{i}/u_{E_{i}}$ is an error scaled by the corresponding
uncertainty. This statistic is closely related to the Birge ratio
$R^{2}$, introduced in 1932 by Birge to test the \emph{statistical
consistency} of residuals of least-squares fits.\citep{Birge1932}
Applying a modern metrological formulation,\citep{Kacker2008} one
gets
\begin{align}
R^{2} & =\frac{1}{\nu}\sum_{i=1}^{M}\left(\frac{E_{i}}{u_{E_{i}}}\right)^{2}
\end{align}
where $\nu$ is the number of degrees of freedom of the error set:
if the errors are independent random variables, $\nu=M$; if they
are residuals from a fit, $\nu=M-p$, where $p$ is the number of
fit parameters. For a consistent set of errors and uncertainties,
on should have $R^{2}\simeq1$.

For sets of independent errors ($\nu=M$), one has therefore
\begin{equation}
R^{2}=\left\langle Z^{2}\right\rangle 
\end{equation}
and
\begin{align}
\mathrm{Var}(Z) & =\left\langle Z^{2}\right\rangle -\left\langle Z\right\rangle ^{2}\\
 & \simeq R^{2}
\end{align}
as $\left\langle Z\right\rangle \simeq0$ for unbiased errors.

Note that for normal error distributions, $\nu R^{2}$ has a chi-squared
distribution with $\nu$ degrees of freedom, which enables hypothesis
testing.\citep{Kacker2008}

\section{Calibration of \emph{t}-scores\label{sec:Calibration-of-t-scores}}

\begin{figure}[t]
\noindent \begin{centering}
\includegraphics[height=12cm]{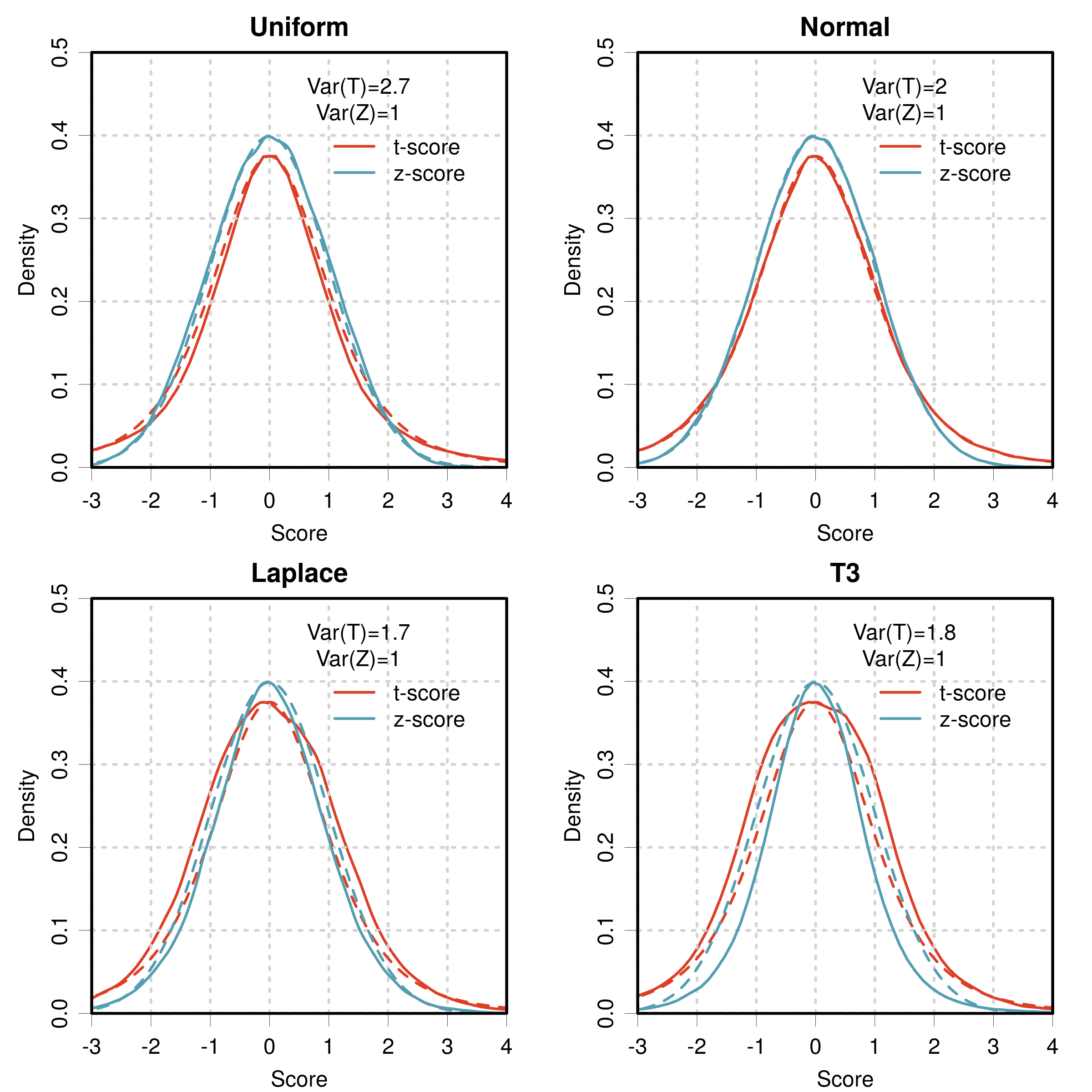}
\par\end{centering}
\caption{\label{fig:A-01}Distributions of \emph{z}-scores and \emph{t}-scores
statistics for several generative distributions. The density plots
(full lines) are generated from $10^{5}$ Monte Carlo realizations
of the mean and standard deviation of sub-samples of size $n=5$.
The dashed lines correspond to the Student's-\emph{t }distribution
with $n-1$ degrees of freedom (red) and the standard normal distribution
(blue). The MC densities have been scaled at the mode of the corresponding
reference distribution. The variances of the samples are given in
the legend of each plot.}
\end{figure}

In order to assess the effect of the generative distribution on the
\emph{t}-score distribution and more particularly on $\mathrm{Var}(T)$,
let us consider a set of distributions covering a large range of shapes
(summarized by their kurtosis value $\kappa$):
\begin{itemize}
\item Beta(1/2,1/2), ($\kappa=1.5$)
\item Uniform between $\pm1$ ($\kappa=1.8$) 
\item Exp4: exponential power ($p=4$) ($\kappa\simeq2.18$) 
\item Normal: standard normal, or Exp2, ($\kappa=3$) 
\item Exp1: exponential power ($p=1$), or Laplace ($\kappa=6$) 
\item T3: Students-$t(\nu=3)$ ($\kappa=\infty$) 
\end{itemize}
Fig.\,\ref{fig:A-01} reports the distributions of the z-scores and
\emph{t}-scores statistics for the mean of samples of $n=5$ random
draws from some of these generative distributions. 

One can check that for the Normal error distribution, the \emph{t}-scores
and \emph{z}-scores have the statistical properties described in Sect.\,\ref{subsec:Small-ensembles}.
For other generative distributions, the \emph{t}- and \emph{z}-score
distributions deviate from the normal references\emph{. }In spite
of this, $\mathrm{Var}(Z)$ is independent of the generative distribution
(and equal to 1 for those calibrated datasets), while $\mathrm{Var}(T)$
depends strongly on the generative distribution. More specifically,
there seems to be a reverse dependence between $\mathrm{Var}(T)$
and the kurtosis of generative distribution, as shown in Table\,\ref{tab:kurt}.
\begin{table}[t]
\noindent \begin{centering}
\begin{tabular}{lcc}
\hline 
Distribution & $\kappa$ & $\mathrm{Var}(T)$\tabularnewline
\hline 
Beta(0.5,0.5) & 1.5 & 5.5\tabularnewline
Uniform(-1,1) & 1.8 & 2.7\tabularnewline
Exp4 & 2.18 & 2.4\tabularnewline
Normal & 3.0 & 2.0\tabularnewline
Exp1 & 6.0 & 1.7\tabularnewline
T3 & $\infty$ & 1.7\tabularnewline
\hline 
\end{tabular}
\par\end{centering}
\caption{\label{tab:kurt}Dependence of $Var(T)$ on the kurtosis of the generative
error distribution for $n=5$.}
\end{table}

\begin{figure}[t]
\noindent \begin{centering}
\includegraphics[height=8cm]{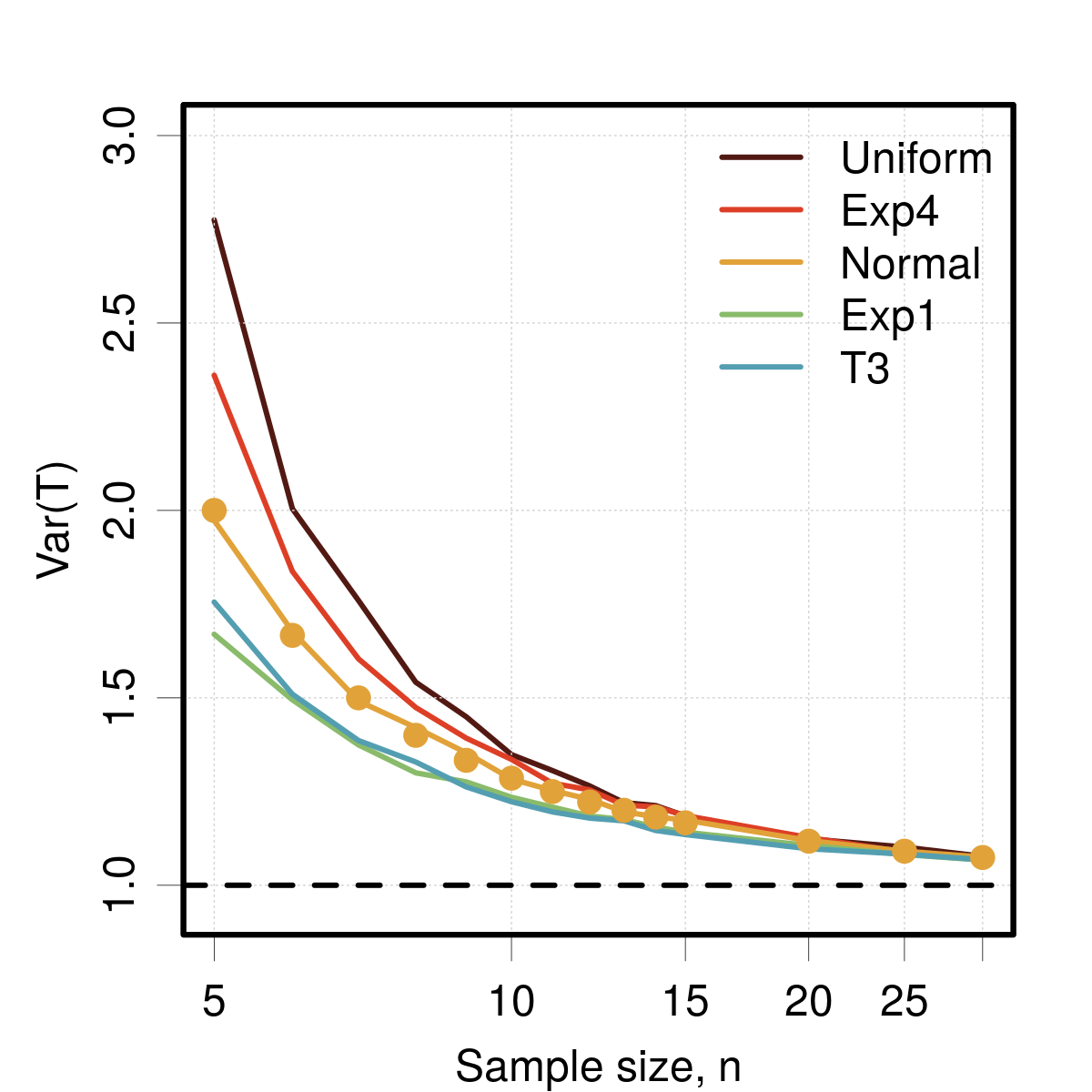}
\par\end{centering}
\caption{\label{fig:A-02}Convergence of $Var(T)$ as a function of sample
size $n$ for a set of generative error distributions. The dots mark
the $(n-1)/(n-3)$ reference points.}
\end{figure}

The impact of the generative error distribution on $\mathrm{Var}(T)$
decreases when the sample size increases (see Fig.\,\ref{fig:A-02}).
For practical purposes, one might consider the uniform and T3 distribution
as extreme cases, as the Beta(0.5,0.5) distribution, displaying a
concentration at the extremities of the variable range is not a very
plausible predictive distribution, and there is not much variation
left beyond $n=10$. One might therefore consider to test \emph{t}-scores
calibration by $\mathrm{Var}(T)\stackrel{?}{=}(n-1)/(n-3)$. For smaller
sample sizes ($n<10$), one should allow for some margin around this
value, within the limits shown in Fig.\,\ref{fig:A-02}. These limits
can be improved if information about the generative error distribution
is available. 

\section{Validation of tightness for small ensembles\label{sec:Validation-of-t-scores}}

Considering the inherent distribution of the standard uncertainties,
testing if the uncertainties provide a good scale for the errors might
be strongly perturbed, notably for heteroscedastic datasets where
it is impossible to disambiguate structural uncertainty variability
from the standard uncertainty statistical noise. 

\subsection{Homoscedastic case}

\begin{figure}[!t]
\noindent \begin{centering}
\includegraphics[height=6cm]{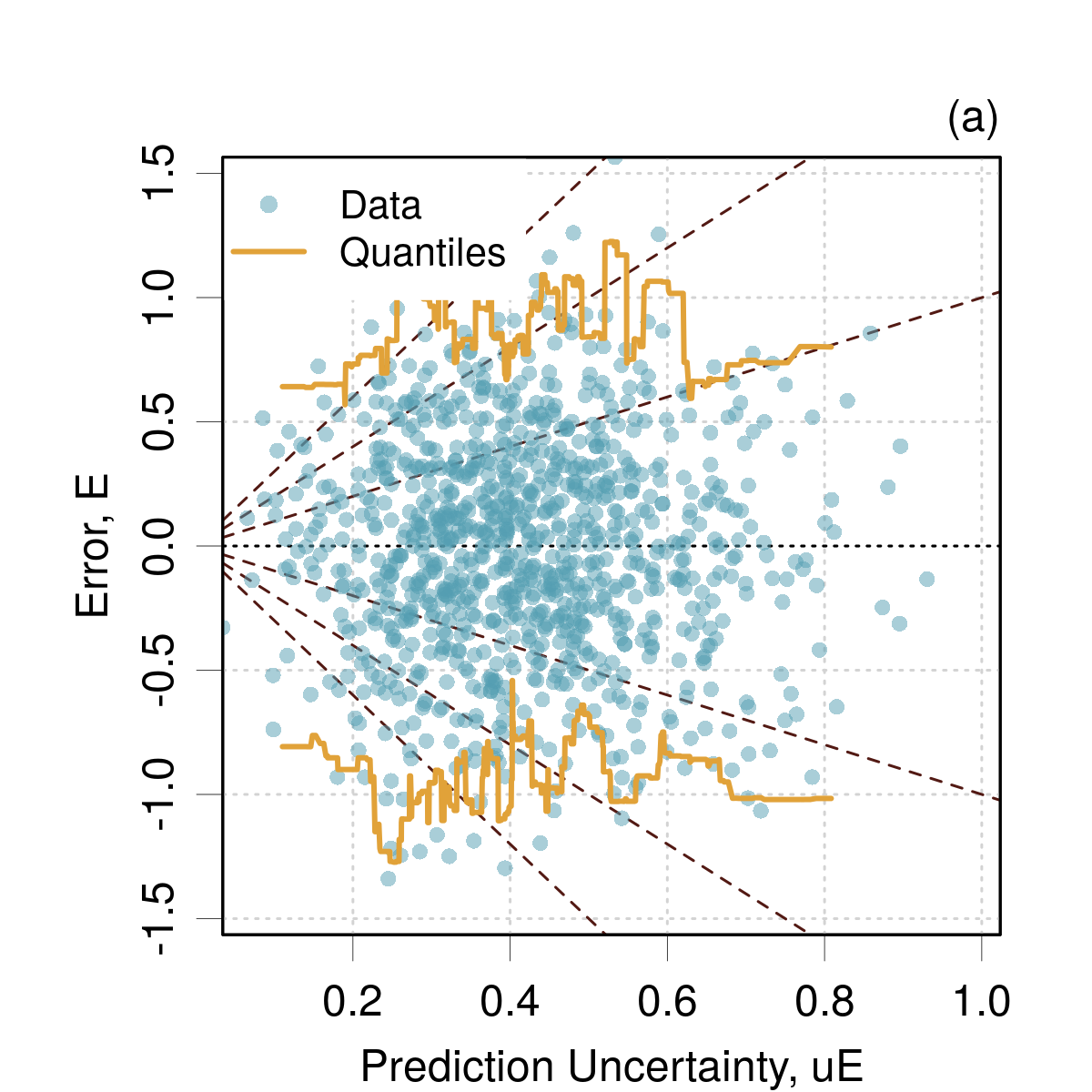}\includegraphics[height=6cm]{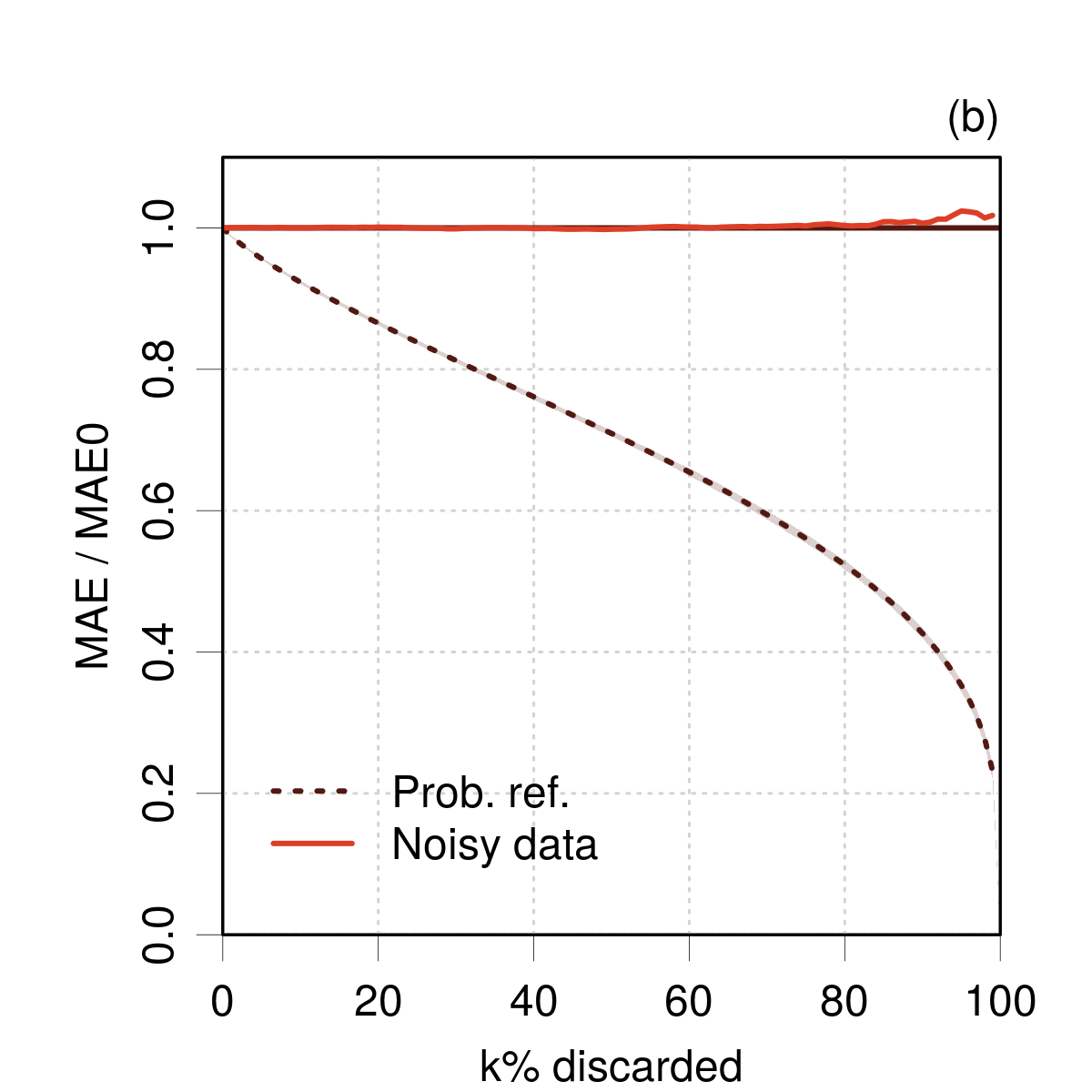}
\par\end{centering}
\noindent \begin{centering}
\includegraphics[height=6cm]{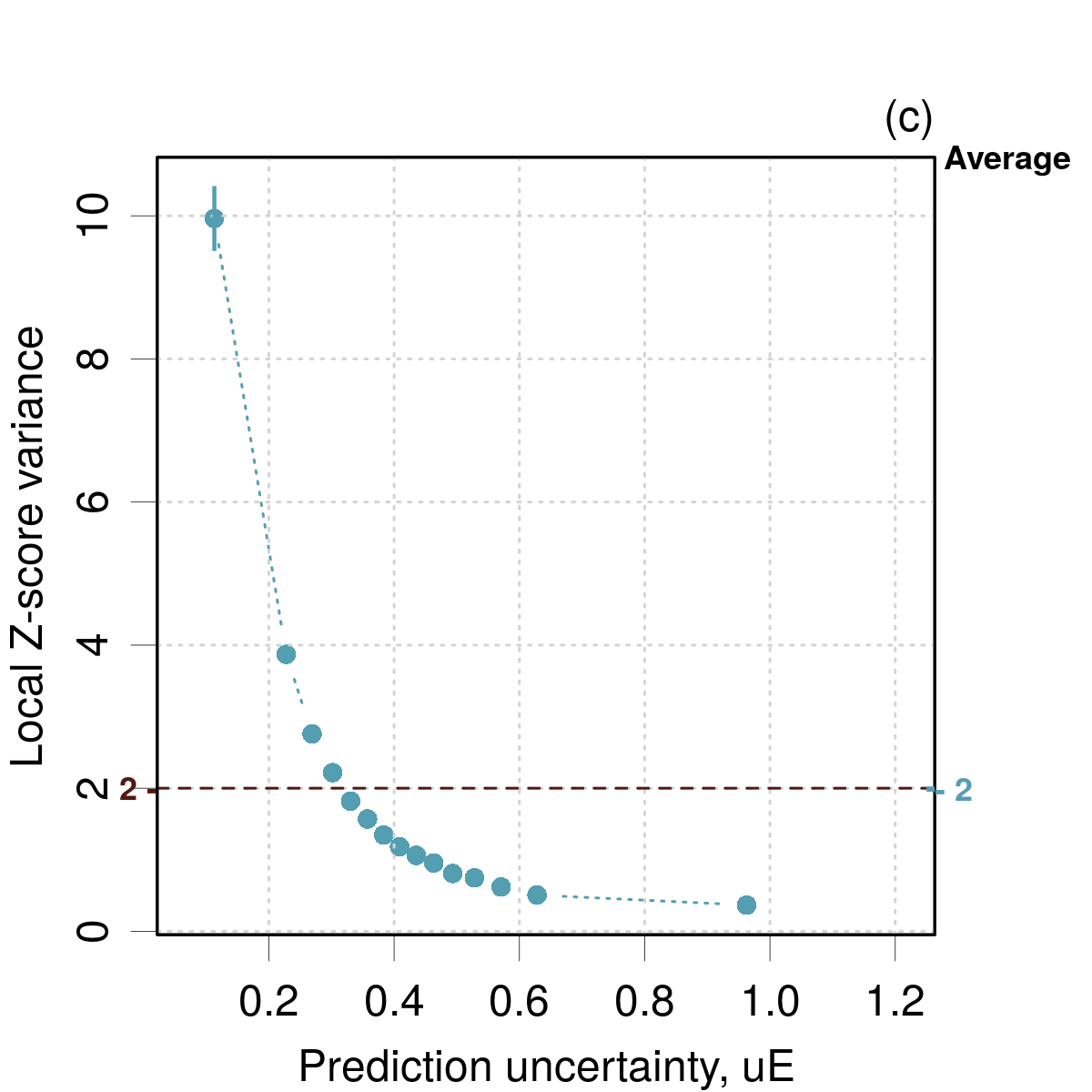}\includegraphics[height=6cm]{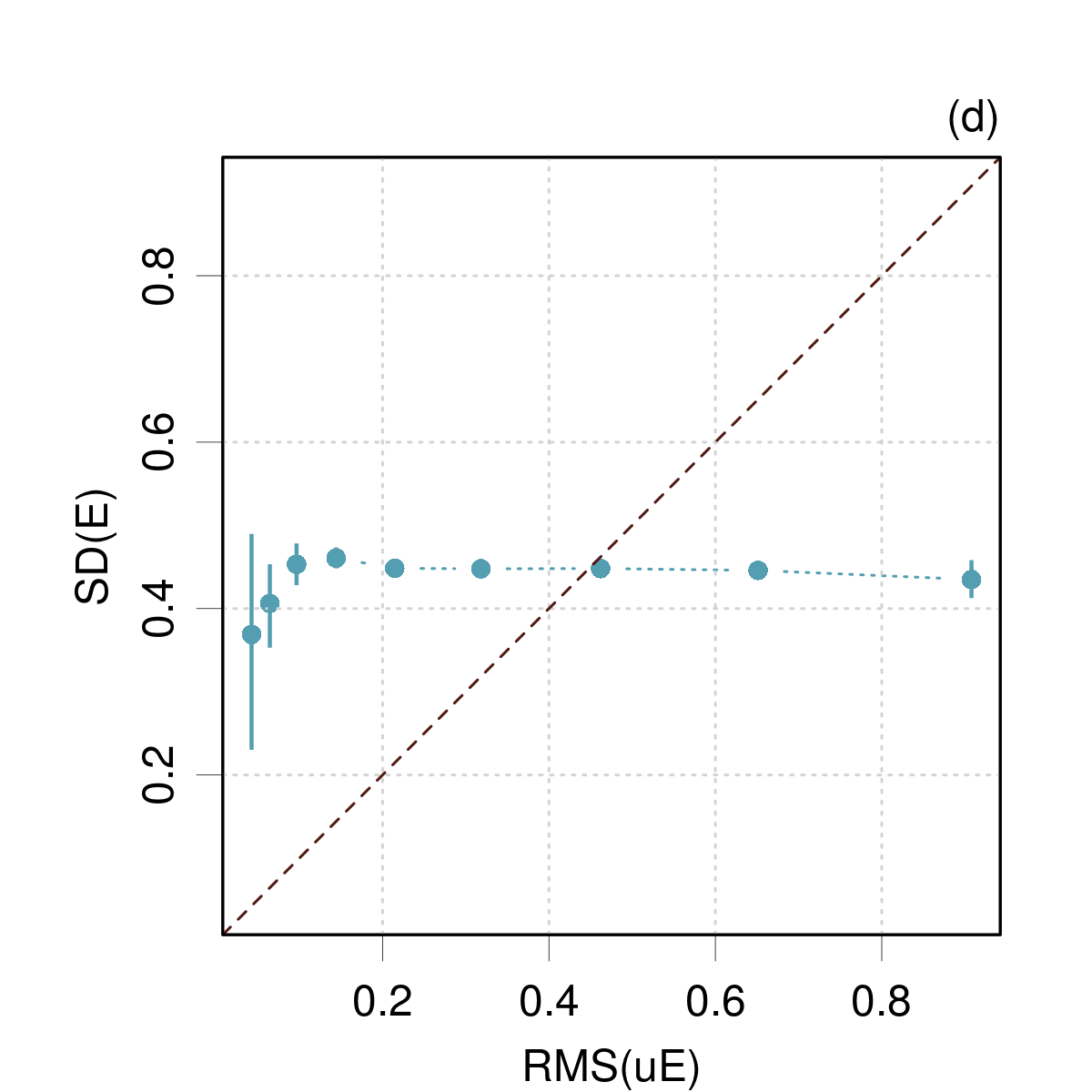}
\par\end{centering}
\caption{\label{fig:A-03}Calibration analysis for a set of sample means and
standard deviations ($n=5$) generated from a normal distribution:
(a) $(uE,E)$ plot; (b) confidence curve; (c) LZV analysis; (d) reliability
diagram. }
\end{figure}
Considering the dataset for the normal generative distribution used
in Fig.\,\ref{fig:A-01}, which theoretically corresponds to an homoscedastic
model, one can still make a $(uE,E)$ plot because of the variability
of the standard deviation {[}Fig.\,\ref{fig:A-03}(a){]}. From this
plot, one would conclude on the absence of tightness, as there is
no correlation between mean values and standard uncertainties, which
is confirmed by the flat confidence curve {[}Fig.\,\ref{fig:A-03}(b){]}.
The LZV analysis {[}Fig.\,\ref{fig:A-03}(c){]} shows that the average
calibration is indeed correct ($\mathrm{Var}(T)=2$ for $n=5$), but
the local analysis with respect to $u_{E}$ is not usable (a local
analysis wrt. the calculated value might still be useful though).
The LZV analysis is confirmed by the reliability diagram {[}Fig.\,\ref{fig:A-03}(d){]}. 

\subsection{Heteroscedastic case}

\begin{figure}[!t]
\noindent \begin{centering}
\includegraphics[height=6cm]{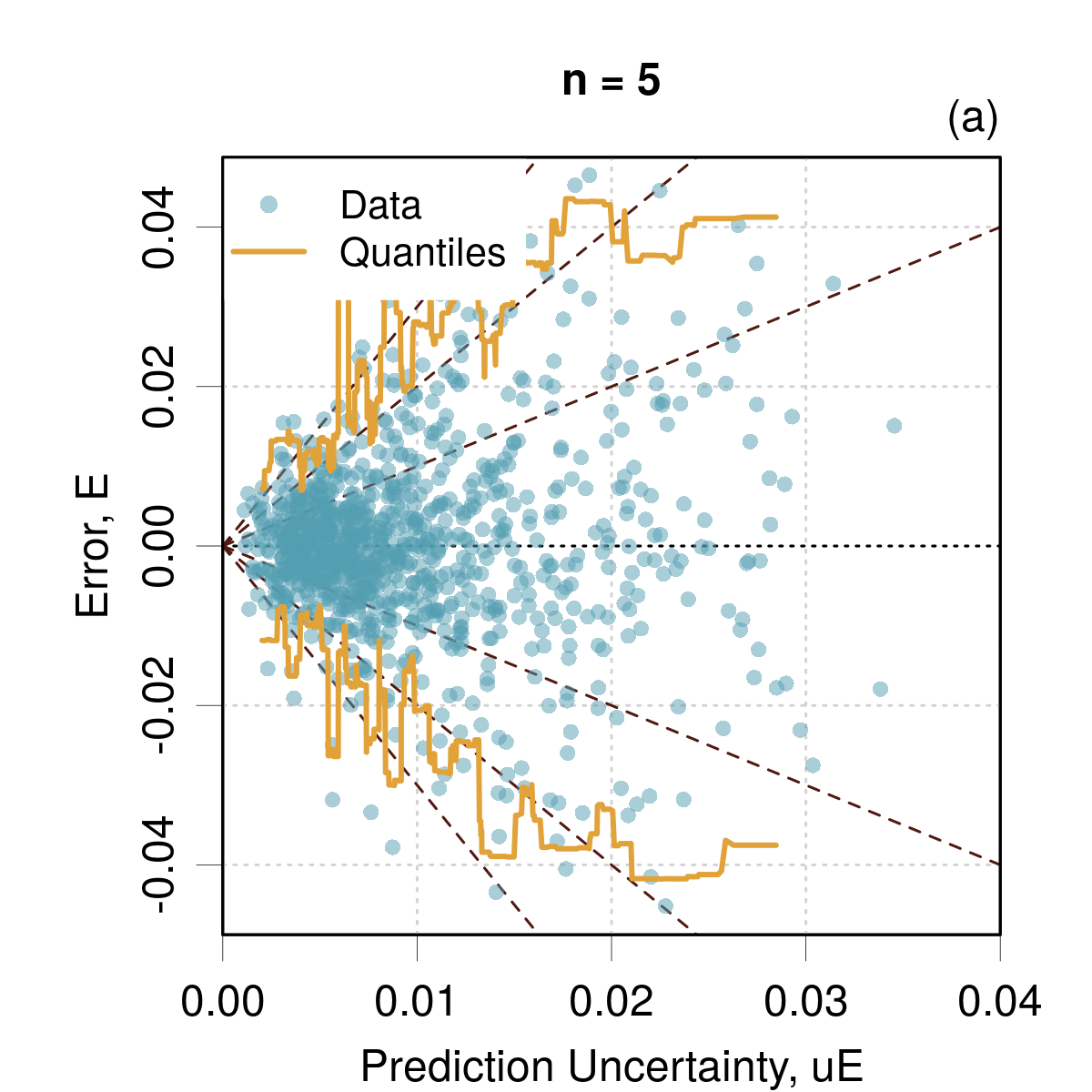}\includegraphics[height=6cm]{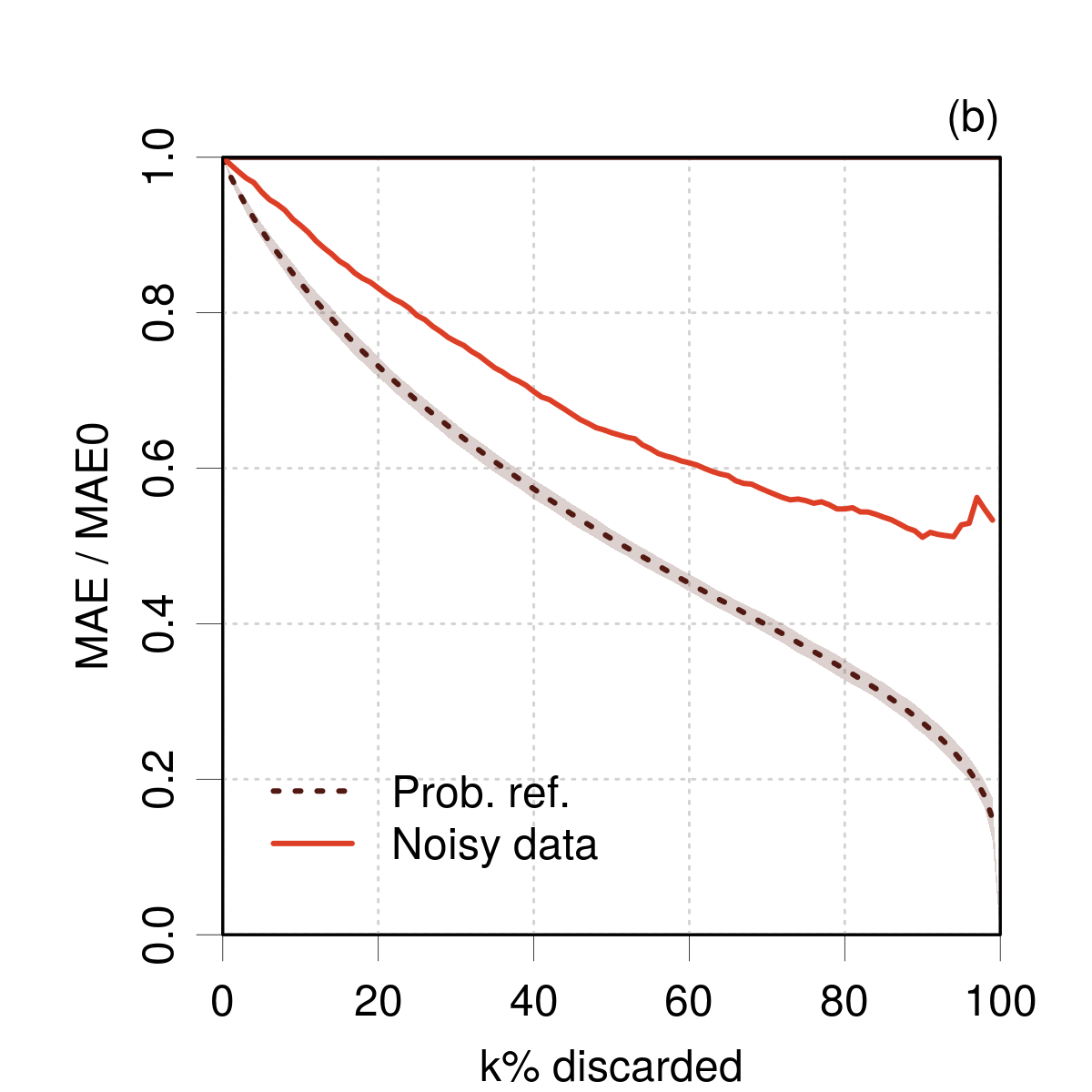}
\par\end{centering}
\noindent \begin{centering}
\includegraphics[height=6cm]{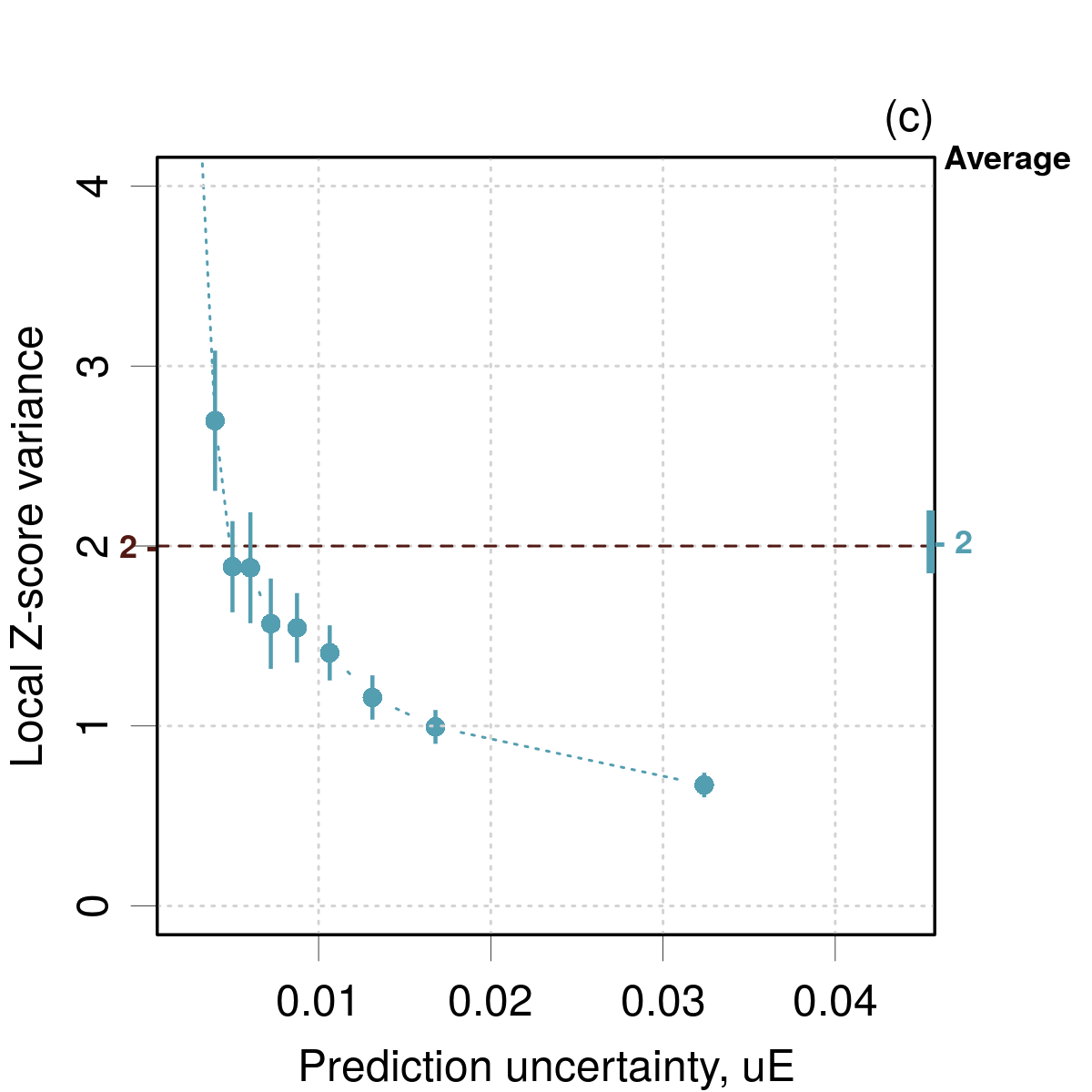}\includegraphics[height=6cm]{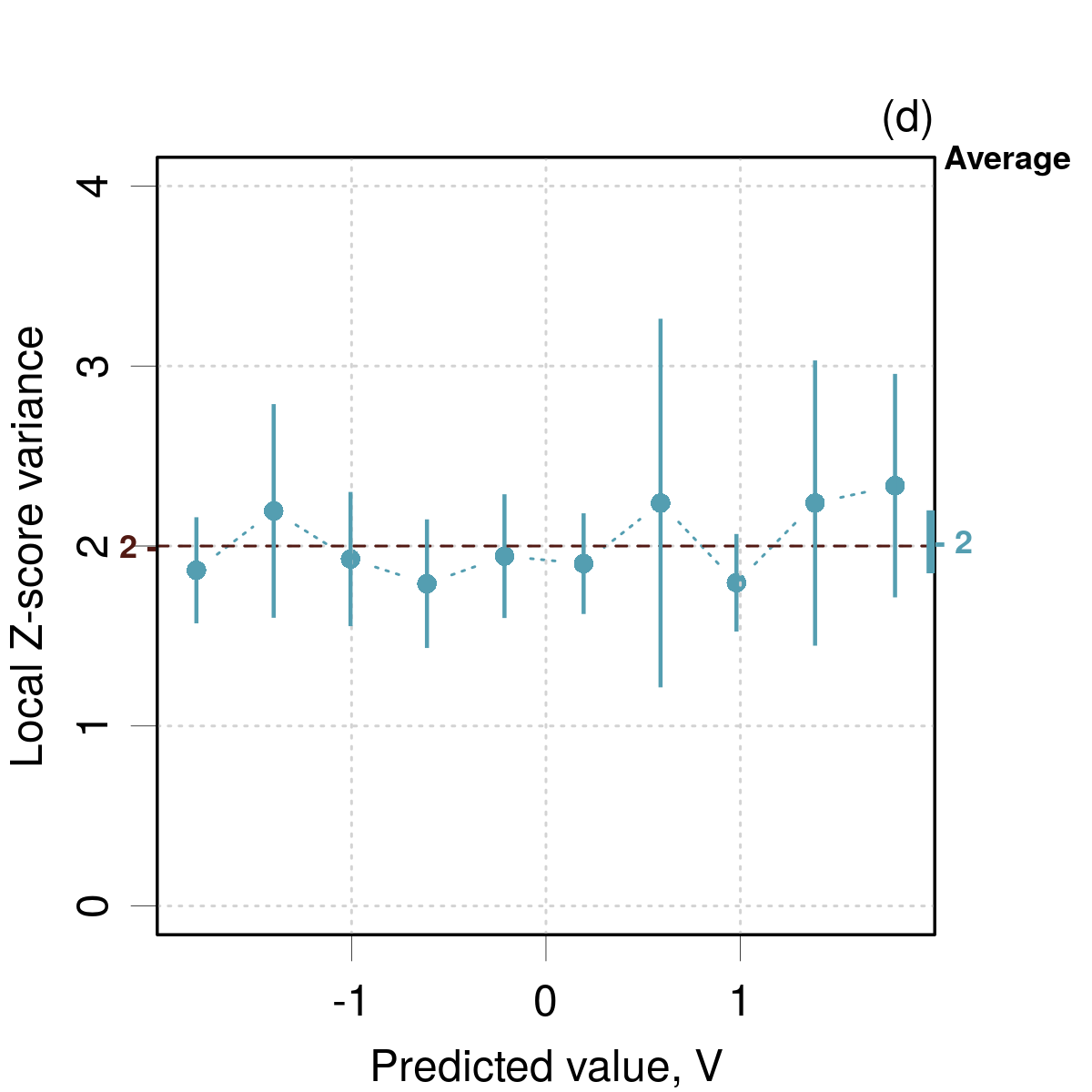}
\par\end{centering}
\caption{\label{fig:A-04}Calibration analysis for a set of sample means and
standard deviations ($n=5$) generated from the SYNT01 dataset (arbitrary
units): (a) $(uE,E)$ plot; (b) confidence curve (Noisy data) compared
to the SYNT01 dataset (Clean data); (c) LZV analysis vs $u_{E}$;
(d) LZV analysis vs $V$. }
\end{figure}
For the SYNT01 dataset, the perturbation introduced by using mean
and standard error of ensembles ($n=5$) can be appreciated in Fig.\,\ref{fig:A-04}.
Comparison of the $(uE,E)$ plot {[}Fig.\,\ref{fig:A-04}(a){]} to
the one for the initial data {[}Fig.\,\ref{fig:01}(b){]} shows that
the dataset is problematic. On the other hand, the confidence curve
is continuously decreasing {[}Fig.\,\ref{fig:A-04}(b){]}, although
it does not match the probabilistic reference. As for the purely noisy
example above, the LZV analysis wrt. $u_{E}$ reveals a correct average
calibration but fails at the local tests {[}Fig.\,\ref{fig:A-04}(c){]}.
However, a LZV analysis wrt. $V$ does not reveal any problem {[}Fig.\,\ref{fig:A-04}(d){]}. 

\begin{figure}[t]
\noindent \begin{centering}
\includegraphics[height=6cm]{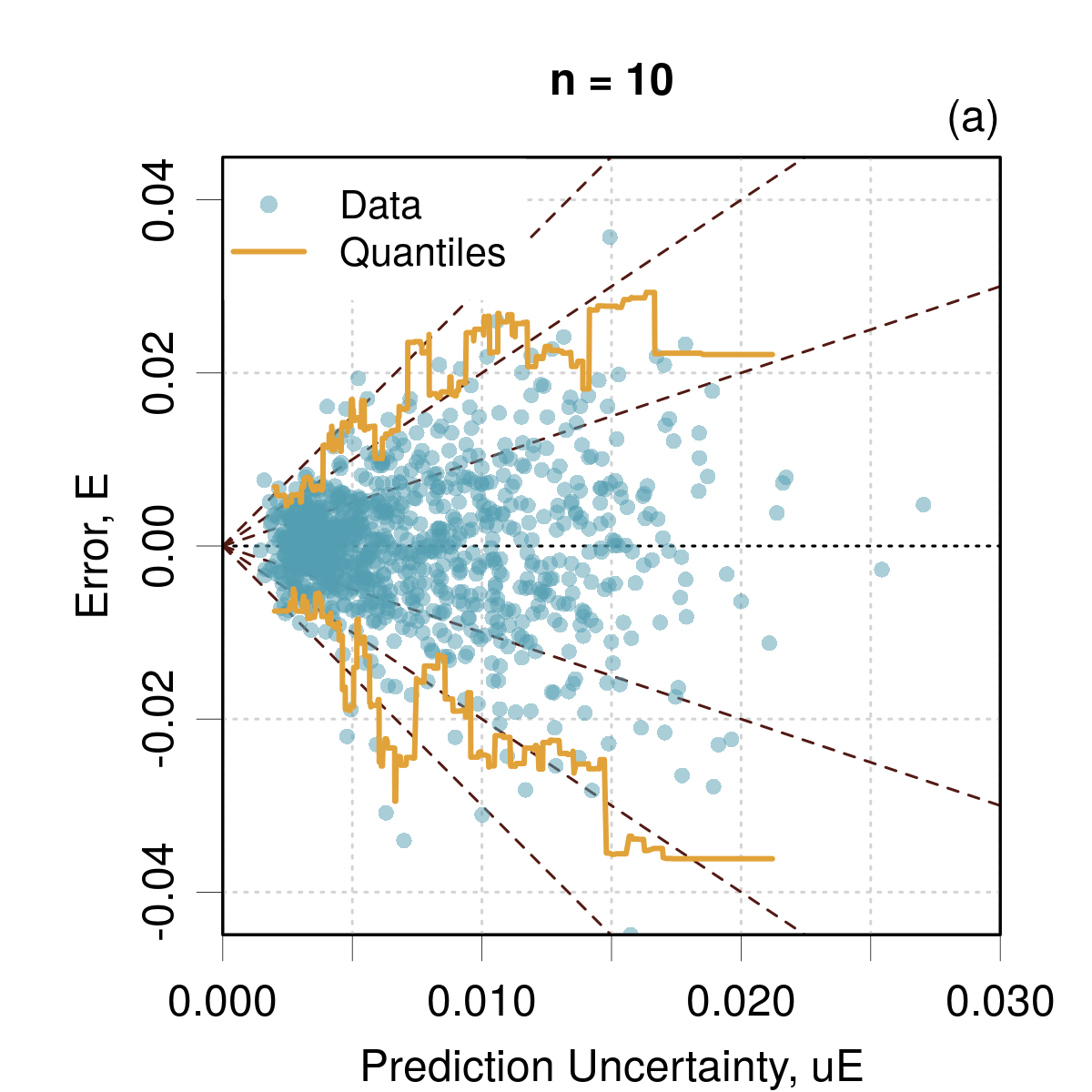}\includegraphics[height=6cm]{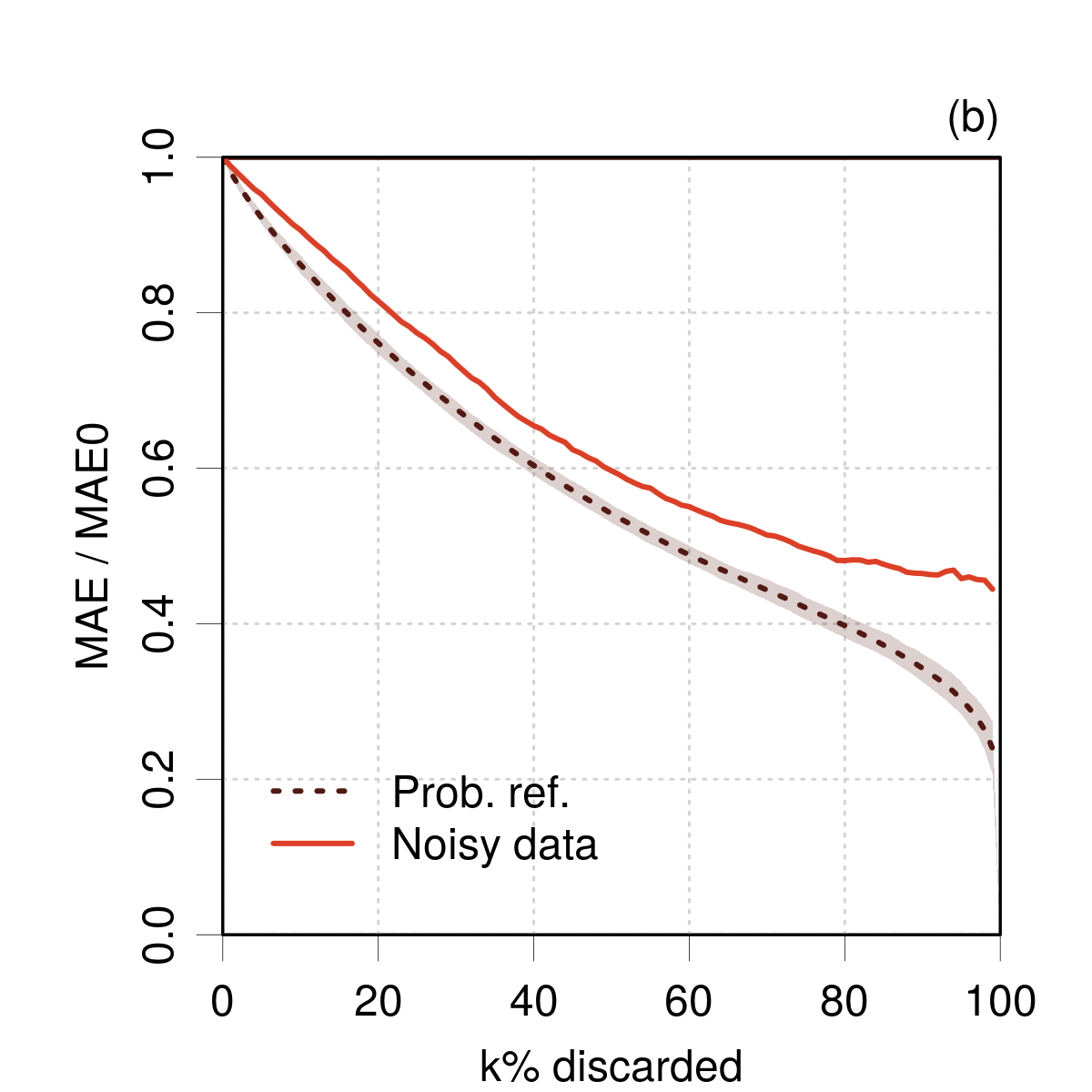}
\par\end{centering}
\noindent \begin{centering}
\includegraphics[height=6cm]{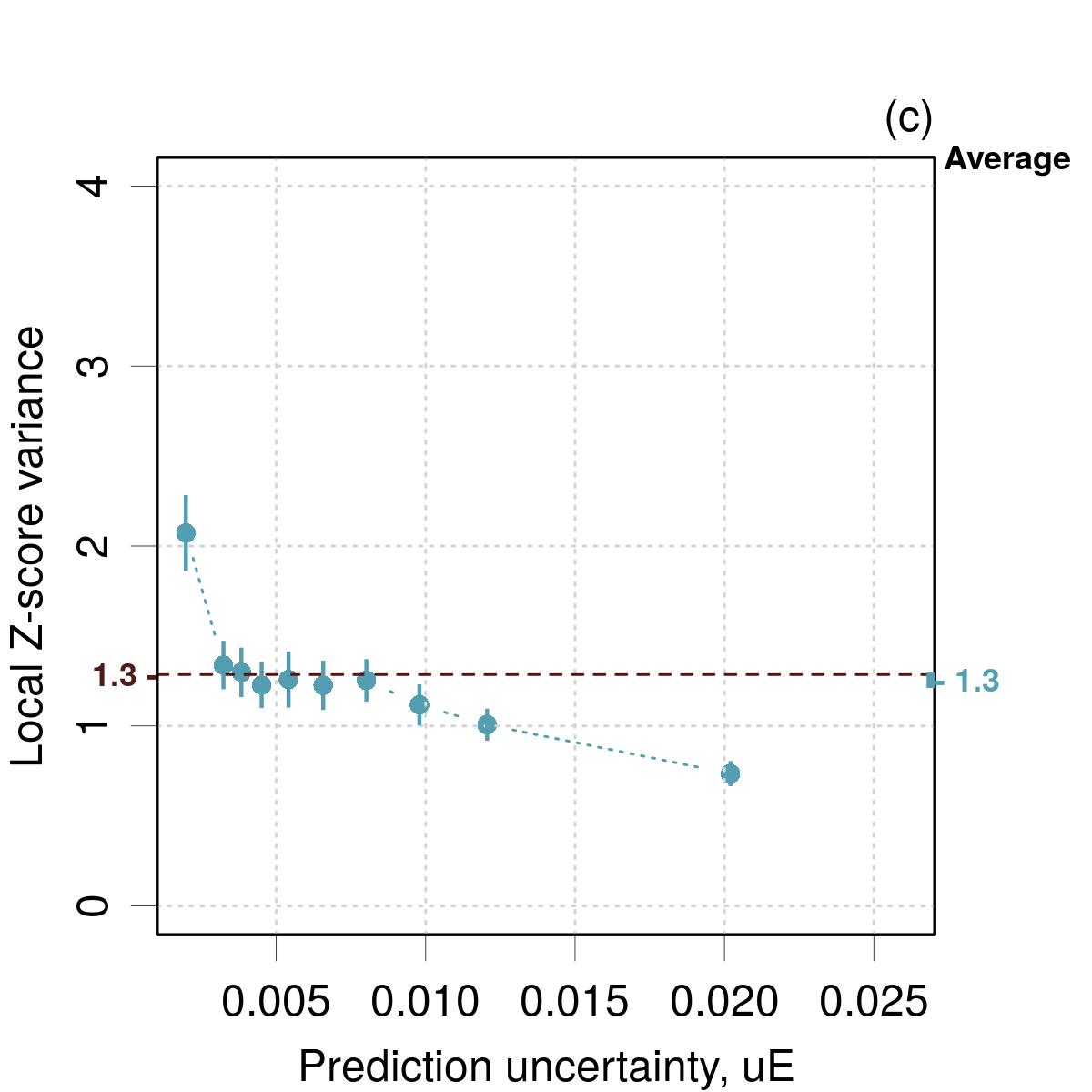}\includegraphics[height=6cm]{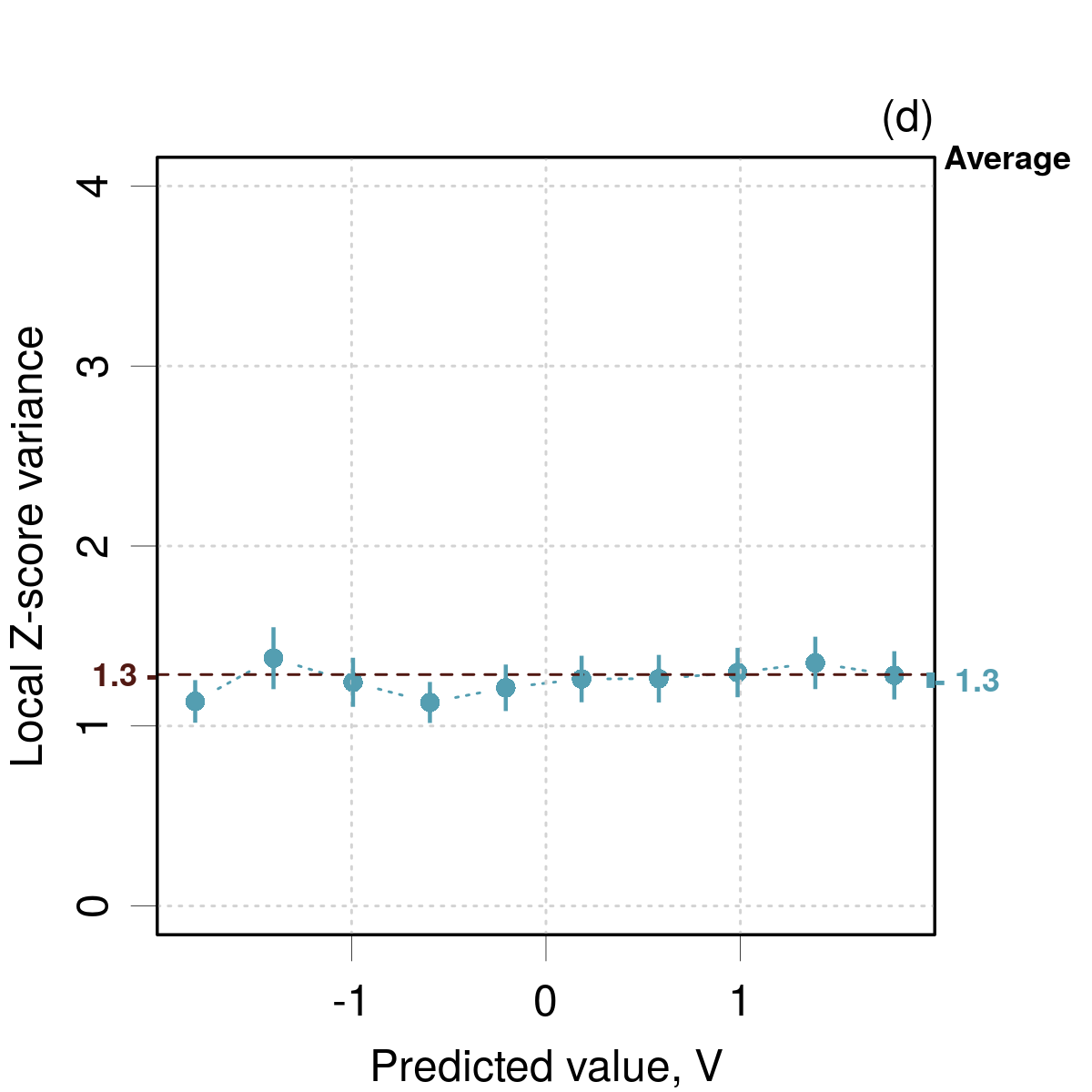}
\par\end{centering}
\caption{\label{fig:A-05}Same as Fig.\,\ref{fig:A-04} for $n=10$.}
\end{figure}

The situation for $n=10$ is slightly improved, however, the confidence
curve and the LZV analysis wrt. $u_{E}$ would still lead to reject
tightness {[}Fig.\,\ref{fig:A-05}{]}. 

\end{document}